\documentclass{aa}

\usepackage{newtxtext,newtxmath}
\usepackage[T1]{fontenc}

\DeclareRobustCommand{\VAN}[3]{#2}
\let\VANthebibliography\thebibliography
\def\thebibliography{\DeclareRobustCommand{\VAN}[3]{##3}\VANthebibliography}

\usepackage{graphicx}	% Including figure files
\usepackage{amsmath}	% Advanced maths commands

\usepackage{etoolbox}
\BeforeBeginEnvironment{tabular}{\begin{center}\footnotesize}
\AfterEndEnvironment{tabular}{\end{center}}
\newcommand\ions[2]{#1$\;${\textsc{#2}}}
\usepackage{xcolor,hyperref}

\newcommand{\DELS}{DELS J0411$-$0907}
\newcommand{\VDES}{VDES J0020$-$3653}
\newcommand{\hb}{H$\beta$}
\newcommand{\ha}{H$\alpha$}

\newcommand{\oiii}{[\ions{O}{iii}] $\lambda\lambda4959,5007$}
\newcommand{\oiiib}{[\ions{O}{iii}] $\lambda4959$}
\newcommand{\oiiia}{[\ions{O}{iii}] $\lambda5007$}
\newcommand{\nii}{[\ions{N}{ii}] ${\lambda\lambda6548,6583}$}
\newcommand{\niia}{[\ions{N}{ii}] ${\lambda6583}$}

\makeatletter
\renewcommand*\aa@pageof{, page \thepage{} of \pageref*{LastPage}}
\makeatother

\begin{document}

   \title{GA-NIFS: Black hole and host galaxy properties of two $z\simeq6.8$ quasars from the NIRSpec IFU}
   \subtitle{}
%\label{firstpage}
%\maketitle
\author{Madeline A. Marshall
\inst{\ref{MM1},\ref{MM2}} \and
Michele Perna\inst{\ref{SA}} \and Chris J. Willott \inst{\ref{MM1}} \and  Roberto Maiolino \inst{\ref{RM1},\ref{RM2},\ref{RM3}} \and Jan Scholtz\inst{\ref{RM1},\ref{RM2}} \and Hannah \"Ubler\inst{\ref{RM1},\ref{RM2}} \and Stefano Carniani\inst{\ref{pisa}} \and Santiago Arribas \inst{\ref{SA}} 
\and Nora L\"utzgendorf\inst{\ref{TB}} 
\and Andrew J. Bunker \inst{\ref{AB}} \and Stephane Charlot \inst{\ref{SC}} 
\and Pierre Ferruit\inst{\ref{PF}} 
\and  Peter Jakobsen\inst{\ref{PJ1},\ref{PJ2}} 
\and Hans-Walter Rix\inst{\ref{HWR}} 
\and Bruno Rodr\'iguez Del Pino\inst{\ref{SA}} \and
Torsten B\"{o}ker\inst{\ref{TB}}, Alex J. Cameron\inst{\ref{AB}} \and Giovanni Cresci\inst{\ref{INAF}} \and Emma Curtis-Lake\inst{\ref{ECL}}  \and Gareth C. Jones\inst{\ref{AB}}  \and Nimisha Kumari\inst{\ref{NK}} \and 
Pablo G. P\'erez-Gonz\'alez\inst{\ref{SA}}
\and  
Sophie L. Reed\inst{\ref{SR}}
}

\institute{% List of institutions
National Research Council of Canada, Herzberg Astronomy \& Astrophysics Research Centre, 5071 west Saanich Road, Victoria, BC V9E 2E7, Canada\label{MM1} \and
ARC Centre of Excellence for All Sky Astrophysics in 3 Dimensions (ASTRO 3D), Australia\label{MM2} \and
Centro de Astrobiolog\'{\i}a (CAB), CSIC-INTA, Ctra. de Ajalvir km 4, Torrej\'on de Ardoz, E-28850, Madrid, Spain \label{SA}\and
Kavli Institute for Cosmology, University of Cambridge, Madingley Road, Cambridge, CB3 0HA, UK\label{RM1}\and
Cavendish Laboratory - Astrophysics Group, University of Cambridge, 19 JJ Thomson Avenue, Cambridge, CB3 0HE, UK\label{RM2}\and
Department of Physics and Astronomy, University College London, Gower Street, London WC1E 6BT, UK\label{RM3}\and
Scuola Normale Superiore, Piazza dei Cavalieri 7, I-56126 Pisa, Italy\label{pisa}\and
European Space Agency, c/o STScI, 3700 San Martin Drive, Baltimore, MD 21218, USA\label{TB}\and
Department of Physics, University of Oxford, Keble Road, Oxford OX1 3RH, UK\label{AB}
\and
Sorbonne Universit\'e, CNRS, UMR 7095, Institut d’Astrophysique de Paris, 98 bis bd Arago, 75014 Paris, France\label{SC}
\and
European Space Agency, ESAC, Villanueva de la Ca\~{n}ada, E-28692 Madrid, Spain\label{PF}\and
Cosmic Dawn Center (DAWN), Copenhagen, Denmark \label{PJ1} \and Niels Bohr Institute, University of Copenhagen, Jagtvej 128, DK-2200, Copenhagen, Denmark \label{PJ2}
\and
Max Planck Institute for Astronomy, Königstuhl 17, 69117 Heidelberg, Germany\label{HWR}
\and
INAF - Osservatorio Astrofisco di Arcetri, largo E. Fermi 5, 50127 Firenze, Italy\label{INAF}\and
Centre for Astrophysics Research, Department of Physics, Astronomy and Mathematics, University of Hertfordshire, Hatfield, AL10 9AB, UK\label{ECL}
\and
AURA for the European Space Agency, Space Telescope Science Institute, Baltimore, Maryland, USA\label{NK}
\and Department of Astrophysical Sciences, Princeton University, 4 Ivy Lane, Princeton, NJ 08544, USA\label{SR}
}
\authorrunning{Marshall et al.}

\abstract
{
{}
{Integral Field Spectroscopy (IFS) with JWST NIRSpec will significantly improve our understanding of the first quasars, by providing spatially resolved, infrared spectroscopic capabilities which cover key rest-frame optical emission lines that have been previously unobservable.}
{Here we present our results from the first two $z>6$ quasars observed as a part of the Galaxy Assembly with NIRSpec IFS (GA-NIFS) GTO program, \DELS\ at $z=6.82$ and \VDES\ at $z=6.86$.}
{By observing the \hb, \oiii, and \ha\ emission lines in these high-$z$ quasars for the first time, we measure accurate black hole masses, 
$M_{\rm{BH}}=1.85\substack{ +2\\-0.8}\times10^9\rm{M}_\odot$ and $2.9\substack{ +3.5\\-1.3}\times10^9\rm{M}_\odot$, 
corresponding to Eddington ratios of $\lambda_{\rm{Edd}}=0.8\substack{ +0.7\\-0.4}$ and $0.4\substack{ +0.3\\-0.2}$ for \DELS\ and \VDES\ respectively. These provide a key comparison for existing estimates from the more uncertain \ion{Mg}{ii} line.
 We perform quasar--host decomposition using models of the quasars’ broad lines, to measure the underlying host galaxies. We also discover multiple emission line regions surrounding each of the host galaxies, which are likely companion galaxies undergoing mergers with these hosts.
We measure the star formation rates, excitation mechanisms, and dynamical masses of the hosts and companions, measuring the $M_{\rm{BH}}/M_{\rm{dyn}}$ ratios at high-$z$ using these estimators for the first time. 
\DELS\ and \VDES\ both lie above the local black hole--host mass relation, and are consistent with the existing observations of $z\gtrsim6$ quasar host galaxies with ALMA. 
We detect ionized outflows in \oiii\ and \hb\ from both quasars, with mass outflow rates of $58\substack{ +44\\-37 }$ and $525\substack{ +75\\-92 }$~$\rm{M}_{\odot}~yr^{-1}$ for \DELS\ and \VDES, much larger than their host star formation rates of $<33$ and  $<54~ \rm{M}_\odot$/yr, respectively.}
   {This work highlights the exceptional capabilities of the JWST NIRSpec IFU for observing quasars in the early Universe.}
}


% These dates will be filled out by the publisher
\date{Accepted XXX. Received YYY; in original form ZZZ}

\maketitle

%%%%%%%%%%%%%%%%%%%%%%%%%%%%%%%%%%%%%%%%%%%%%%%%%%

%%%%%%%%%%%%%%%%% BODY OF PAPER %%%%%%%%%%%%%%%%%%

\section{Introduction}

Observational surveys over the last two decades have revealed several hundreds of high-$z$ quasars \citep[$z\gtrsim6$; e.g.][]{fan_2000,fan_2001,Fan2003,willott_2009,Willott2010,Kashikawa2015,Banados2016,Banados2017,Banados2022,Matsuoka2018,Wang2019,Yang2023}, less than a billion years into the Universe's history. The black holes powering these quasars have been measured to be as massive as $\sim10^{10} \rm{M}_\odot$ \citep{wu_2015}, raising questions about the nature of the first black hole seeds and their extreme growth mechanisms.  
Theories to explain the growth of these first quasars generally require massive black hole seeds, sustained accretion at the Eddington limit, or phases of accretion at super-Eddington rates \citep[e.g.][]{Volonteri2005,Volonteri2015,Sijacki2009}. 
Indeed, some quasars have been measured to have accretion rates above the Eddington limit \citep[e.g.][]{Banados2017,Pons2019a,Farina2022}, although when considering the uncertain black hole mass measurements and large scatter in the adopted scaling relations, \citet{Farina2022} found no clear evidence for a significant population of super-Eddington high-$z$ quasars.
Thus accurate measurements of the black hole mass and accretion rate for high-$z$ quasars provide vital constraints for these theories.

Current high-$z$ quasar black hole mass measurements rely on calibrations that can be quite uncertain.
The \ha\ and \hb\ lines are generally considered the most reliable single-epoch indicators of black hole mass, as they are used in many reverberation mapping studies \citep[e.g.][]{Wandel1999,Kaspi2000}. However, these reverberation mapping studies are performed for low-$z$, low-luminosity active galactic nuclei (AGN). At higher-$z$, and for luminous quasars, extrapolations are required, and so at high-$z$ even these best estimates are uncertain. More critically, for high-$z$ quasars these lines are redshifted into infrared wavelengths $>2.5\mu$m, beyond the range observable from the ground (for sources at $z\gtrsim4$ for \hb\ and $z\gtrsim3$ for \ha).
Thus black hole mass measurements for high-$z$ quasars instead have typically relied on the \ion{Mg}{ii} and \ion{C}{iv} lines.
Reverberation mapping studies based on \ion{Mg}{ii} and \ion{C}{iv} are more uncertain \citep[e.g.][]{Trevese2014,Kaspi2021}, and the resulting scaling relations required to obtain single-epoch black hole mass estimates are affected by luminosity-dependent biases and provide highly uncertain mass measurements, for example due to significant blueshifts caused by outflows \citep[e.g.][and references within]{shen2008,Shen2012,Peterson2009,Coatman2016,Farina2022}. With the James Webb Space Telescope \citep[JWST;][]{Gardner2006,Gardner2023,McElwain2023} and the Near-Infrared Spectrograph \citep[NIRSpec;][]{Jakobsen2022,Boeker2023}, \hb\ is now observable up to $z=9.8$, and \ha\ up to $z=7.0$ (and beyond with the Mid-Infrared Instrument, MIRI), allowing us to more accurately measure the black hole masses and Eddington ratios of high-$z$ quasars with the same estimator at all redshifts \citep[e.g.][]{Eilers2022,Yang2023a}. 
This will improve our understanding of the first black holes and their growth rates by answering questions such as: are these black holes truly so massive, and are they accreting above the Eddington limit? 

Significant questions also exist surrounding the nature of the galaxies in which these quasars reside. These hosts can be measured in the sub-mm, tracing their cold gas and dust, which reveals a varying population of objects with a wide range of dynamical masses ($10^{10}$--$10^{11}M_\odot$), dust masses ($10^7$--$10^9M_\odot$) and sizes (<1--5 kpc), and star formation rates \citep[SFRs, $10$--$2700 M_\odot/$yr; e.g.][]{Bertoldi2003,Walter2003,Walter2004,Riechers2007,Wang2010,Wang2011,Venemans2015,Venemans2017a,Willott2017,Trakhtenbrot2017,Izumi2018,Neeleman2021,Walter2022}.
However, their host galaxies have not been observed in the rest-frame UV/optical even with the resolution of the Hubble Space Telescope \citep[HST;][]{Mechtley2012,Decarli2012,McGreer2014,Marshall2019c}, due to the intense emission from the quasar concealing the small, fainter galaxy underneath  \citep[e.g.,][]{schmidt_1963,mcleod_1994,dunlop_2003,hutchings_2003}. 

The JWST not only allows us to access the near infrared spectral range with unprecedented sensitivity, but also provides high angular resolution and a stable point-spread function (PSF).
With these improvements over HST, simulations predict that the host galaxies of these high-$z$ quasars will be detectable for the first time \citep{MarshallBTpsfMC}, and indeed the first signs of host emission have been detected with the Near-Infrared Camera \citep[NIRCam;][]{Ding2022}.
Thus JWST and the NIRSpec Integral Field Unit \citep[IFU;][]{Boeker2022} will permit the characterization of the spatially resolved spectral properties of the hosts, measuring their stellar properties, interstellar medium (ISM) structure and kinematics, excitation mechanisms, and stellar-based SFRs. This will significantly advance our understanding of quasars, their host galaxies, and the rapid growth of early supermassive black holes.

Based on JWST and NIRSpec IFU observations, in this paper we use the standard \ha\ and \hb\ single-epoch black hole mass estimators to derive accurate black hole mass measurements for two $z>6$ quasars, \DELS\ and \VDES.
\DELS\ (VHS J0411-0907) at $z=6.82$ was discovered independently by \citet{Pons2019a} and \citet{Wang2019}. \DELS\ is the most distant of the Pan-STARRS quasars, and is the brightest quasar in the J-band above $z=6.7$ \citep{Pons2019a}.
With a relatively low \ions{Mg}{ii} black hole mass of $(6.13\pm0.51)\times10^8 \rm{M}_\odot$, and a bolometric luminosity of $(1.89\pm0.07)\times10^{47} \rm{erg~ s} ^{-1}$, \citet{Pons2019a} measured an Eddington ratio of $2.37\pm0.22$, implying that this quasar could be a highly super-Eddington quasar undergoing extreme accretion. 
The quasar \VDES\ at $z=6.86$ was discovered by \citet{Reed2019}.
\VDES\ has a larger \ions{Mg}{ii} black hole mass than \DELS, measured as $(16.7\pm3.2)\times10^8 \rm{M}_\odot$ by \citet{Reed2019}, and rederived as $25.1\times10^8 \rm{M}_\odot$ by \citet{Farina2022} using the \citet{Shen2011} \ions{Mg}{ii}--black hole mass relation. With a bolometric luminosity of $L_{\rm{Bol}} = (1.35\pm0.03) \times 10^{47}\rm{erg~s}^{-1}$, these correspond to Eddington ratios of $0.62\pm0.12$ and 0.43 respectively, suggesting that this quasar is powered by a larger supermassive black hole with less extreme accretion than \DELS.
However, these black hole mass and accretion measurements are based on the \ion{Mg}{ii} line; robust \ha\ and \hb\ mass estimates will better constrain the accretion of these quasars.

This paper is outlined as follows. In Section \ref{sec:data} we describe the JWST NIRSpec observations, and our data reduction and analysis techniques. In Section \ref{sec:results} we present our results, with the black hole properties in Section \ref{sec:BHproperties}, the host galaxy properties in Section \ref{sec:HostProperties}, and the outflow properties in Section \ref{sec:outflows}. We present a discussion in Section \ref{sec:discussion}, before concluding with a summary of our findings in Section \ref{sec:conclusions}.

Throughout this work we adopt the WMAP9 cosmology \citep{Hinshaw2013} as included in \textsc{AstroPy} \citep{Astropy2013}, with $H_0=69.32$ km / (Mpc s), $\Omega_m=0.2865$, and $\Omega_\Lambda=0.7134$.

\section{Data, Reduction, and Analysis}
\label{sec:data}
\subsection{Observations}

\DELS\ and \VDES\ were observed as part of the Galaxy Assembly with NIRSpec Integral Field Spectroscopy (GA-NIFS)  Guaranteed Time Observations (GTO) program, contained within program \#1222 (PI Willott).
This work focuses on the observations of these two quasars with the NIRSpec IFU \citep{Boeker2022}, which provides spatially resolved imaging spectroscopy over a 3'' × 3'' field of view with 0.1'' × 0.1'' spatial elements.
The IFU observations were taken with a grating/filter pair of G395H/F290LP. This results in a spectral cube with spectral resolution $R\sim2700$ over the wavelength range 2.87--5.27$\mu$m.
The observations were taken with a NRSIRS2 readout pattern with 25 groups, using a 6-point medium cycling dither pattern, resulting in a total exposure time of 11029 seconds.
We constrained the complete position angle (PA) window available for observing these targets with JWST to a more limited PA range, to minimise the leakage of light through the micro-shutter assembly (MSA) from bright sources. 
Additionally, these quasars have been observed with NIRSpec fixed-slit spectroscopy at shorter wavelengths (with G140H /F070LP, G235H/F170LP), which will be analysed in a separate paper. 

\DELS\ was observed on Oct 1, 2022, at a PA of 61.893 degrees. \VDES\ was observed on Oct 16, 2022, at a PA of 160.169 degrees. Unfortunately, only three out of the six planned dithers were observed due to a spacecraft guiding failure. The final three dithers were observed on Nov 17, 2022, at a position angle of 188.177. 
With half of the \VDES\ observations observed at a different PA, combining these exposures and their subsequent data analysis was more challenging than the \DELS\ observations. 
Due to an error in the reference files of the pipeline \citep[see][for details]{Perna2023}, we performed a bona fide astrometric registration matching the peak pixel locations of the \ha\ BLR collapsed images. We applied a constant offset in RA and Dec to the pipeline Stage 1 products of the second set of observations, which allowed the pipeline to correctly combine all six dithers in a final data cube.

Due to an uncertainty in the specified astrometry of our observations, we assume that the quasar is centred on the known positions from the literature:
04h 11m 28.63s, $-$09d07m49.8s for \DELS\ \citep{Wang2019} and
00h 20m 31.472s, $-$36d53m41.82s for \VDES\ \citep{Reed2019}. 
These were the targeted positions for our IFU observations, which have been found to have a typical pointing accuracy of $\sim0.1''$ compared to the requested pointing \citep{Rigby2022}.

\subsection{Data Reduction}

The data reduction was made with the JWST calibration pipeline version 1.8.2 \citep{Bushouse2022}, using the context file {\sc jwst\_1023.pmap}. All of the individual raw images were first processed for detector-level corrections using the {\sc Detector1Pipeline} module of the pipeline (Stage 1 hereinafter). The individual products (count-rate images) were then calibrated through {\sc Calwebb\_spec2} (Stage 2 hereinafter), where WCS-calibration, flat-fielding, and the flux-calibrations are applied to convert the data from units of count rate to flux density. The individual Stage 2 images were then resampled and coadded onto a final data cube through the {\sc Calwebb\_spec3} processing (Stage 3 hereinafter). A number of additional steps and corrections in the pipeline code have been applied to improve the data reduction quality. In particular, 

\begin{itemize}
    
    \item The individual count-rate frames were further processed at the end of Stage 1, to correct for a different zero-level in the dithered frames. For each image, we therefore subtracted the median value, computed considering the entire image, to get a base-level consistent with zero counts per second.
    
    \item We further processed these count-rates to subtract the $1/f$ noise \citep{Kashino2022}. This correlated vertical noise is modelled in each column (i.e. along the spatial axis)  with a low-order polynomial function, after removing all bright pixels (e.g. associated with the observed target) with a sigma-clipping. The modelled $1/f$ noise is then subtracted before proceeding with the Stage 2 of the pipeline.  

    \item The {\sc outlier\_detection} step of the third and last step of the pipeline is required to identify and flag all remaining cosmic-rays and other artefacts that result in a significant number of spikes in the reduced data. 
    In pipeline version 1.8.2, this step tends to identify too many false positives that compromises the data quality significantly. We therefore implement a modified version of the {\sc lacosmic} algorithm \citep{Dokkum2001} in the JWST pipeline, which is applied at the end of Stage 2 (details in D'Eugenio et al. in prep). 

    \item We applied the {\sc cube\_build} step to produce combined data cubes with a spaxel size of 0.1\arcsec, using the Exponential Modified Shepard Method (EMSM) weighting. This provides high signal-to-noise at the spaxel level yet slightly lower spatial resolution with respect to other weightings; the loss in resolution is compensated by the fact that these cubes are less affected by oscillations in the extracted spectra due to resampling effects, for example \citep[see][and Appendix \ref{sec:AppDrizzle}]{Law2023,Perna2023}. 
    
\end{itemize}

We modified the {\sc cube\_build} script to fix an error affecting the drizzle algorithm, as implemented in the pipeline version 1.8.5 which was not released at the time of data reduction\footnote{{\sc cube\_build} code changes in \url{https://github.com/spacetelescope/jwst/pull/7306}}. Finally, we also modified the  {\sc photom} script, applying the corrections implemented in the same unreleased pipeline version\footnote{{\sc photom} code changes in \url{https://github.com/spacetelescope/jwst/pull/7319}}, which allows us to infer more reasonable flux densities (i.e. a factor $\sim100$ smaller with respect to those obtained with standard pipeline 1.8.2). To test this calibration we also reduced the data using an external flux calibration \citep[details in][]{Perna2023}, and obtained consistent results.

A complete description of the data reduction process and the tests we have performed to ensure its robustness can be found in \citet{Perna2023}.

\subsection{Data Analysis}

\subsubsection{Quasar and Host Integrated Line Fitting}
\label{sec:QuasarLineFitting}

To obtain our integrated quasar spectrum, we integrate our data cube across an aperture of radii 0.35'' for the wavelengths blue-ward of the detector gap, and 0.45'' red-ward of the gap. These apertures were chosen by examining the PSF structure around \ha\ and \hb. These aperture radii contain the central peak of the PSF, but not the ring of reduced flux that occurs prior to the second radial PSF peak. This allows us to maximise our signal to noise, while containing the majority of the quasar flux. 

We compare the flux in this aperture to the total flux in the image (to radius 1.5'') at the peak \ha\ and \hb\ wavelengths, finding an average fraction for both quasars of 81.4\% at \hb, and 86.6\% at \ha. We use these percentages as our flux correction, multiplying the integrated aperture flux blue-ward of the detector gap by 1.228 and the integrated flux red-ward of the detector gap by 1.155.

To fit our quasar spectra we use the Markov chain Monte Carlo (MCMC)-based technique of \textsc{QubeSpec}\footnote{\url{https://github.com/honzascholtz/Qubespec}}.
Our quasar spectra contain \hb\ and \ha\ emission from the BLR, which are blended with the narrow galaxy \hb, \oiii, and \ha\ lines (galaxy narrow: `GN'), and also with broader lines from the galaxy associated with a kinematically distinct component (e.g. an outflow; galaxy broad: `GB'). We note that we see no evidence for any \nii\ emission in our cubes, from either the quasar or the host galaxies and companions, and so we do not include an \nii\ line in our fit. This likely indicates a low metallicity for our sources.

For the \ha\ and \hb\ lines, we find that the BLR signal cannot be well fit with a single Gaussian. We instead fit the BLR \ha\ and \hb\ lines with a broken power law, 
\begin{equation}
f(x)=
\begin{cases}
(x/x_{\rm{break}})^{-\alpha_1}: x<x_{\rm{break}}\\(x/x_{\rm{break}})^{-\alpha_2}: x>x_{\rm{break}},
\end{cases}
\end{equation}
that is convolved with a Gaussian with standard deviation $\sigma$ \citep[as in e.g.][]{Nagao2006,Cresci2015}.
In our spectral fit we assume that the galaxy lines are composed of a narrow component (GN), as well as a broad component (GB), which is typically blue-shifted and therefore generates a wing in the total profile, which could be attributed to ionized gas outflows. For the two kinematic components, GN and GB, we fit each of the \oiii\ and \hb\ lines with a Gaussian with velocity offset and line widths equal in velocity space, as each of these lines likely arise from the same physical region with similar kinematics. 
We tie the amplitude of \oiiib\ to be 3 times less than that of \oiiia, for both GN and GB components. The quasar continuum is modelled as a power law.

The \ha\ line is fit separately after the \oiii--\hb\ complex. The GN \ha\ component is constrained to have the same velocity as the \oiii\ lines. For the \ha\ GN line width, the \ha\ GB velocity offset and line width, and the \ha\ BLR velocity and slopes $\alpha_1$ and $\alpha_2$, the initial guess is the respective \hb\ fit property, with normally distributed priors surrounding that value. This is again because the lines likely arise from the same physical region with similar kinematics. We also choose an initial guess for the \ha\ GN, GB, and BLR peak values to be 3.5 times that of the respective \hb\ values, with normally distributed priors around that value, which helps the fit to converge on a \ha\ model that has realistic flux ratios relative to \hb.
Despite the \hb--\oiii\ and the \ha\ complexes being fit independently, the agreement among the various kinematic properties was excellent, particularly  for the host components (see Table \ref{tab:QSOLines}). However, for \VDES\ we note that the \ha\ BLR full width at half maximum (FWHM) is $3092\substack{ +39\\-1}$ km/s, which is significantly lower than the \hb\ FWHM of $4461\substack{ +233\\-182}$ km/s. This discrepancy may be due to the \ha\ line being inaccurately measured as it lies close to the edge of the wavelength coverage of the detector; we are more confident in the \hb\ line measurement. There are also fit degeneracies which could be affecting this measurement. 

\begin{figure*}
\begin{center}
\includegraphics[scale=0.6]{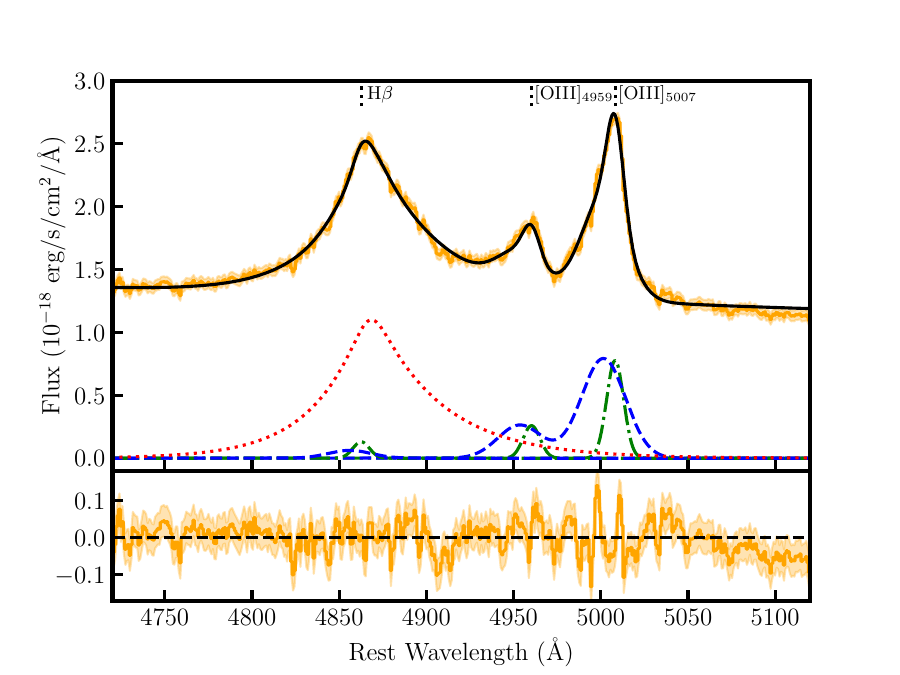} \hspace{-0.8cm}
\includegraphics[scale=0.6]{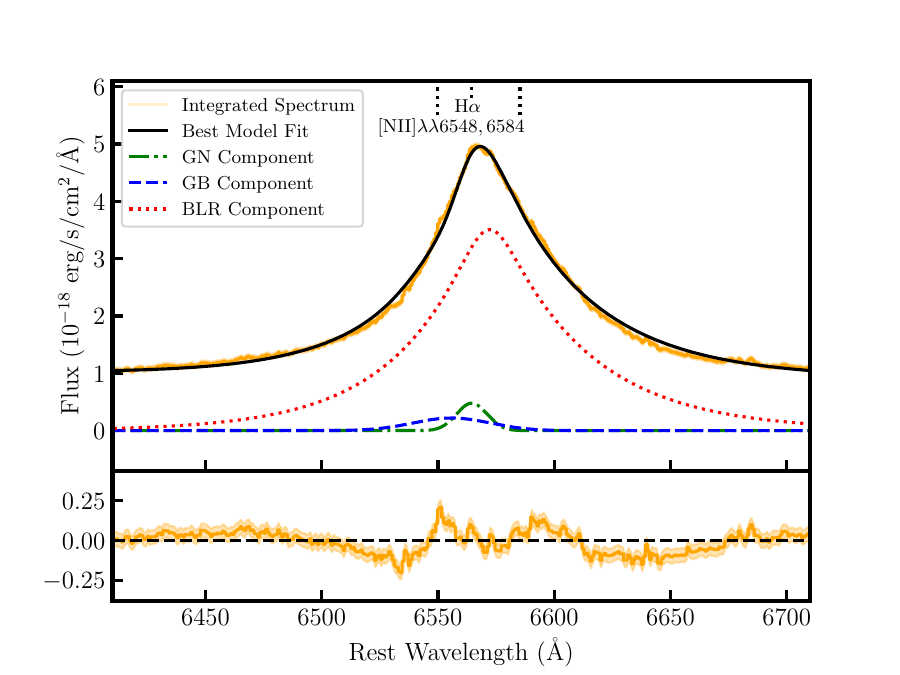}
\vspace*{-7cm}  % Tune this to the image height.
\begin{center}
\large{\DELS}
\end{center}
\vspace*{6cm}

\includegraphics[scale=0.6]{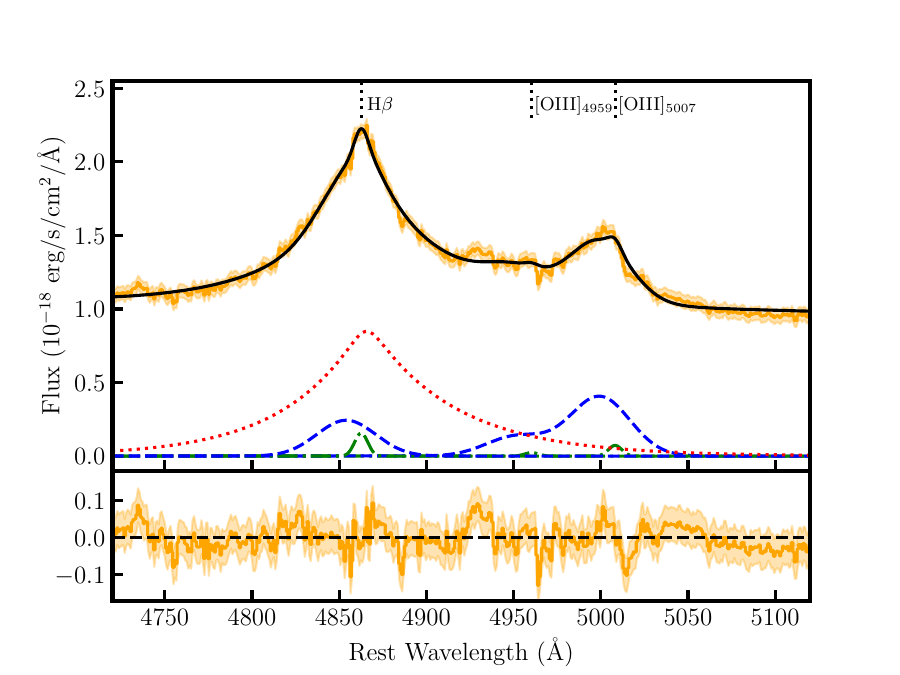} \hspace{-0.8cm}
\includegraphics[scale=0.6]{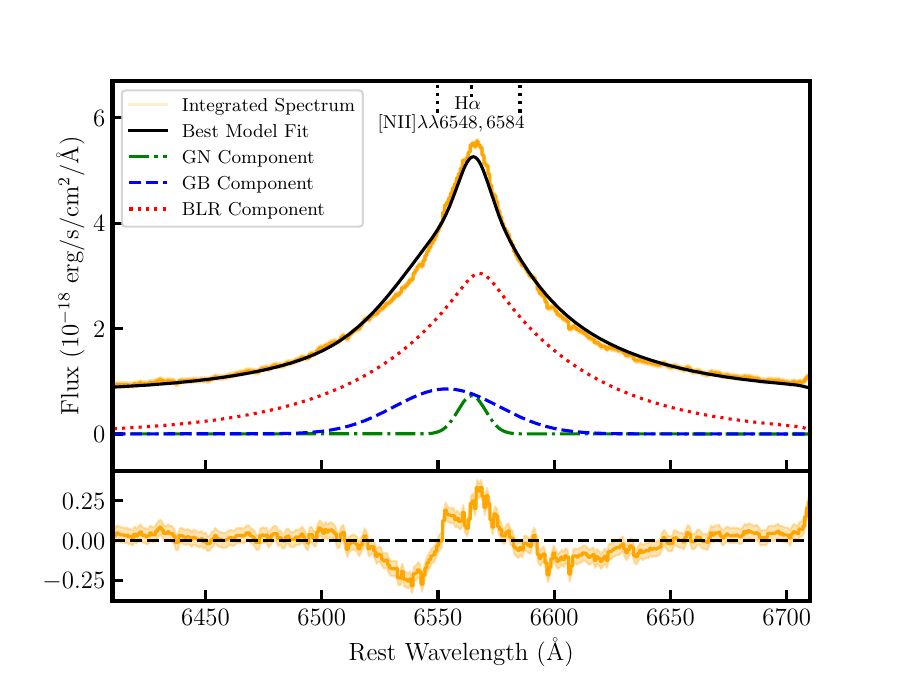}
\vspace*{-7cm}  % Tune this to the image height.
\begin{center}
\large{\VDES}
\end{center}
\vspace*{6cm}

\caption{The integrated quasar spectrum for \DELS\ (top) and \VDES\ (bottom) (orange curve), showing the regions around \hb\ (left) and \ha\ (right).  The integrated spectrum is integrated over a radius of 0.35'' for the wavelengths blue-ward of the detector gap, and 0.45'' red-ward of the gap, corresponding to the apertures selected for our quasar fitting (Section \ref{sec:QuasarLineFitting}). No flux correction has been applied. 
The best model fit (black) is shown alongside the narrow GN line components (green dot-dashed), the broader GB components (blue dashed lines), and the broken power law component for the BLR (red dotted lines). 
The lower panels show the residual of the model fit. }
\label{fig:QuasarFit}
\end{center}
\end{figure*}

We show the full integrated quasar spectrum for \DELS\ and \VDES\ in Figure \ref{fig:ContinuumSubtraction}.
In Figure \ref{fig:QuasarFit} we show the \hb\ and \ha\ regions of these integrated spectra, alongside the best-fitting model.
The velocity offset and FWHM for each of the model components are given in Table \ref{tab:QSOLines}, and the resulting BLR line fluxes and luminosities are given in Table \ref{tab:BHmasses}.

\begin{table*}
\caption{Measured line velocity offsets and velocity dispersions (relative to the quoted redshift) from the integrated quasar spectra. The FWHM have been corrected for instrumental broadening (FWHM$_{\rm{inst,H\beta}}=115$ km/s, FWHM$_{\rm{inst,H\alpha}}=85$ km/s, \citealt{Jakobsen2022}).}
\begin{tabular}{lllllll}
\hline 
\hline 
& GN ([\ions{O}{iii}], \hb) & GB ([\ions{O}{iii}], \hb) & BLR, \hb\ & GN, \ha\ & GB, \ha\ & BLR, \ha\ \\
\hline 
\hline
\DELS\ \\

\hline  
$V_{r}$ (km/s) & 0 [z=6.818] & $-404\pm24$ & $317\substack{ +69\\-64 }$ & 0 & $-398\pm5$ & $338\pm7$ \\
FWHM (km/s) & $685 \pm32$ & $1784 \pm39$ & 
      $3404\substack{ +55\\-155 }$ & $700\pm10$ & 
      $1780\pm6$ & $2794\substack{ +1\\-39 }$ \\

\hline
\VDES\ \\

\hline
      $V_{r}$ (km/s) & 0 [z=6.855] & $-530\pm42$ & $181\substack{ +65\\-73 }$ & 0 & $-492\pm5$ & $159\pm8$ \\
FWHM (km/s) & $541 \substack{ +110\\-84 }$ & $2439 \pm73$ & 
      $4461\substack{ +233\\-182 }$ & $646\pm9$ & 
      $2429\pm5$ & $3092\substack{ +39\\-1}$ \\
\hline
\end{tabular} 

\footnotesize{}
\label{tab:QSOLines}
\end{table*}

\subsubsection{Continuum Subtraction}
\begin{figure*}
\begin{center}
\includegraphics[scale=0.8]
{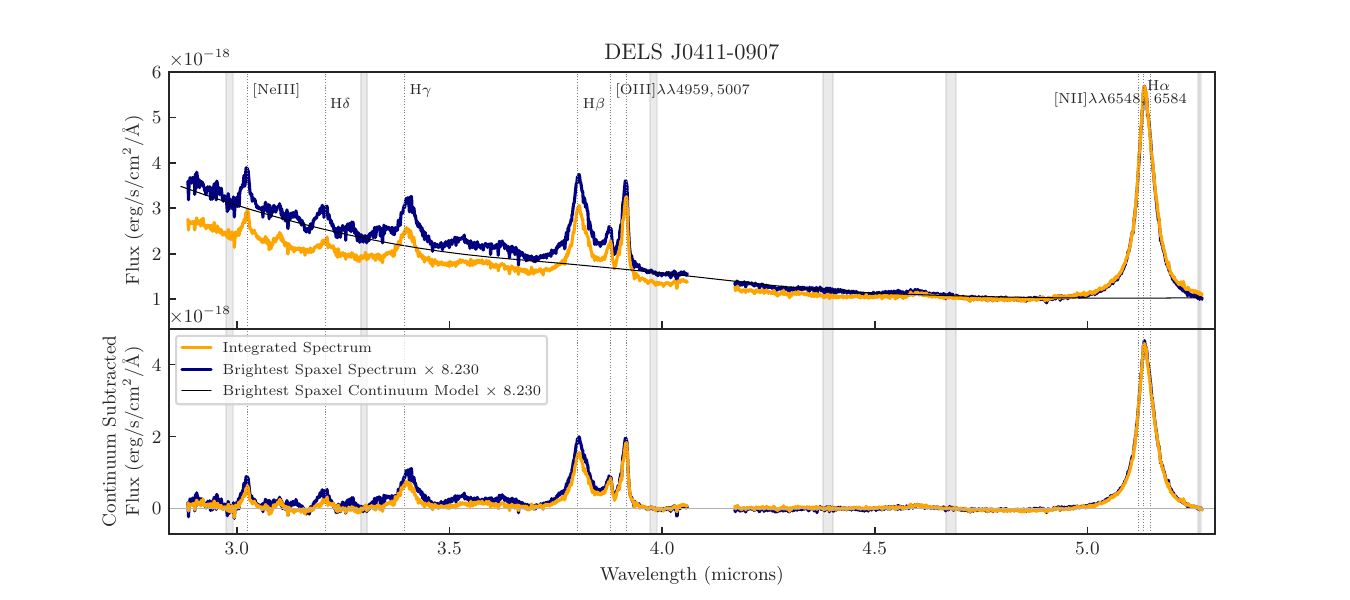}
\includegraphics[scale=0.8]{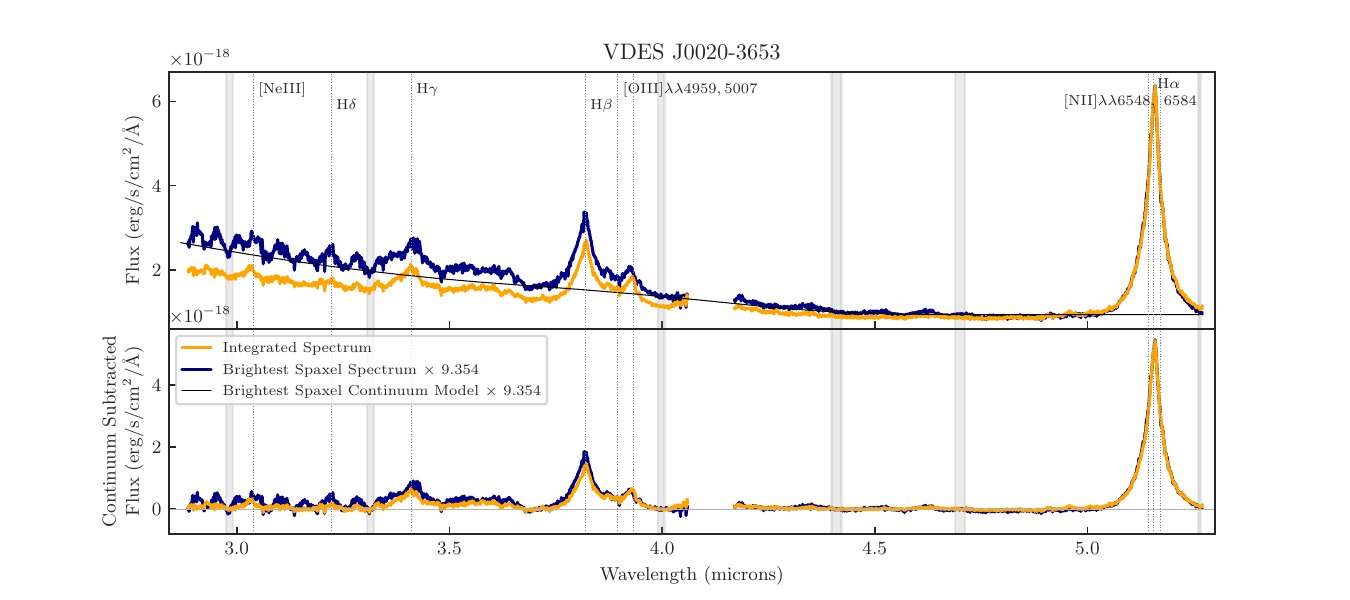}
\caption{The quasar spectrum for \DELS\ (top) and \VDES\ (bottom), showing the spectrum from the brightest spaxel (navy), alongside the integrated spectrum (orange). The integrated spectrum is integrated over a radius of 0.35'' for the wavelengths blue-ward of the detector gap, and 0.45'' red-ward of the gap, corresponding to the apertures selected for our quasar fitting (Section \ref{sec:QuasarLineFitting}). Our flux corrections of 81.4\% and 86.6\% have been applied to the integrated spectrum. 
The spectrum of the brightest spaxel is increased by a constant scaling factor, 8.230 for \DELS\ and 9.354 for \VDES, such that the peak of \ha\ is equal for both spectra in the plots.
The lower panel in each shows the spectra after continuum subtraction. For the brightest spaxel spectrum, the continuum model subtracted from the spectrum is shown in black in the top panel. For the continuum-subtracted integrated spectrum in the lower panel, the integration is performed on the continuum-subtracted cube (i.e. the continuum subtraction was first performed on each of the spaxels individually).
The dotted grey vertical lines depict lines of interest at the redshift of the quasar ($z=6.854$ for \VDES\ and $z=6.818$ for \DELS), and the grey regions show the continuum windows. The detector gap from 4.06-4.17$\mu$m has been masked. We note that some minor oscillations are present in the single spaxel spectrum (navy), which can be particularly noticed below 3.3$\mu$m in the single spaxel spectrum. These are caused by spatial undersampling of the PSF, which is a known issue when considering spectra in individual NIRSpec IFU spaxels; we describe this in detail in \citet{Perna2023}. 
When measuring the quasar properties in Section \ref{sec:BHproperties}, we use the integrated spectrum (orange), thus removing these single pixel effects. For the quasar subtraction, we find the best performance using the single spaxel spectrum (navy), however these oscillations do not significantly affect our results which directly focus on the regions around \ha\ and \hb.}
\label{fig:ContinuumSubtraction}
\end{center}
\end{figure*}

To further analyse our quasar data cubes, we first subtract the continuum emission from the spectra in each spaxel, by considering the flux in the continuum windows at rest frame 3790--3810\AA, 4210--4230\AA, 5080--5100\AA,
5600--5630\AA\ and 5970--6000\AA, following \citet{Kuraszkiewicz2002} and \citet{Kovacevic2010}; these windows are selected to be uncontaminated by emission lines and blended iron emission. We also
added an additional continuum window at the red side of \ha\ at the edge of the spectral range, at 5.260--5.266 $\mu$m in the observed frame, as well as a window near the blue edge of the spectral range, at 2.975--2.990 $\mu$m in the observed frame, to better constrain the continuum. 
We calculate the signal-to-noise ratio (SNR) at each wavelength in each spaxel, taking the noise to be the `ERR' cube produced by the pipeline.
For each spaxel in which the median SNR within these continuum windows is greater than 2, we interpolate between these continuum windows using \textsc{scipy}'s \textsc{interp1d} function using a spline interpolation of second order; this is found to be more robust than fitting a power law curve in each spaxel. In each of these spaxels the interpolation is subtracted from the spectra to create our continuum-subtracted cube.
Figure \ref{fig:ContinuumSubtraction} shows the original and continuum-subtracted spectrum of the brightest quasar spaxel.

We note that for \VDES\ there may be some broad \ha\ flux in this reddest continuum window in the brightest quasar spaxels, as at this redshift the IFU spectral coverage does not extend to the pure continuum region red-ward of \ha.
However, we find that this window is necessary for finding a reasonable fit to the continuum around \ha. 
To reduce any potential effect of this on our results, we use the full, non-continuum subtracted cube in Section \ref{sec:QuasarLineFitting} when we fit our integrated quasar model to investigate the broad line properties. With our QubeSpec modelling we fit the emission lines as well as the continuum in the form of a power law. As the integrated quasar spectrum is well described by a power law, this approach does not require the continuum window strategy used here. In contrast, within individual spaxels the continuum is not well described by a power law due to the spatial variation of the PSF with wavelength, and so this continuum window approach is more reliable. 
For our spatially resolved analysis we are primarily interested in the narrow line emission, which is further from the edge of the detector, and so this should be less biased by this issue. However, we do caution that at this high redshift, the \ha\ measurement of \VDES\ could be generally susceptible to edge effects from the detector and an imperfect continuum subtraction.

\subsubsection{Quasar Subtraction}
\label{sec:QuasarSubtraction}
In order to fully understand our quasar--host systems, we decompose the observed emission into its various components. 
The aim is to subtract the bright unresolved flux from the quasar, revealing the extended narrow line emission from the host galaxy.

The full cube contains broad line emission from the quasar's broad-line region (BLR) surrounding the black hole, for example from the \ha\ and \hb\ lines. The BLR is spatially unresolved at these redshifts, and thus this broad line emission is spread spatially according to the PSF of the instrument. In our BLR Quasar Subtraction method (see below), we subtract this BLR component from the cube. The remaining flux in the cube is narrow line emission, for example from \oiii\ as well as the \ha\ and \hb\ lines which also have a strong broad component. This narrow emission is emitted both by the quasar's narrow line region (NLR), photoionized by the AGN and shocks, as well as gas throughout the extended host galaxy (and any companion galaxies) which is photoionized by star formation.
While the NLR can be spatially extended, significant NLR emission is produced by the spatially unresolved central region surrounding the AGN, and so this component will clearly trace out the PSF shape. To understand the true spatial extent and kinematics of the extended narrow line emission, and not be biased by the bright unresolved NLR emission, we subtract this narrow unresolved flux from our cube. This is our Narrow \& Broad (N\&B) Quasar Subtraction, as detailed below.\\

To separate the host and quasar emission, we use the code \textrm{QDeblend3D} \citep{Husemann2013,Husemann2014}, which uses the relative strength of quasar broad lines in each spaxel to map out the spatial PSF. 
This strategy allows us to use the specific quasar observation to create an accurate PSF model, instead of relying on an external observation of a known point source (i.e. star) or a theoretical PSF model. The spatial PSF varies with the precise position of the source on the detectors, the general position on the sky, and time, so matching the PSF of our specific observations with a PSF observation or model is less precise than the broad line model we can create here. This methodology has successfully been used for IFU observations of lower-redshift quasar host galaxies \citep[e.g.][]{Husemann2013,Husemann2014,Vayner2014,Kakkad2020}.

As the NIRSpec PSF varies with wavelength, we opt to perform the quasar subtraction at two different wavelength regimes. 
For the spectra blueward of the detector gap, we use the \hb\ line for the subtraction, and for the spectra redward of the detector gap we use the \ha\ line for subtraction.
This allows for the best subtraction for the wavelength region around each of these two lines. 

\begin{figure*}
\begin{center}
\includegraphics[scale=0.8]{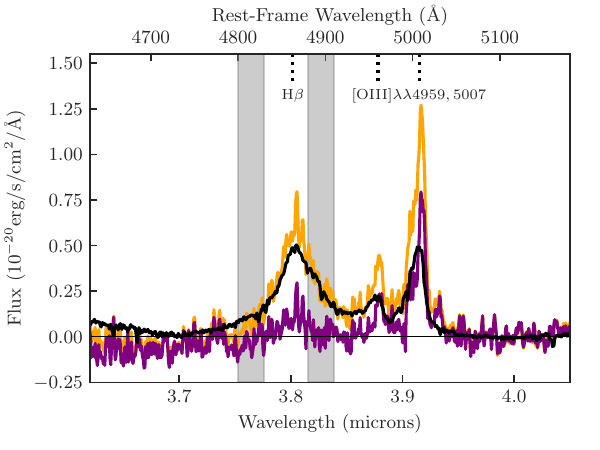}
\includegraphics[scale=0.8]{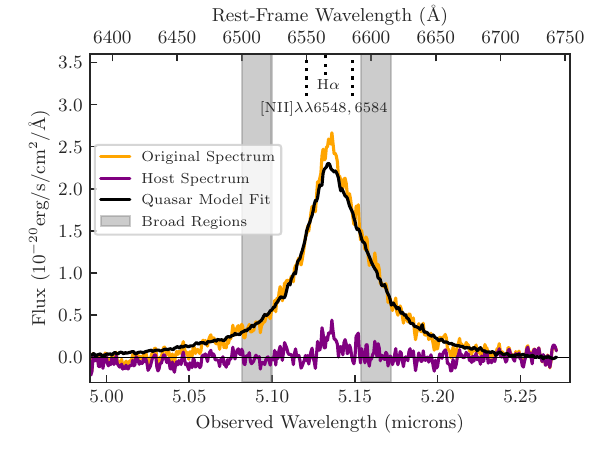}
\vspace*{-6.8cm}  %DIF >  Tune this to the image height.
\begin{center}
\large{N\&B Quasar Subtraction}
\end{center}
\vspace*{6cm}

\includegraphics[scale=0.8]{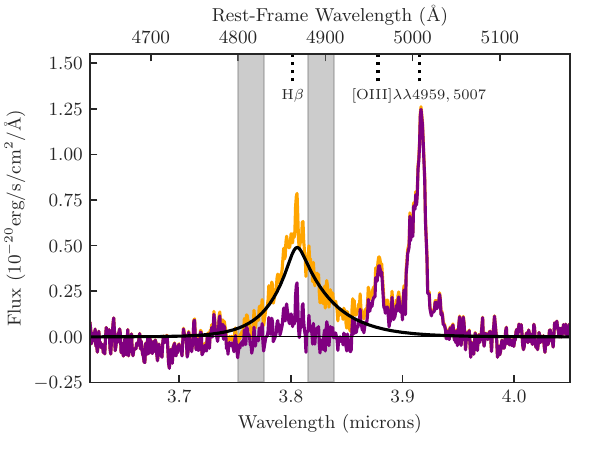}
\includegraphics[scale=0.8]{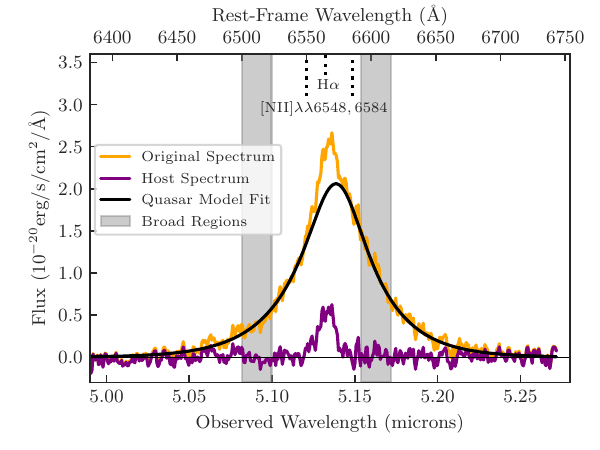}
\vspace*{-6.8cm}  %DIF >  Tune this to the image height.
\begin{center}
\large{BLR Quasar Subtraction}
\end{center}
\vspace*{6cm}
\caption{Visualisation of the quasar subtraction using the broad \hb\ (left) and \ha\ (right) lines from \DELS. This shows the spectrum in spaxel (23,24), with the quasar peak located at spaxel (26, 24), in image coordinates. The orange spectrum is the original continuum-subtracted spectrum in the spaxel, containing line flux from both the quasar and host galaxy. The black line is the quasar model spectrum. The top panels show the choice that the quasar model spectrum is that of the central brightest quasar spaxel, which contains both narrow and broad emission---the N\&B subtraction. The bottom panel shows the choice that the quasar model spectrum is the BLR component from our best-fit model to the quasar spectrum (Figure \ref{fig:QuasarFit})--the BLR subtraction.
Using the broad spectral windows marked in grey, the quasar model (black) is scaled to fit the combined spectrum (orange). The purple spectrum shows the residual to that fit, which is the host spectrum. }
\label{fig:QuasarSubtraction}
\end{center}
\end{figure*}

\textit{Step 1, Spatial PSF model:} We determine the spatial PSF shape by measuring the relative flux of the quasar BLR at each spatial location. 
The BLR is spatially unresolved and thus this emission will trace the PSF shape of the instrument. 
%Additionally, they will have the highest ratio of quasar to host galaxy flux of any wavelength in the spectrum. 
To measure this relative flux we use the continuum-subtracted cube. For \ha, we take the mean of the flux between rest-frame 6500--6523\AA\ and 6592--6615\AA, either side of the peak of \ha. 
These broad spectral windows are free from any narrow lines, which would bias the measurement; they are marked in Figure \ref{fig:QuasarSubtraction}, which shows an example of this quasar subtraction process. For \hb, the equivalent is performed for emission between rest frame 4800--4830\AA\ and 4880--4910\AA.
Once this broad line flux is measured in each spatial pixel, it is normalised to that of the central brightest quasar spaxel, giving a 2D fractional map of the relative flux of the spatially unresolved BLR, that is, the 2D PSF. 

To identify spaxels with significant quasar flux from the PSF model, we calculate the SNR in each individual wavelength channel across the selected broad spectral window for each spaxel, using the `ERR' extension as the noise. For our final PSF shape, we select spaxels in which at least three consecutive wavelength channels have $\rm{SNR}>10$, and then expand this region spatially by selecting any adjacent spaxels with $\rm{SNR}>2$ over two consecutive wavelength channels, following the \textsc{find\_signal\_in\_cube} algorithm of \citet{SunGithub} \citep[see also][]{Sun2018}.
To reduce the effect of any artefacts and companion galaxies on the measured PSF shape, we cut the PSF at 10$\sigma$, where $\sigma=\rm{FWHM}/(2\sqrt{2\log{2}})$ and the FWHM is the diffraction limit of the telescope at the observed wavelength of \hb\ and \ha, 0.146'' and 0.201'' respectively. This ensures we are only performing the quasar subtraction in regions that have a broad line signal, while ensuring we extract as much of the quasar as possible. 
The resulting masked PSF shapes are shown in Figure \ref{fig:PSFs}, for both the \ha\ and \hb\ subtractions for both quasars.
Measuring the FWHM of the resulting \hb\ PSFs gives 0.18'' for \VDES\ and 0.16'' for \DELS, and 0.23'' for \ha\ for both \VDES\ and \DELS, consistent with expectations for the PSF size.

\textit{Step 2, Quasar spectral model:} 
To perform the quasar subtraction we require a 3D quasar model cube to subtract from the initial quasar and host cube. To create a 3D cube for the quasar emission, alongside the PSF shape (spatial x,y dimensions) we also need a model for the quasar spectrum (spectral $\lambda$ dimension), for which we consider two approaches.
The first approach is to use the observed spectrum of the brightest spaxel, the centre of the quasar emission, under the assumption that the host line emission is negligible. We refer to this as the `N\&B Quasar Subtraction', as this quasar cube includes both the narrow (N) and broad (B) components of the central spectrum, and so both components will be subtracted.
Our second approach is to use a model of the BLR component of the quasar spectrum. We take this model from our fit to the integrated quasar spectrum (the red curve in Figure \ref{fig:QuasarFit}, with values quoted in Table \ref{tab:QSOLines}). This quasar model only includes the BLR component of the fit, and so we refer to this as the `BLR Quasar Subtraction' as only the BLR emission is subtracted from the initial cube. Our process for integrating and fitting the quasar spectrum is described in detail in Section \ref{sec:QuasarLineFitting}. 

In our quasar subtraction, the spatial variation of the quasar's BLR flux is used to determine the PSF shape. However, the N\&B subtraction subtracts both the broad and narrow emission that are present in the central quasar spaxel.
These emission lines are emitted from the central, unresolved NLR, which is spread according to the PSF shape.
This N\&B subtraction results in a clear view of the narrow line flux in the underlying, spatially extended host galaxy, without contamination from the bright lines from the central NLR region that have been smeared by the PSF shape. This is ideal for measurements of the spatial distribution and kinematics of emission line regions, as well as measurements of the SFRs which would otherwise be significantly biased by the large amount of AGN-photoionized flux in the centre.
However, this method uses the spectrum from the individual brightest quasar spaxel, and so can be subject to observational artefacts that would be removed when integrating over a larger region, for example oscillations in the spectra \citep[see][]{Perna2023}. We chose to use the individual spaxel spectrum and not an integrated spectrum for the model as we found that this resulted in the best quasar subtraction.
This model also assumes that the host contribution to the line flux in this central spaxel is negligible, while there may be some small contribution at the position of the optical lines analysed in this work.

In contrast, the BLR quasar subtraction method allows us to view the complete narrow line emission maps, including the contribution of the central unresolved NLR. Our BLR subtraction technique subtracts only the BLR quasar flux, with all of the narrow flux from the central core remaining. Much of this remaining narrow line flux is spread according to the PSF shape, which significantly contaminates any spatial analysis of the extended host region.
This approach is also model dependent, varying based on the BLR model fit to the integrated quasar spectrum. With their various advantages and disadvantages, we use both approaches throughout this work.

\textit{Step 3, Subtracting the 3D PSF model:} 
For each of these choices for the model quasar spectrum, we scale the quasar spectrum by the 2D PSF to create a 3D quasar cube. We then subtract this quasar cube from the original cube, revealing our estimate of the host galaxy's 3D spectrum. We visually inspect the quasar model fit and residual for a range of spaxels to confirm our subtraction is robust; an example for one spaxel is shown in Figure \ref{fig:QuasarSubtraction}.

\subsubsection{Host Line Maps}
\label{sec:Maps}
To estimate the noise for our host galaxy maps, we consider that the quasar subtraction introduces additional noise to the resulting host cube that should be accounted for. 
Following a Monte Carlo process, we generate 200 different realisations of the continuum-subtracted cube with normally-distributed noise, where $\sigma$ is taken from the `ERR' extension from the original cube. We then perform the \ha\ and \hb\ quasar subtraction on each of these 200 cube realizations.
To create our modified noise cube, we calculate the standard deviation of these 200 realisations of the quasar subtracted cube at each wavelength in each spaxel.

For the \hb--\oiii\ line complex the values in these noise maps are up to 34\% higher than the original noise maps in the central spatial regions of the quasar for the N\&B quasar subtraction, and 4\% higher for the BLR subtraction. For the \ha\ line, the values in the noise map are up to 47\% higher than the original noise map for the N\&B quasar subtraction, and 10\% higher for the BLR subtraction. The N\&B subtraction introduces more noise than the BLR subtraction, as the N\&B quasar model spectrum contains noise while the BLR quasar spectrum is a smooth model and the model uncertainties are not considered in this calculation.
%%FIX
The quasar subtraction adds additional uncertainty, particularly where more quasar flux has to be subtracted.

To create maps of the \ha, \hb, and \oiii\ emission of the host galaxy, we fit the spectrum in each spaxel as a series of Gaussians, using \textsc{AstroPy's} Levenberg-Marquardt algorithm.
We assume that the narrower galaxy lines are composed of a narrow component (GN) as well as a broader wing (GB; see Section \ref{sec:QuasarLineFitting}).
For the GN and GB components, we fit each of the four lines as a Gaussian with mean velocity and line widths equal in velocity space. 
We tie the amplitude of \oiiib\ to be 3 times less than that of \oiiia, for both the GN and GB components. These theoretical flux ratio constraints provide a good fit for the host galaxy \oiii\ lines for both the N\&B and BLR quasar subtraction, which gives further confidence in our subtraction techniques.
As noted above, we see no evidence for any \nii\ emission in our cubes, and do not include those lines in our fit.

When the spectrum can be well fit with two Gaussians, the Gaussian with the smallest velocity dispersion $\sigma$ is assigned as the GN component, while the largest is assigned to the GB component. When only one Gaussian is preferred, if it has $\sigma\leq33$\AA, or approximately 600 km/s, it is assigned as a GN component, otherwise as a GB component.

To determine the flux of each of the GN and GB line components we integrate the individual Gaussian model. 
The SNR is calculated by integrating the noise cube, calculated from the quasar subtraction algorithm, across the same spectral region. Using these noise maps we search for spaxels with $\rm{SNR}>5$, and then expand this region spatially by selecting any adjacent spaxels with $\rm{SNR}>1.5$, following the \textsc{find\_signal\_in\_cube} algorithm of \citet{SunGithub}. In the maps we only display spaxels meeting this criterion.
The resulting maps are shown in Figures \ref{fig:DELSMapsQSO} and \ref{fig:VDESMapsQSO} for the N\&B subtraction for \DELS\ and \VDES\ respectively, with the BLR subtraction maps for both quasars shown in Figure \ref{fig:MapsBroad} to show the differences between the two subtraction methods.

To create velocity maps, we calculate the median velocity $v_{50}$, a non-parametric measure for the line centre.
For the velocity dispersion maps, we calculate $w_{80}$, the width containing 80\% of the line flux. This is a non-parametric measure, which is approximately equal to the FWHM for a Gaussian line \citep[$w_{80}=1.088$ FWHM; see e.g.][]{Zakamska2014}.
The velocity dispersion maps are derived after removing the instrumental width in quadrature at each spaxel, where for NIRSpec in the G395H/F290LP configuration, FWHM$_{\rm{inst,H\beta}}\simeq115$ km/s, and FWHM$_{\rm{inst,H\alpha}}\simeq85$ km/s for the redshift of these quasars \citep{Jakobsen2022}.
These velocity and velocity dispersion maps are also shown in Figures \ref{fig:DELSMapsQSO} and \ref{fig:VDESMapsQSO} for the N\&B subtraction, with the BLR subtraction maps also presented in Figure \ref{fig:MapsBroad}.

\section{Results}
\label{sec:results}

\subsection{Black Hole Properties}
\label{sec:BHproperties}

\subsubsection{Redshift}
\label{sec:redshift}
We show the integrated quasar spectrum for \DELS\ and \VDES\ in Figure \ref{fig:QuasarFit}, with the parameters from our spectral fit given in Table \ref{tab:QSOLines}.

We derive the quasar redshift based on the galaxy narrow (GN) \oiiia\ emission. For \DELS, we find a redshift of $z=6.818\pm0.001$. %confidence interval based on MCMC fit as well as difference between FC and QS cubes.
This is broadly in agreement with the Lyman-$\alpha$ redshift of \DELS, $z_{\rm{Ly}\alpha}=6.81\pm0.03$ \citep{Pons2019a,Wang2019}, and the \ions{Mg}{ii} redshift, $z_{\rm{Mg\textsc{ii}}}=6.824$ \citep{Pons2019a}.

For \VDES, the quasar redshift based on the narrow (GN) \oiiia\ emission is $z=6.855\pm0.002$.
In comparison, the Lyman-$\alpha$ redshift of \VDES\ is $z_{\rm{Ly}\alpha}=6.86\pm0.01$ and the \ions{Mg}{ii} redshift is $z_{\rm{Mg\textsc{ii}}}=6.834\pm0.0004$  \citep{Reed2019}. For \VDES, there is a significant blueshift of \ions{Mg}{ii} compared to the \oiiia\ and Lyman-$\alpha$ lines.

\subsubsection{Black Hole Masses}
\label{sec:BHmasses}
\begin{table}
\caption{The fluxes and luminosities measured for the 5100\AA\ continuum, \ha\ and \hb\ BLR lines, alongside the derived black hole masses and Eddington ratios using these properties. 
The errors for the black hole masses and Eddington ratios include the fit uncertainties as well as the 0.43 dex uncertainty due to the intrinsic scatter in the scaling relations.
We note that all flux uncertainties are measurement uncertainties and do not include any intrinsic uncertainty in the flux level from the data calibration.}
\begin{tabular}{lll}
\hline 
\hline 
 & \DELS\hspace{-0.3cm} & \VDES\hspace{-0.3cm} \\ 
\hline
\hline
$F_{\rm{5100}} (10^{-18} \rm{erg/s/cm}^2/$\AA)  & $1.40\pm0.02$& $1.18\pm0.01$\\
$F_{\rm{H\alpha}} (10^{-16} \rm{erg/s/cm}^2)$ & $26.4\substack{ +0.2\\-0.3 }$&$25.9\substack{ +0.2\\-0.2 }$\\
$F_{\rm{H\beta}} (10^{-16} \rm{erg/s/cm}^2)$ & $8.0\pm0.3$&$8.3\substack{ +0.5\\-0.4 }$\\
\hline
$\lambda L_{\rm{5100}} (10^{46} \rm{erg/s})$ & $3.2\pm0.2$&$2.7\pm0.2$\\
$L_{\rm{H\alpha}} (10^{44} \rm{erg/s})$ & $14.9\substack{ +0.1\\-0.2 }$&$14.8\pm0.1$\\
$L_{\rm{H\beta}} (10^{44} \rm{erg/s})$ & $4.5\pm0.2$ & $4.7\pm0.3$\\
\hline
$M_{\rm{BH,H\beta,5100}} (10^9 M_\odot)$ (G) & $2.0\substack{ +3.4\\-1.3 }$& $3.1\substack{ +5.3\\-2.0 }$\\
$M_{\rm{BH,H\beta,5100}} (10^9 M_\odot)$ (V) & $1.7\substack{ +2.8\\-1.1 }$& $2.7\substack{ +4.5\\-1.7 }$\\
$M_{\rm{BH,H\beta}} (10^9 M_\odot)$ (G)  & $1.3\substack{ +2.1\\-0.8 } $& $2.3\substack{ +3.9\\-1.5 } $\\
$M_{\rm{BH,H\beta}} (10^9 M_\odot)$ (V) & $2.5\substack{ +4.3\\-1.6 }$& $4.5\substack{ +7.7\\-2.9 }$\\
$M_{\rm{BH,H\alpha}} (10^9 M_\odot)$ & $0.9\substack{ +1.6\\-0.6 } $ & $1.1\substack{ +1.9\\-0.7 } $ \\
\hline
$L_{\rm{Bol}}/L_{\rm{Edd,H\beta,5100}}$ (G) & $0.7\substack{ +0.4\\-0.3 }$ & $0.3\substack{ +0.2\\-0.1 }$ \\
$L_{\rm{Bol}}/L_{\rm{Edd,H\beta,5100}}$ (V) & $0.9\substack{ +0.5\\-0.4 }$ & $0.4\pm0.2$ \\
$L_{\rm{Bol}}/L_{\rm{Edd,H\beta}}$ (G) & $1.2\substack{ +0.7\\-0.5 }$ & $0.5\substack{ +0.3\\-0.2 }$ \\
$L_{\rm{Bol}}/L_{\rm{Edd,H\beta}}$ (V) & $0.6\substack{ +0.3\\-0.2 }$ & $0.2\pm0.1$ \\
$L_{\rm{Bol}}/L_{\rm{Edd,H\alpha}}$ & $1.6\substack{ +1.0\\-0.7 }$ & $0.9\substack{ +0.6\\-0.4 }$ \\
\hline
\end{tabular} 

\footnotesize{The two sets of $M_{\rm{BH,H\beta,5100}}$ and $M_{\rm{BH,H\beta}}$ and corresponding Eddington ratios refer to the calculations using Equations \ref{eq:MBH_5100} and  \ref{eq:MBH_GB} for the \citet{Greene2005} and \citet{Vestergaard2006} coefficients; these are marked with a `G' and `V' respectively.}
\label{tab:BHmasses}
\end{table}

Black hole masses can be estimated using the virial relation:
\begin{equation}
M_{\rm{BH}} = \frac{f R \Delta V^2}{G}
\end{equation}
where $R$ is the size of the BLR, $\Delta V$ is the width of the emission line, $G$ is the gravitational constant and $f$ is a scale factor.
To robustly determine the BLR size $R$, reverberation mapping studies measure the time delay $\tau$ of variability in the emission line to the continuum region, with $R=c\tau$, where $c$ is the speed of light. Empirical correlations have been found between $\tau$ and the continuum luminosity of the quasar \citep[e.g.][]{Kaspi2000}, with $R\propto L^\gamma$, allowing the black hole mass to be estimated from single-epoch measurements.

For this work we use the empirically-derived relation:
\begin{equation}
\label{eq:MBH_5100}
M_{\rm{BH,H\beta,5100}\AA} = a ~\left( \frac{\lambda L_{\rm{5100}\AA}}{10^{44} \rm{erg~ s}^{-1}} \right)^{b} \left( \frac{\rm{FWHM}_{\rm{H\beta}}}{10^{3} \rm{km~ s}^{-1}} \right)^{2} \rm{M}_\odot,
\end{equation}
where \citet{Greene2005} calibrate the constants $a$ and $b$ to be $a=(4.4\pm0.2) \times10^6$, $b=0.64\pm0.02$, while alternatively \citet{Vestergaard2006} derive $a=(8.1\pm0.4)\times10^6$, $b=0.50\pm0.06$.
For the quasars under study we measure the continuum-luminosity at rest-frame 5100\AA, $L_{\rm{5100}\AA}$, from our integrated quasar spectrum, taking the mean value between 5090--5110\AA\ to account for noise variations, and using the best-fit redshift of the quasar \oiii\ emission (Section \ref{sec:redshift}). Using this and our \hb\ broad line FWHM, we calculate the black hole masses using Equation \ref{eq:MBH_5100} for both the \citet{Greene2005} and \citet{Vestergaard2006} correlations, which we present in Table \ref{tab:BHmasses}.

However, the continuum luminosity can be difficult to measure due to contamination by \ions{Fe}{ii} lines and, for lower luminosity quasars, the host galaxy itself. To circumvent this, \citet{Greene2005} and \citet{Vestergaard2006} derived empirical relations for estimating the black hole mass based solely on the \hb\ emission lines:
\begin{equation}
\label{eq:MBH_GB}
M_{\rm{BH,H\beta}} = a ~\left( \frac{ L_{\rm{H\beta}}}{10^{42} \rm{erg~ s}^{-1}} \right)^{b} \left( \frac{\rm{FWHM}_{\rm{H\beta}}}{10^{3} \rm{km~ s}^{-1}} \right)^{2} \rm{M}_\odot
\end{equation}
where 
\citet{Greene2005} estimate $a=(3.6~\pm~0.2) ~\times~ 10^6$, $b=0.56~\pm~0.02$, while \citet{Vestergaard2006} derive $a=(4.7~\pm~0.3)~\times~ 10^6$, $b=0.63~\pm~0.06$.
\citet{Greene2005} also derived a correlation based on the \ha\ broad line:
\begin{equation}
\begin{split}
M_{\rm{BH,H\alpha}} &= ~\big(2.0~\substack{+0.4\\-0.3}\big) ~\times~ 10^6 \\
&~\times~ \left( \frac{L_{\rm{H\alpha}}}{10^{42} \rm{erg~ s}^{-1}} \right)^{0.55\pm0.02} \left( \frac{\rm{FWHM}_{\rm{H\alpha}}}{10^{3} \rm{km~ s}^{-1}} \right)^{2.06\pm0.06} \rm{M}_\odot.
\end{split}
\label{eq:MBH_HA}
\end{equation}
Using our measured \hb\ and \ha\ broad line properties, we calculate the black hole masses using Equations \ref{eq:MBH_GB} and \ref{eq:MBH_HA} for both the \citet{Greene2005} and \citet{Vestergaard2006} correlations. These are also listed in Table \ref{tab:BHmasses}.

As these empirical black hole mass relations have large scatter, we take the error in these relations to be 0.43 dex, the quoted scatter from \citet{Vestergaard2006}, and add this in quadrature with our measurement uncertainty from the MCMC fit.\\

\textbf{\DELS:}

For \DELS, we find that the various scaling relations (Equations \ref{eq:MBH_5100}--\ref{eq:MBH_HA}) give a black hole mass between 0.9 and $2.5\times10^9 \rm{M}_\odot$ (Table \ref{tab:BHmasses}). These measures are all consistent within the uncertainties. 
We note that while the \hb--L$_{5100\AA}$ black hole mass estimated by the \citet{Greene2005} and \citet{Vestergaard2006} relations are similar, at 2.0 and $1.7\times10^9 \rm{M}_\odot$ respectively, the pure \hb\ black hole mass estimates are quite different between the two versions of the relation. The \citet{Greene2005} \hb\ relation gives a low black hole mass of $1.3\times10^9 \rm{M}_\odot$, while the \citet{Vestergaard2006} relation gives $2.5\times10^9\rm{M}_\odot$, twice as large.
This variation between the two scaling relations could be due to the brightness of this quasar, which is brighter than the quasars considered in those studies. 
We also note that the \citet{Greene2005} \ha\ black hole mass for \DELS\ is $0.9\times10^9\rm{M}_\odot$, the lowest of these estimates, yet this is consistent with the other estimates within the uncertainties.
As the pure \hb\ and \ha\ relations involve a further calibration from $L_{5100\AA}$ to $L_{\rm{H\alpha/H\beta}}$, we opt to take the average of the two $M_{\rm{BH,H\beta,5100}\AA}$ measurements as our best black hole mass estimate: $\left(1.85\substack{ +2\\-0.8}\right)\times10^9\rm{M}_\odot$. We reiterate that all of these black hole mass estimates are consistent with each other within the uncertainties.

In comparison, from the FWHM of the \ion{Mg}{ii} line and the rest-frame 3000\AA\ luminosity measured from Magellan FIRE spectra \citep{Pons2019a}, the \DELS\ BH mass derived using the \citet{Vestergaard2009} relation is $M_{\rm{BH,Mg\textsc{ii}}}= (0.61 \pm 0.05)\times10^9 \rm{M}_\odot$. This is much lower than our measurements. This quoted error considers only measurement uncertainties, and does not include the  additional $\sim0.55$ dex uncertainty from the scatter in the scaling relation \citep{Vestergaard2009}, which would be $M_{\rm{BH,Mg\textsc{ii}}}=\left(0.6\substack{ +1.6\\-0.4}\right)\times10^9 \rm{M}_\odot$. 
Alternatively, applying the \citet{Shen2011} relation to the same \ion{Mg}{ii} line FWHM and rest-frame 3000\AA\ luminosity from \citet{Pons2019a}, following \citet{Farina2022}, we derive $M_{\rm{BH,Mg\textsc{ii}}}= 0.95\times10^9 \rm{M}_\odot$ for \DELS.
The \ion{Mg}{ii} line of this quasar was also observed with Keck NIRES, from which the black hole mass was estimated as $M_{\rm{BH,Mg\textsc{ii}}}=(0.95\pm0.09)\times10^9 \rm{M}_\odot$ using the \citet{Vestergaard2009} relation \citep{Yang2021a}, with an additional $\sim0.55$ dex uncertainty from the scatter in the scaling relation--$M_{\rm{BH,Mg\textsc{ii}}}=\left(1.0\substack{ +2.4\\-0.7}\right)\times10^9 \rm{M}_\odot$.
While these \ion{Mg}{ii} black hole masses are all lower than our best estimates, when including the large uncertainty from the scatter in the scaling relations, all of the measurements are consistent.

In summary, the best estimate for the black hole mass of \DELS\ from NIRSpec/IFU, $M_{\rm{BH}}=\left(1.85\substack{ +2\\-0.8}\right)\times10^9\rm{M}_\odot$, is larger than the existing measurements from \ion{Mg}{ii}, $M_{\rm{BH,Mg\textsc{ii}}}=0.6$--$1\times10^9\rm{M}_\odot$. Our measurements use the \hb\--5100\AA\ black hole mass relation with scatter of $\sim0.43$ dex, compared with the larger $\sim0.55$ dex scatter in the \ion{Mg}{ii} relation, highlighting the benefit of using these Balmer lines as black hole mass estimators.

\textbf{\VDES:}

For \VDES, we find that the various scaling relations give a wide range of black hole mass estimates, between 1.1 and $4.5\times10^9 \rm{M}_\odot$. These estimates are all consistent within the uncertainties. 
We note that the lowest measurement is the \ha\ mass measurement of $1.1\times10^9 \rm{M}_\odot$. For \VDES, the \ha\ line falls at the edge of the wavelength coverage of the detector, which could cause issues with this measurement. Thus the \hb\ measurements of 2.3-- $4.5\times10^9 \rm{M}_\odot$ are likely more reliable, although all measurements are consistent. As the pure \hb\ and \ha\ relations involve a further calibration from $L_{5100\AA}$ to $L_{\rm{H\alpha/H\beta}}$, we opt to take the average of the two $M_{\rm{BH,H\beta,5100}\AA}$ measurements as our best black hole mass estimate: $\left(2.9\substack{ +3.5\\-1.3}\right)\times10^9\rm{M}_\odot$. 
We note again that the \citet{Greene2005} and \citet{Vestergaard2006} black hole masses are quite different, with the \citet{Greene2005} estimate larger for the \hb--L$_{5100\AA}$ relation, while the \citet{Vestergaard2006} mass is twice as large for the pure \hb\ relation. 
This variation could again be due to the brightness of this quasar, which is brighter than the quasars considered in those studies. 
We reiterate that all of these black hole mass estimates are consistent within the uncertainties.

In comparison, \citet{Reed2019} measured a BH mass from the \ion{Mg}{ii} line and the rest-frame 3000\AA\ luminosity of $M_{\rm{BH,Mg\textsc{ii}}}=(1.67\pm0.32)\times10^9 \rm{M}_\odot$, using the \citet{Vestergaard2009} relation. Accounting for the additional $\sim0.55$ dex scatter in the \ion{Mg}{ii} black hole mass relation, this is $M_{\rm{BH,Mg\textsc{ii}}}=\left(1.7\substack{ 4.3\\-1.2}\right)\times10^9 \rm{M}_\odot$.
Applying the \citet{Shen2011} relation to the same \ion{Mg}{ii} line FWHM and rest-frame 3000\AA\ luminosity from \citet{Reed2019}, \citet{Farina2022} derived a larger $M_{\rm{BH,Mg\textsc{ii}}}=2.51\times10^9 \rm{M}_\odot$. These estimates are in agreement with our estimated black hole mass of $\left(2.9\substack{ +3.5\\-1.3}\right)\times10^9\rm{M}_\odot$.

For our two quasars, we find that \VDES\ has a \hb--L$_{5100\AA}$ black hole mass that is consistent with the existing \ion{Mg}{ii} black hole mass estimates, while the \hb--L$_{5100\AA}$ black hole mass for \DELS\ is larger than the \ion{Mg}{ii} estimates, albeit consistent within the large uncertainties. 
\citet{shen2008} found that for $\sim8000$ low-$z$ SDSS quasars, $\log(M_{\rm{BH,H\alpha}}/M_{\rm{BH,MgII}})$ follows a Gaussian distribution with mean of 0.034 and dispersion of 0.22 dex, generally indicating a good agreement yet with some quasars found to have significantly different values of $\log(M_{\rm{BH,H\alpha}}/M_{\rm{BH,MgII}})$ up to $\pm 1.5$.
While high-$z$ quasars may follow a different distribution, our quasars are consistent with this \citet{shen2008} scenario where \ion{Mg}{ii} and \hb\ mass estimates can be consistent, although for some quasars these estimates can vary significantly. This highlights the need for obtaining \hb\ and \ha\ masses of a larger sample of high-$z$ quasars with JWST, to more
accurately measure black hole mass in the early Universe \citep[see also][]{Yang2023}.

\subsubsection{Eddington Ratios}

We now consider the Eddington Ratio $\lambda_{\rm{Edd}} = L_{\rm{Bol}}/L_{\rm{Edd}}$, the ratio of the quasars' bolometric to Eddington luminosity.

The quasars' bolometric luminosities have been estimated from their rest-frame 3000\AA\  luminosities using the conversion~ $L_{\rm{Bol}} = 5.15 \times \lambda L_{3000}$ \citep{Shen2011}. \citet{Pons2019a} calculated $L_{\rm{Bol}} = (1.89 \pm 0.07) \times 10^{47}\rm{erg~s}^{-1}$ for \DELS, and 
\citet{Reed2019} calculated $L_{\rm{Bol}} = (1.35 \pm 0.03) \times 10^{47}\rm{erg~s}^{-1}$ for \VDES.

The Eddington luminosity, the theoretical maximum luminosity an object could have while balancing radiation pressure and gravity, is typically assumed to be
\begin{equation}
\begin{split}
L_{\rm{Edd}} &= \frac{4\pi G m_{p}c M_{\rm{BH}}}{\sigma_T}= 1.26\times10^{38} \left(\frac{M_{\rm{BH}}}{\rm{M}_\odot}\right) \rm{erg~s^{-1}},
\end{split}
\end{equation}
where $G$ is the gravitational constant, $m_{p}$ the proton mass, $c$ the speed of light, and $\sigma_T$ the Thomson scattering cross-section.
Using the various black hole mass estimates from Section \ref{sec:BHmasses}, we calculate the corresponding Eddington ratios which are listed in Table \ref{tab:BHmasses}. 

From our best estimate of the black hole mass for \DELS, $M_{\rm{BH}}=\left(1.85\substack{ +2\\-0.8}\right)\times10^9\rm{M}_\odot$, we estimate an Eddington ratio of $\lambda_{\rm{Edd}}=0.8\substack{ +0.7\\-0.4}$, with our full range of estimates from 0.6-1.6.
In comparison, the \citet{Pons2019a} estimate of $\lambda_{\rm{Edd}}$ based on \ion{Mg}{ii} black hole mass is $\lambda_{\rm{Edd, Mg\textsc{ii}}}=2.37 \pm 0.22$ for \DELS, including only measurement uncertainties and not the scatter in the \ion{Mg}{ii} black hole mass relation.
As discussed in Section \ref{sec:BHmasses}, instead re-deriving the \ion{Mg}{ii} black hole mass from the \citet{Pons2019a} measurements using the \citet{Shen2011} relation,  we calculate $\lambda_{\rm{Edd, Mg\textsc{ii}}}=1.6$. With independent spectra,
\citet{Yang2021a} calculate $\lambda_{\rm{Edd, Mg\textsc{ii}}}=1.3 \pm 0.2$, again including only measurement uncertainties.

These \ion{Mg}{ii}-derived ratios are all above the Eddington limit, with the \citet{Pons2019a} measurement implying an extremely rapidly accreting quasar. However, our measurements show that the Eddington ratio is $\lambda_{\rm{Edd}}=0.8\substack{ +0.7\\-0.4}$, a more moderate accretion rate.
In summary, our results indicate that for \DELS\ the \ion{Mg}{ii} measurements underestimate the black hole mass, and thus overestimate the accretion rate of the black hole. 

\VDES\ has a lower bolometric luminosity than \DELS, and a larger black hole mass, resulting in a lower Eddington ratio for this quasar. From our best estimate of the black hole mass for \VDES, $M_{\rm{BH}}=\left(2.9\substack{ +3.5\\-1.3}\right)\times10^9\rm{M}_\odot$, we estimate an Eddington ratio of $\lambda_{\rm{Edd}}=0.4\substack{ +0.3\\-0.2}$, with our full range of estimates from 0.2--0.5, except for the large \ha\ estimate of $0.9\substack{ +0.6\\-0.4}$, which could be affected by its proximity to the edge of the detector.
In comparison, the estimate of $\lambda_{\rm{Edd}}$ based on the \ion{Mg}{ii} BH mass was $\lambda_{\rm{Edd,\ion{Mg}{ii}}}=0.62\pm0.12$ for \VDES\ \citep{Reed2019}, or alternatively $\lambda_{\rm{Edd, \ion{Mg}{ii}}}=0.43$ as re-calculated by \citet{Farina2022}. 
Our derived Eddington ratios are consistent with both values, which all suggest a sub-Eddington accretion rate for this quasar.

\subsection{Host Galaxy Properties}
\label{sec:HostProperties}

\begin{figure*}
\begin{center}
\includegraphics[scale=0.7]{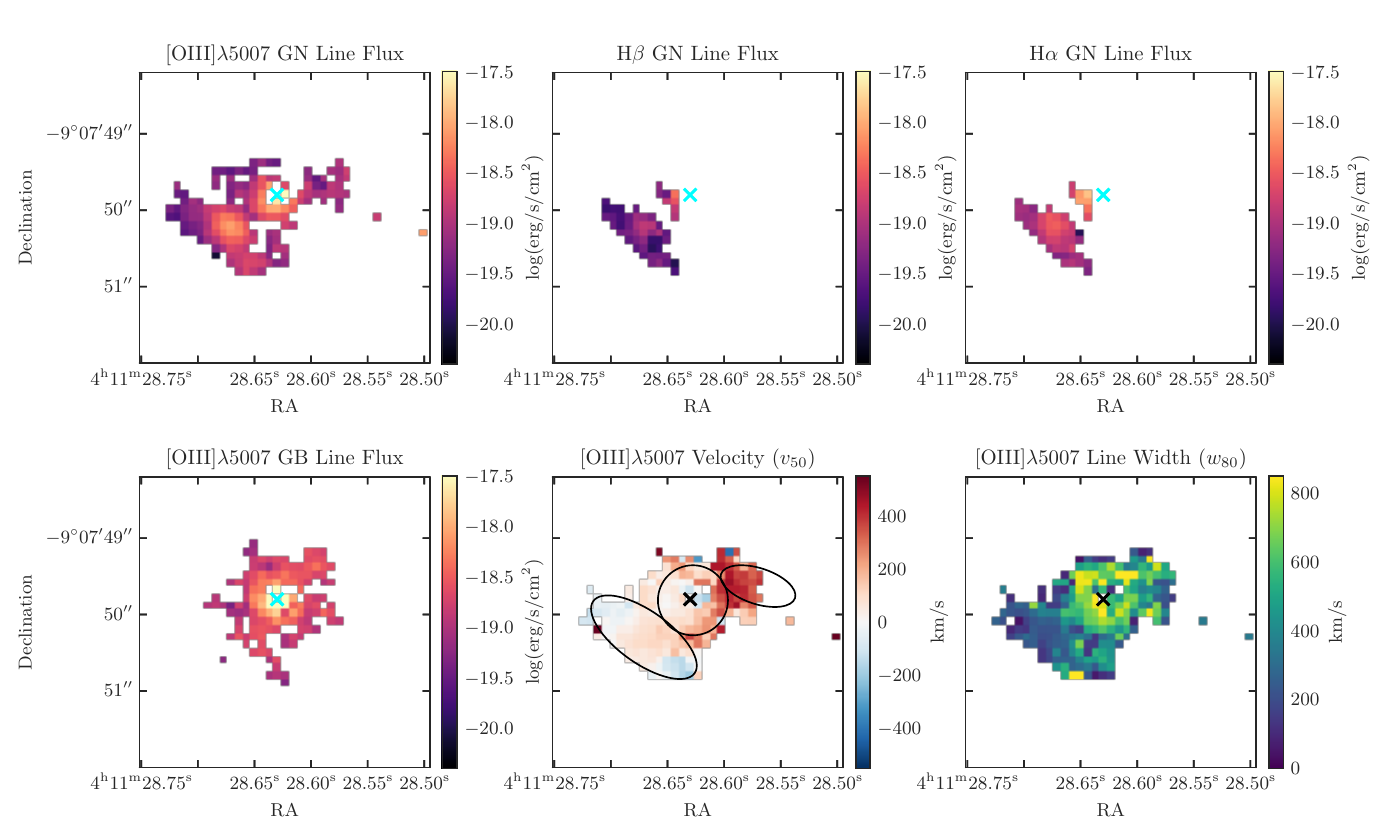}
\caption{Flux and kinematic maps for \DELS, after the subtraction of the quasar emission using the N\&B subtraction technique. The top panels show the flux of the narrow component of the \oiiia, \hb, and \ha\ lines (GN), as fit by a Gaussian. 
The bottom left panel shows the flux of the broader GB component of the \oiiia\ line, a second Gaussian with a larger $\sigma$. 
The lower middle and right panels show our kinematic maps, showing the non-parametric central velocity of the line ($v_{50}$; middle) and the line width ($w_{80}$; right). As these are non-parametric, this combines both the GN and GB components. 
We describe our method for creating these maps in Section \ref{sec:Maps}.
The black ellipses in the lower middle panel depict the regions as in Figure \ref{fig:DELSRegions}.}
\label{fig:DELSMapsQSO}
\end{center}
\end{figure*}

\begin{figure*}
\begin{center}
\includegraphics[scale=0.7]{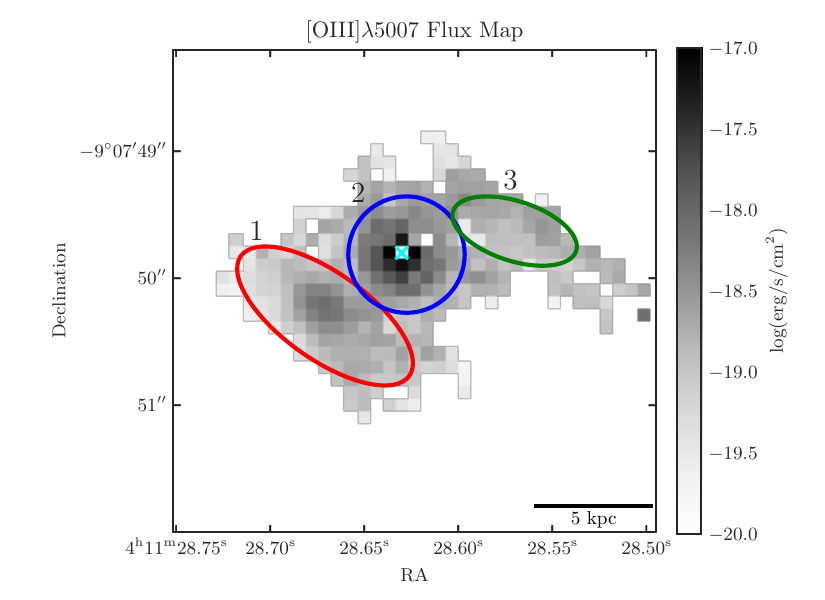}

\hspace{-1cm}
\includegraphics[scale=0.7]{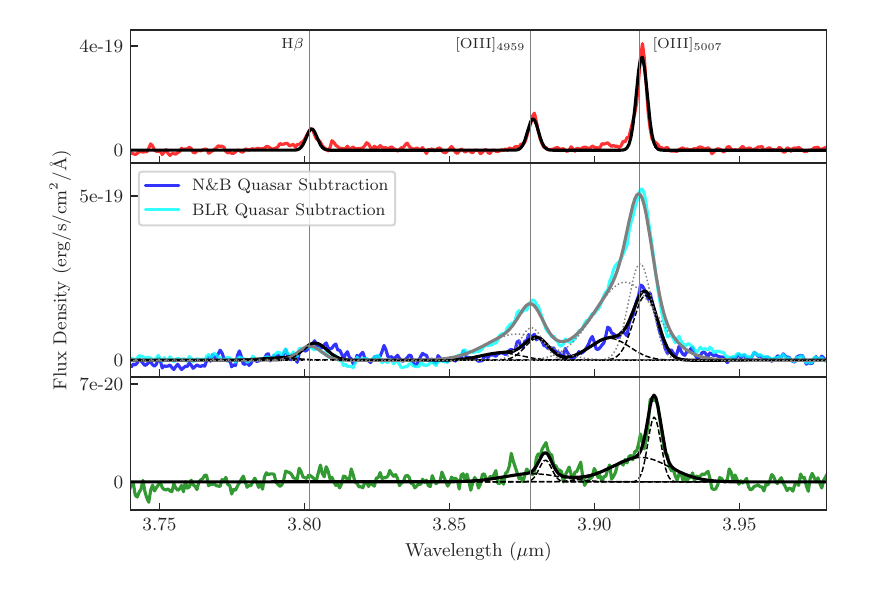}
\hspace{-0.5cm}
\includegraphics[scale=0.7]{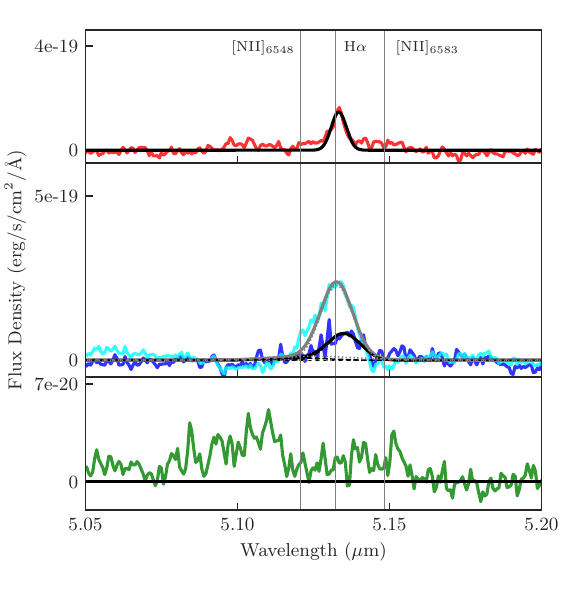}\hspace{-1cm}
\caption{Top: The \oiiia\ flux map for \DELS\ after N\&B quasar subtraction, with three regions of interest marked. Bottom: The integrated spectrum for the three regions (coloured lines), along with our best fit model in black. The middle panel, for Region 2, also shows the integrated spectra from the BLR quasar subtraction, in cyan, with the best fit model in grey. 
This region is the host of the quasar (position marked as a cyan cross in the top panel), and so the residual spectrum in this region depends heavily on the type of quasar subtraction. For Regions 1 and 3 the spectrum for both subtraction methods look equivalent, and so the BLR-subtracted spectra are not shown. 
For Region 2, we exclude the central 9 pixels surrounding the quasar peak, as these are highly corrupted by the quasar subtraction and introduce significant noise and artefacts.
This means that we will slightly underestimate the total flux in this region, however the fluxes will be significantly more reliable than if these 9 most corrupted spaxels were included.
In Region 1, the spectral lines can be described by a single narrow Gaussian (GN), whereas in Regions 2 and 3 there is also a broader component fit with a second Gaussian (GB); these separate components can be seen as the dashed black lines for the N\&B quasar subtraction, and dotted grey lines for the BLR quasar subtraction.}
\label{fig:DELSRegions}
\end{center}
\end{figure*}

\subsubsection{ISM Structure and Kinematics}
\label{sec:Structure}

In this section we study the spatial structure of the emission line regions present within our data cubes. 
To verify the spatial structures that we see, we produce a second set of data cubes by sub-sampling to a pixel scale of 0.05''/pix, and combining the various dithers with a different weighting scheme (see Appendix \ref{sec:AppDrizzle}), giving higher spatial resolution data cubes. Unfortunately, due to PSF effects the quasar subtraction does not produce successful results using these cubes. As a result, we only use these cubes to qualitatively examine the spatial morphologies in greater detail, with all quantitative results instead using the 0.1''/pix cubes. We show the \oiiia\ maps, as measured from these 0.05''/pix data cubes in Figure \ref{fig:OIII0p05Cube}, and discuss this further in Appendix \ref{sec:AppDrizzle}.

\textbf{\DELS:}

\begin{table*}
\caption{Fluxes and velocities for the \ha, \hb\ and \oiii\ galaxy line components, 
from the integrated spectra over the regions shown in Figure \ref{fig:DELSRegions} and \ref{fig:VDESRegions} for \DELS\ (upper panels) and \VDES\ (lower panels). The values in braces give the integrated SNR of each line.
We present the results from both the N\&B quasar subtraction and the BLR quasar subtraction.
We fit the narrow line spectra with two components, a true narrow line (GN) as well as a galaxy `broad' wing (GB) component likely associated with outflows; note that this `broad' component is not from the quasar BLR, which has been subtracted from these spectra. In regions where no broader GB line component is detected, we fit only a single GN component.
Upper limits for undetected lines are 3$\sigma$ limits on the line flux, taken from the ERR array. Uncertainties on the velocity and velocity dispersion assume an uncertainty of $\pm1$ wavelength element, $\pm6.65$\AA\ or 51 km/s.
The velocity dispersions have been corrected for an instrumental broadening of FWHM$_{\rm{inst}}=115$ km/s. }
\begin{tabular}{lrrrrrrrr}
\hline 
\hline 
 & $F_{\rm{H\beta}} $ 
 & $F_{\rm{[O\textsc{iii}]{\lambda4959}}}$  
 & $F_{\rm{[O\textsc{iii}]{\lambda5007}}}$  
 & $F_{\rm{H\alpha}}$ 
 & $F_{\rm{[N\textsc{ii}]{\lambda6584\AA}}} $  
 & $V_r$  & $V_\sigma$ \\
  & \multicolumn{5}{c}{$(10^{-18}\rm{erg/s/cm}^2)$ [SNR]} & (km/s) & (km/s)\\

\hline
\hline
\DELS\\
\hline
\multicolumn{2}{l}{N\&B QUASAR SUBTRACTION}\\
\hline

Region 1, GN & 3.8$\pm$0.2 [23] & 5.6$\pm$0.2 [34] & 16.8$\pm$0.2 [92.4] & 9.0$\pm$0.4 [21] & <1.0 &65$\pm$51 &135$\pm$51 \\
Region 2, GN & 5.2$\pm$1.1 [4.8] & 6.8$\pm$0.9 [7.2] & 20.5$\pm$1.1 [19] & 11.0$\pm$2.3 [4.7] & <5.1 &163$\pm$51 &314$\pm$51 \\
Region 2, GB & <5.5 & 3.8$\pm$1.6 [2.5] & 11.6$\pm$1.7 [6.8] & <8.6 & <8.5 & -783$\pm$51 &526$\pm$51 \\
Region 3, GN & <0.4 & 0.8$\pm$0.1 [6.7] & 2.5$\pm$0.1 [18]& <0.8 & <0.8 & 396$\pm$51 &161$\pm$51 \\
Region 3, GB & <1.9 & 1.5$\pm$0.6 [2.4] & 4.4$\pm$0.6 [6.8] & <3.5 & <3.5 & 38$\pm$51 &764$\pm$51 \\

\hline
\multicolumn{2}{l}{BLR QUASAR SUBTRACTION}\\
\hline

Region 1, GN & 3.4$\pm$0.2 [20] & 5.6$\pm$0.2 [34] & 16.9$\pm$0.2 [92.9] & 9.4$\pm$0.4 [22] & <1.0  &64$\pm$51 &135$\pm$51 \\
Region 2, GN & 4.2$\pm$1.0 [4.1] & 9.8$\pm$0.9 [11] & 29.6$\pm$1.0 [28.7] & 30.1$\pm$2.3 [13] & <4.9  &22$\pm$51 &303$\pm$51 \\
Region 2, GB & <8.2 & 20.9$\pm$2.4 [8.8] & 62.8$\pm$2.5 [25] & <12.6 & <12.6 & -374$\pm$51 &807$\pm$51 \\
Region 3, GN & <0.4 & 0.8$\pm$0.1 [6.6] & 2.5$\pm$0.1 [18]& <0.8 & <0.8 & 392$\pm$51 &162$\pm$51 \\
Region 3, GB & <1.9 & 1.6$\pm$0.6 [2.7] & 5.0$\pm$0.6 [7.6] & <3.5 & <3.5 & 7$\pm$51 &764$\pm$51 \\
\hline
\hline
\VDES\\
\hline
\multicolumn{2}{l}{N\&B QUASAR SUBTRACTION}\\
\hline

Region 1, GN & 2.1$\pm$0.2 [11] & 5.4$\pm$0.2 [27] & 16.2$\pm$0.3 [82] & 7.9$\pm$0.6 [22] & <1.4 &-30$\pm$51 &319$\pm$51 \\
Region 2, GN & 2.3$\pm$1.2 [2.6] & 4.8$\pm$1.0 [5.3] & 14.4$\pm$1.0 [16] & 17.4$\pm$2.8 [11] & <5.9 &149$\pm$51 &369$\pm$51 \\
Region 2, GB & <6.7 & 2.4$\pm$1.9 [1.4] & 7.4$\pm$1.9 [4.3] & <11.5 & <11.2 & -877$\pm$51 &698$\pm$51\\
Region 3, GN & 1.2$\pm$0.2 [8.9] & 1.5$\pm$0.2 [11] & 4.6$\pm$0.2 [33] & 5.7$\pm$0.5 [19] & <1.1 &232$\pm$51 &250$\pm$51 \\
Region 4, GN & 1.4$\pm$0.2 [13] & 2.1$\pm$0.2 [19] & 6.2$\pm$0.2 [56] & 5.6$\pm$0.4 [21] & <0.9 &367$\pm$51 &189$\pm$51 \\
\hline

\multicolumn{2}{l}{BLR QUASAR SUBTRACTION}\\
\hline

Region 1, GN & 2.9$\pm$0.3 [11] & 5.8$\pm$0.3 [23] & 17.3$\pm$0.3 [63] & 8.3$\pm$0.7 [12] & <1.4  &-40$\pm$51 &332$\pm$51 \\
Region 2, GN & 4.5$\pm$0.9 [5.2] & 4.6$\pm$0.7 [6.5] & 13.7$\pm$0.7 [19] & 17.9$\pm$2.1 [8.5] & <4.0  &15$\pm$51 &263$\pm$51 \\
Region 2, GB & 16.0$\pm$2.9 [5.6] & 14.3$\pm$2.6 [5.6] & 43.1$\pm$2.6 [17] & <14.8 & <14.6 & -460$\pm$51 &942$\pm$51\\
Region 3, GN & 1.3$\pm$0.2 [6.1] & 1.5$\pm$0.2 [7.8] & 4.6$\pm$0.2 [23] & 5.7$\pm$0.5 [12] & <1.1  &227$\pm$51 &252$\pm$51 \\
Region 4, GN & 1.4$\pm$0.2 [8.6] & 2.1$\pm$0.2 [12] & 6.2$\pm$0.2 [34] & 5.6$\pm$0.4 [14] & <0.9  &367$\pm$51 &189$\pm$51 \\

\hline
\end{tabular} 
\footnotesize{}
\label{tab:HostFlux}
\end{table*}

\begin{table*}
\caption{Flux ratios, SFRs, dust properties, and dust-corrected SFRs for the emission-line regions in \DELS\ (upper rows) and \VDES\ (lower rows). The upper section lists the derived values from the N\&B quasar subtraction, and the lower section lists the values from the BLR quasar subtraction. Line ratios are quoted as values if both lines are detected, limits if one line is detected, and not quoted if neither are detected.
We note that we constrain $F_{\rm{[O\textsc{iii}]{\lambda5007}}}/F_{\rm{[O\textsc{iii}]{\lambda4959}}}=3$ in all regions.
The uncertainties on SFR consider only the flux measurement uncertainty, and no systematic uncertainty introduced by the conversion Eq. \ref{eq:Kenn}.
We quote the Region 2 SFR values as upper limits as the AGN photoionization will contribute, however these are calculated from a measured \ha\ flux, not a \ha\ limit, hence the quoted uncertainties on those values.
}
\begin{tabular}{lrrrrrrrrr}
\hline 
\hline 
 & $\frac{F_{\rm{H\alpha}}}{F_{\rm{H\beta}}}$ 
 & $\frac{F_{\rm{[O\textsc{iii}]{\lambda5007}}}}{F_{\rm{H\beta}}}$
 & $\frac{F_{\rm{[N\textsc{ii}]{\lambda6584\AA}}}}{F_{\rm{H\alpha}}}$ 
 & SFR 
 & E(B-V) 
 & $A_{\rm{H}\alpha}$
 & SFR$_{\rm{dust~ corr}}$ \\
 
 & & & &  \tiny{$(\rm{M}_\odot/yr)$} & (mag) & (mag) & \tiny{$(\rm{M}_\odot/yr)$}& \\

 \hline \hline
\multicolumn{2}{l}{\DELS}\\
 \hline 
\multicolumn{3}{l}{N\&B QUASAR SUBTRACTION}\\
\hline

Region 1, GN &2.4$\pm$0.2 & 4.5$\pm$0.2 & <0.1 &27$\pm$1 &&&&\\
Region 2, GN &2.1$\pm$0.6 & 4.0$\pm$0.9 & <0.5 &<33$\pm$7 & & & &\\
Region 2, GB &- & >2.1 &- & \\
Region 3, GN &- & >6.3 &- & & & & &\\
Region 3, GB &- & >2.4 &- & \\

\hline
\multicolumn{3}{l}{BLR QUASAR SUBTRACTION}\\
\hline

Region 1, GN &2.8$\pm$0.2 & 5.0$\pm$0.2 & <0.1 &29$\pm$1 & & &\\
Region 2, GN &7.1$\pm$1.8 & 7.0$\pm$1.7 & <0.2 &<91$\pm$7 & & & &\\
Region 2, GB &- & >7.6 &- & & & & &\\
Region 3, GN &- & >6.3 &- & & & & &\\
Region 3, GB &- & >2.7 &- & & & & &\\

 \hline \hline 
 \multicolumn{2}{l}{\VDES}\\
 \hline 
\multicolumn{4}{l}{N\&B QUASAR SUBTRACTION}\\

\hline

Region 1, GN &3.8$\pm$0.5 & 7.7$\pm$0.9 & <0.2 &24$\pm$2 & 0.2$\pm$0.1 & 0.8$\pm$0.4 & 51$\substack{ +50\\-22 }$ &\\
Region 2, GN &7.5$\pm$3.9 & 6.2$\pm$3.1 & <0.3 &<54$\pm$9 & & & &\\
Region 2, GB &- & >1.1 &- & & & & &\\
Region 3, GN &4.8$\pm$0.9 & 3.9$\pm$0.7 & <0.2 &18$\pm$2 & 0.4 $\pm$0.2& 1.5$\substack{ +0.5\\-0.6 }$ & 70$\substack{ +118\\-40 }$ &\\
Region 4, GN &4.0$\pm$0.5 & 4.4$\pm$0.5 & <0.2 &17$\pm$1 & 0.3 $\pm$0.1& 1.0$\substack{ +0.4\\-0.4 }$ & 42$\substack{ +42\\-18 }$ &\\

\hline
\multicolumn{3}{l}{BLR QUASAR SUBTRACTION}\\

\hline
Region 1, GN &2.9$\pm$0.3 & 6.0$\pm$0.5 & <0.2 &26$\pm$2 & 0.0$\pm$0.1 & 0.0$\pm$0.3 & 27$\substack{ +15\\-8 }$ &\\
Region 2, GN &3.9$\pm$0.9 & 3.0$\pm$0.6 & <0.3 &<55$\pm$6 & & & &\\
Region 2, GB &<0.9 &2.7$\pm$0.5 & - & & & & &\\
Region 3, GN &4.6$\pm$0.8 & 3.7$\pm$0.6 & <0.2 &18$\pm$2 & 0.4 $\pm$0.1& 1.3$\substack{ +0.5\\-0.6 }$ & 60$\substack{ +91\\-33 }$ &\\
Region 4, GN &4.0$\pm$0.5 & 4.4$\pm$0.5 & <0.2 &17$\pm$1 & 0.3 $\pm$0.1& 1.0$\substack{ +0.4\\-0.4 }$ & 42$\substack{ +42\\-18 }$ &\\
\hline

\end{tabular} 
\footnotesize{}
\label{tab:HostProperties}
\end{table*}

Our quasar-subtracted cube for \DELS\ reveals three extended narrow-line emission regions, as shown in Figure \ref{fig:DELSRegions} and labelled as Regions 1, 2, and 3. These three structures can be seen in both the 0.1''/pix and 0.05''/pix cubes. The integrated spectra for each of the regions are also shown in Figure \ref{fig:DELSRegions}.
Our kinematic maps for the subtracted cube, showing the flux of the lines, their relative velocity, and linewidth in each spaxel are shown in Figure \ref{fig:DELSMapsQSO} for the N\&B quasar subtraction and in Figure \ref{fig:MapsBroad} for the BLR quasar subtraction.
We fit the emission lines in the integrated spectra of each of the three regions with either one or two Gaussians, depending on whether a broader wing component is seen,
using the spectral fitting method used to construct our kinematic maps described in Section \ref{sec:Maps}.
For Region 2, we exclude the central 9 pixels surrounding the quasar peak, as these are highly corrupted by the quasar subtraction and introduce significant noise and artefacts.
The extracted flux values and velocities are given in Table \ref{tab:HostFlux} and the flux ratios and our derived physical properties are listed in Table \ref{tab:HostProperties}. 

Region 1 is a companion galaxy to the southeast of the quasar. The ellipse defining this region in Figure \ref{fig:DELSRegions} is offset $0.78''$ from the quasar position. In WMAP9 cosmology at $z=6.818$, $1''=5.446$ kpc, and so this offset is 4.26 kpc, quasar-to-centre. 
For reference, the region ellipse, which we fit by eye, has $a=0.814''=4.4$ kpc, $b=0.341''=1.9$ kpc and a PA of 54.6 degrees. By fitting a 2D Gaussian to the spatial distribution of the \oiiia\ flux in this region, we find this region can be best described with a $\sigma_a=(0.173\pm0.007)''=0.94\pm0.04$ kpc, an axis ratio of $0.70\pm0.04$, and a PA of $54\pm4$ degrees.
We note that this is an unconvolved Gaussian fit to the observed shape, we have not considered the PSF extent or shape to determine an intrinsic size. 

From the integrated spectrum, this region has a relative velocity shift of $+65\pm51$ km/s with respect to the quasar's peak \oiiia\ emission at $z=6.818$. The GN \hb, \oiii, and \ha\ lines are all detected with a peak SNR $>5$, with a line $\sigma$ of $135\pm51$ km/s. This region does not contain significant GB flux, with the lines well-fit by a single narrow Gaussian component.
Considering instead the cube where only the BLR component of the quasar has been subtracted, the \oiii\ properties of Region 1 are relatively unaffected, while the \hb\ and \ha\ fluxes slightly differ; these values are also given in Tables \ref{tab:HostFlux} and \ref{tab:HostProperties}. 
This region is spatially offset from the quasar such that narrow flux from the central spaxel, smeared by the PSF, does not contribute strongly. However, some quasar flux is present, hence the slightly different results for the different quasar subtractions.

Region 2 is the host galaxy of the quasar. 
The circle defining this region in Figure \ref{fig:DELSRegions} has a radius $r=0.458''$ or 2.5 kpc, for reference. By fitting a 2D Gaussian to the spatial distribution of the \oiiia\ flux in this region, we find this can be best described with a $\sigma_a=0.14\pm0.02''=0.75\pm0.09$ kpc, an axis ratio of $0.81\pm0.14$, and a PA of $83\pm20$ degrees.
The GN \hb, \oiii, and \ha\ lines are detected with a SNR $\geq17$, while the brighter \oiii\ lines also exhibit a broader (GB) wing, detected with SNR $\geq9$. 
The GN line is offset from the quasar at $z=6.818$ by $+163\pm51$ km/s, with a line $\sigma$ of $314\pm51$ km/s, while the broader wing is offset by $-783\pm51$ km/s with a line $\sigma$ of $526\pm51$ km/s. 

These line properties for Region 2 vary significantly when considering the BLR quasar subtracted cube, that has removed only the BLR model of the quasar, instead of the N\&B quasar subtracted cube, in which unresolved NLR flux from the centre of the galaxy is also removed.
The line properties for the BLR subtracted cube are presented in Table \ref{tab:HostFlux}, alongside the N\&B subtracted properties. The integrated spectrum of this region is also shown for the BLR-subtracted cube in Figure \ref{fig:DELSRegions}, and clearly contains significantly more flux than for the N\&B quasar subtracted cube.
This central region is coincident with the quasar, and so this BLR subtracted cube contains the unresolved narrow line flux from the NLR surrounding the quasar, spread across the image according to the PSF shape. This can also be seen in the flux maps for this BLR-subtracted cube, presented in Figure \ref{fig:MapsBroad}, which show the bright central narrow line emission.
In the BLR-subtracted cube there is significant GB flux, offset by $-374\pm51$ km/s from the quasar's GN \oiiia\ peak, with a line $\sigma$ of $807\pm51$ km/s. 

Region 3 sits to the northwest of the quasar. 
The ellipse we have fit by eye to define this region in Figure \ref{fig:DELSRegions} is offset $0.90''=4.9$ kpc from the quasar centre-to-centre, with $a=0.509''=2.8$ kpc, $b=0.235''=1.3$ kpc and a PA of 72.1 degrees. 
This region is not well fit by a 2D Gaussian, so we do not constrain its $\sigma$ and axis ratio as for the other regions.
From the integrated spectrum of this region there are significant detections of \oiii\ GN components and a broader \oiiia\ GB component, with a marginal detection of an \oiiib\ GB component. No \ha\ or \hb\ is significantly detected.
The GN line component is offset from the quasar at $z=6.818$ by $+396\pm51$ km/s, with a line $\sigma$ of $161\pm51$ km/s, while the broader GB wing is offset by $+38\pm51$ km/s with a line $\sigma$ of $764\pm51$ km/s. 
With its larger relative velocity and line width, and non-Gaussian spatial distribution, Region 3 is likely a tidal tail.
As for Region 1, the BLR subtracted cube produces consistent results to the N\&B quasar-subtracted cube in this region, as Region 3 is spatially offset enough that the narrow-line flux from the quasar has negligible contribution.\\

\begin{figure*}
\begin{center}
\includegraphics[scale=0.7]{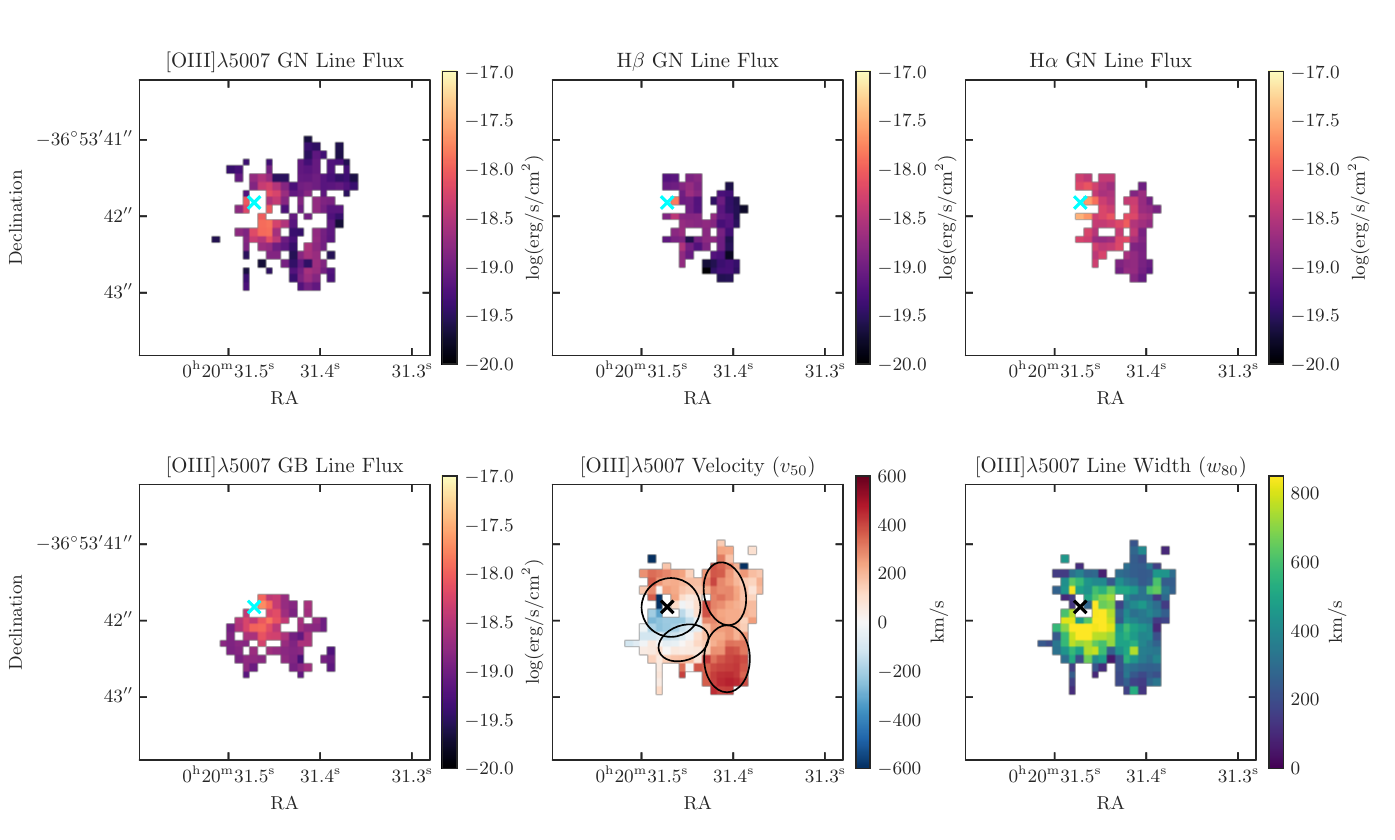}
\caption{Kinematic maps for \VDES, after the subtraction of the quasar emission using the N\&B subtraction technique. The top panels show the flux of the narrow component of the \oiiia, \hb, and \ha\ lines (GN), as fit by a Gaussian. 
The bottom left panel shows the flux of the broader GB component of the \oiiia\ line, a second Gaussian with a larger $\sigma$. 
The lower middle and right panels show our kinematic maps, showing the non-parametric central velocity of the line ($v_{50}$; middle) and the line width ($w_{80}$; right).
As these are non-parametric, this combines both the GN and GB components. 
We describe our method for creating these maps in Section \ref{sec:Maps}.
The black ellipses in the lower middle panel depict the regions as in Figure \ref{fig:VDESRegions}.}
\label{fig:VDESMapsQSO}
\end{center}
\end{figure*}

\begin{figure*}
\begin{center}
\includegraphics[scale=0.7]{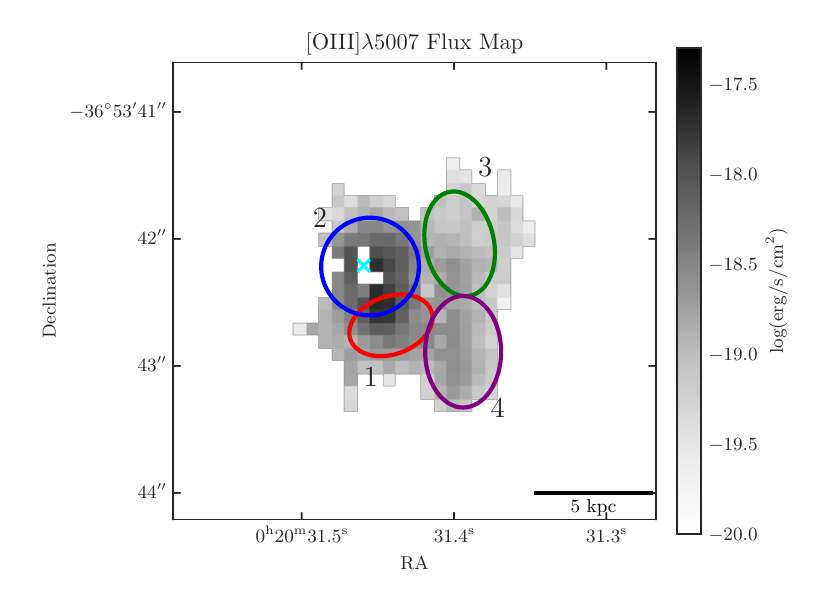}

\hspace{-1cm}
\includegraphics[scale=0.7]{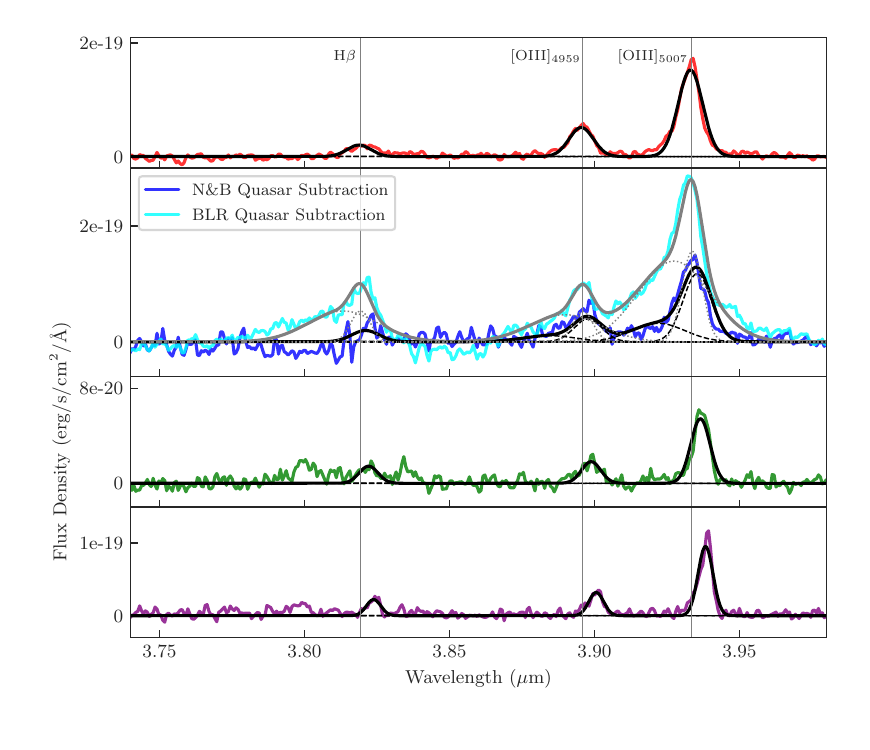}
\hspace{-0.5cm}
\includegraphics[scale=0.7]{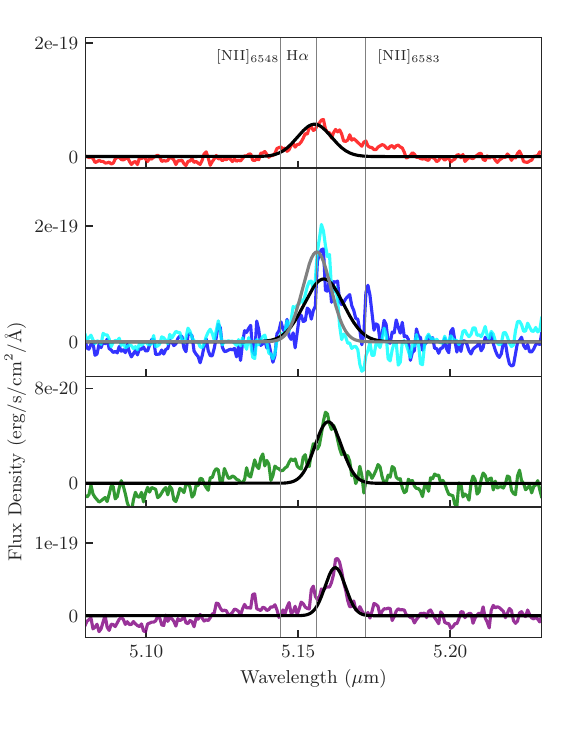}\hspace{-1cm}
\caption{Top: The \oiiia\ flux map for \VDES\ after N\&B quasar subtraction, with four regions of interest marked. Bottom: 
The integrated spectrum for the four regions (coloured lines), along with our best fit model in black. The middle panel, for Region 2, also shows the integrated spectra from the BLR quasar subtraction, in cyan, with the best fit model in grey. This region is the host of the quasar (position marked as a cyan cross in the top panel), and so the residual spectrum in this region depends heavily on the type of quasar subtraction. 
We note that to account for the overlapping area between Regions 1 and 2, we assign the flux in those spaxels to Region 1 only.
For Region 2, we exclude the central 9 pixels surrounding the quasar peak, as these are highly corrupted by the quasar subtraction and introduce significant noise and artefacts. This means that we will slightly underestimate the total flux in this region, however the fluxes will be significantly more reliable than if these 9 most corrupted spaxels were included.
In Regions 1, 3, and 4, the spectral lines can be described by a single narrow Gaussian (GN), whereas in Region 2 there is also a broader component which we fit with a second Gaussian  (GB); these separate components can be seen as the dashed black lines for the N\&B quasar subtraction, and dotted grey lines for the BLR quasar subtraction.
We note that the emission around 3.8$\mu$m in Regions 3 and 4 is likely an artefact in the cube. We do not see any significant detection of \nii.}
\label{fig:VDESRegions}
\end{center}
\end{figure*}

\textbf{\VDES:}

Our quasar-subtracted cube for \VDES\ also reveals three extended narrow-line emission regions, with their \oiiia\ flux and integrated spectra shown in Figure \ref{fig:VDESRegions}.  
The \oiiia\ maps made using the higher spatial resolution 0.05''/pix cubes (Figure \ref{fig:OIII0p05Cube}) expose a fourth emission line region, separated by only $\sim0.37''$ or 2 kpc (quasar peak to companion peak) to the south-southwest of the quasar. This is so close to the quasar host that in the 0.1''/pix cubes this appears to be one elongated host galaxy. Using the spatial structure from the higher resolution cube, we define four region ellipses from which we extract the spectra in each region. Thus, while the 0.1''/pix cubes can barely distinguish the host and this nearby companion galaxy, using the additional spatial information from the higher resolution cubes we are able to more accurately measure the host galaxy properties from the 0.1''/pix cubes, excluding the additional flux from the nearby companion.
The extracted flux values from each of the four region ellipses are given in Table \ref{tab:HostFlux}, with their flux ratios and derived physical properties in Table \ref{tab:HostProperties}. 
Our kinematic maps for the subtracted cube, showing the flux of the lines, their relative velocity, and linewidth in each spaxel are shown in Figure \ref{fig:VDESMapsQSO} for the N\&B quasar subtraction and Figure \ref{fig:MapsBroad} for the BLR quasar subtraction.

Region 1 is a companion galaxy to the south-southwest of the quasar, with such a small separation that it is barely resolvable in the 0.1''/pix cube. The ellipse defining this region in Figure \ref{fig:VDESRegions} is offset $0.52''$ from the quasar position. In WMAP9 cosmology at $z=6.854$, $1''=5.429$ kpc, and so this offset is 2.8 kpc, quasar-to-centre. The ellipse has $a= 0.338'' = 1.8$ kpc, $b= 0.228'' = 1.2$ kpc and a PA of 109 degrees.
As this region is blended with the host emission at this small separation, the emission does not follow a 2D Gaussian profile and so we do not quote the best fitting parameters as for the other regions.
From the integrated spectrum, this region has a relative velocity shift of $-30\pm51$ km/s with respect to the quasar's peak \oiiia\ emission at $z=6.854$. 
The GN \hb, \oiii, and \ha\ lines are all detected with a SNR $\geq11$, with a line $\sigma$ of $319\pm51$ km/s. These lines are well-fit by a single Gaussian component, with no secondary GB flux detected.
When using the BLR subtracted cube, the flux is measured to be slightly larger due to contamination by the narrow-line flux from the quasar at this separation. The flux seen around \niia\ in Region 1 is likely remaining broad \ha\ flux from the quasar.

Region 2 is the host galaxy of the quasar. 
For reference, the circle defining this region in Figure \ref{fig:VDESRegions} has a radius $r=0.385''=2.1$ kpc.
By fitting a 2D Gaussian to the spatial distribution of the \oiiia\ flux in this region, we find this can be best described with a $\sigma_a=(0.18\pm0.02)''=(1.0\pm0.1)$ kpc, an axis ratio of $0.66\pm0.12$, and a PA of $26\pm10$ degrees.
The GN \oiii, \hb\ and \ha\ lines are detected with a SNR $\geq2.6$. The brighter \oiii\ lines also exhibit a broader (GB) wing, detected with SNR $\geq1.4$. 
The GN line is offset from the quasar at $z=6.854$ by $+149\pm51$ km/s, with a line $\sigma$ of $369\pm51$ km/s, while the broader wing is offset by $-877\pm51$ km/s with a large line $\sigma$ of $698\pm51$ km/s.

As for the host region in \DELS, the \VDES\ line properties for Region 2 vary significantly when considering the BLR quasar subtracted cube. The line properties for the BLR subtracted cube are presented in Table \ref{tab:HostFlux}, alongside the N\&B subtracted properties, and the integrated spectrum for Region 2 for the BLR subtraction is 
also shown in Figure \ref{fig:VDESRegions}.
As for \DELS, this central region is coincident with the quasar, and so this BLR subtracted cube contains the unresolved narrow line flux from the NLR surrounding the quasar, spread across the image according to the PSF shape.
This can also be seen in the kinematic maps for this cube, presented in Figures \ref{fig:VDESMapsQSO} and \ref{fig:MapsBroad}, which show the bright central emission.

Region 3 is likely a second companion galaxy to the northwest of the quasar. 
The ellipse defining this region in Figure \ref{fig:VDESRegions} is offset $0.77''=4.17$ kpc from the quasar centre-to-centre, with $a=0.417''=2.3$ kpc, $b=0.271''=1.5$ kpc and a PA of 11 degrees. 
By fitting a 2D Gaussian to the spatial distribution of the \oiiia\ flux in this region, we find that this can be best described with a $\sigma_a=(0.4\pm0.1)''=(2.0\pm0.5)$ kpc, an axis ratio of $0.6\pm0.2$, and a PA of $193\pm5$ degrees.
In this region there are significant detections of GN \oiii, \hb\ and \ha\ components, with SNR $\geq8.9$.
The GN line component is offset from the quasar at $z=6.854$ by $+232\pm51$ km/s, with a line $\sigma$ of $250\pm51$ km/s. No broader GB component is detected. 
The properties of Region 3 are relatively unaffected when considering instead the BLR quasar subtracted cube; these are also given in Tables \ref{tab:HostFlux} and \ref{tab:HostProperties}. This region is spatially offset from the quasar such that no significant narrow flux from the central region contributes.

Region 4 is a companion galaxy to the southwest of the quasar. The ellipse defining this region in Figure \ref{fig:VDESRegions} is offset $1.03''=5.61$ kpc from the quasar position, quasar-to-centre. 
For reference, the region ellipse that we have fit by eye has $a=0.439''=2.4$ kpc, $b=0.299''=1.6$ kpc and a PA of 0 degrees. By fitting a 2D Gaussian to the spatial distribution of the \oiiia\ flux in this region, we find that this region can be best described with a $\sigma_a=0.43\pm0.08''=2.3\pm0.4$ kpc, an axis ratio of $0.47\pm0.13$, and a PA of $-4\pm2$ degrees. We note again that this is an unconvolved Gaussian fit to the observed shape.
From the integrated spectrum, this region has a relative velocity shift of $+367\pm51$ km/s with respect to the quasar's peak \oiiia\ emission at $z=6.854$. 
The GN \hb, \oiii, and \ha\ lines are all detected with a SNR $\geq13$, with a line $\sigma$ of $189\pm51$ km/s. These lines are well-fit by a single Gaussian component, with no GB flux detected.
As for Region 3, the BLR subtracted cube has consistent results to the N\&B quasar-subtracted cube, as Region 4 is spatially offset enough that the narrow-line flux from the quasar has negligible contribution.

\subsubsection{Limits on Star Formation Rates and Excitation Mechanisms}
\label{sec:SFRs}

Using our integrated spectrum for each region, we estimate the SFR using the GN \ha\ line flux.
Assuming no extinction, a solar abundance, and a Salpeter IMF, the SFR can be estimated as:
\begin{equation}
\textrm{SFR} = 5.37\times 10^{-42} \frac{L_{\rm{H}\alpha}}{(\textrm{erg/s})} \rm{M}_\odot / \rm{yr}
\label{eq:Kenn}
\end{equation}
\citep[][note that these values are 68\% of those calculated assuming the \citealt{kennicutt_1998} relation]{Kennicutt2012}. We use this equation to calculate the SFR for each of the regions in which \ha\ is detected. The resulting SFRs for \DELS\ and \VDES\ are given in Table \ref{tab:HostProperties}.

This equation assumes that all of the \ha\ emission is from star formation; for our host galaxies, however, we note that even for our N\&B quasar subtracted cubes we expect some AGN-photoionized narrow line flux to contribute.
The N\&B quasar subtraction removes all of the GN component that is emitted from the central unresolved region surrounding the quasar, although spatially resolved GN flux from both the NLR and star-formation remains. Thus, while the N\&B quasar subtracted cubes provide the most accurate measurement of the SFRs of our emission-line regions, in comparison to the BLR subtracted cubes, for our host components these are upper limits. 
For comparison, we also include the SFRs calculated using the BLR quasar subtracted cube in 
Table \ref{tab:HostProperties}, which includes all quasar NLR emission, showing the significant improvement that the narrow-line quasar subtraction can have on being able to measure host galaxy properties.
For \DELS, the host SFR measured from the N\&B subtraction is $<33 ~\rm{M}_\odot$/yr, whereas the BLR subtraction can only constrain the SFR to $<91 ~\rm{M}_\odot$/yr, a much larger upper limit.
The host galaxy of \VDES\ has SFR constraints of  $<55~\rm{M}_\odot$/yr, consistent for both the BLR and N\&B subtractions, likely due to the noisy subtraction around \ha.

In the absence of dust extinction, theoretically the ratio of \ha\ to \hb\ flux is $F_{\rm{H}\alpha}/F_{\rm{H}\beta} \simeq 2.86$, estimated using a temperature
$T = 10^4$ K and an electron density $n_e = 10^2
\rm{cm}^{-3}$
for Case B recombination \citep{Osterbrock1989,Dominguez2013}.
Hence, the amount of dust can be estimated from these two lines via $E(B-V) = 1.97 \log_{10}[ (F_{\rm{H}\alpha}/F_{\rm{H}\beta}) / 2.86]$ following \citet{Dominguez2013}, and so $A_{\rm{H}\alpha} = (3.33 \pm 0.80) \times E(B-V )$ assuming the \citet{Calzetti2000} reddening law. For companion regions where both \ha\ and \hb\ are significantly detected and $F_{\rm{H\alpha}}/F_{\rm{H\beta}}>2.86$, we calculate $E(B-V)$, $A_{\rm{H}\alpha}$ and a dust-corrected SFR, which are given in Table \ref{tab:HostProperties}.
As the host regions contain NLR quasar-photoionized flux which can have significantly higher \ha/\hb\ ratios than the $2.86$ assumed here, we do not calculate dust-corrected SFR limits for the host galaxies.

For \DELS, the companion galaxy Region 1 has a SFR of $27\pm1 ~\rm{M}_\odot$/yr, as measured from the N\&B quasar subtracted spectrum. As the $F_{\rm{H\alpha}}/F_{\rm{H\beta}}$ ratio is $<2.86$, we do not estimate the dust corrected values. These flux ratios are unlikely to be truly $<2.86$, with this more likely being an issue with the quasar subtraction technique and its removal of flux, as the N\&B subtraction has $F_{\rm{H\alpha}}/F_{\rm{H\beta}}=2.4\pm0.1$ while the BLR subtraction has $F_{\rm{H\alpha}}/F_{\rm{H\beta}}=2.8\pm0.1$. The BLR quasar subtraction results in slightly more \ha\ flux, corresponding to a SFR of $29\pm1~\rm{M}_\odot$/yr, as well as a lower \hb\ flux which also contributes to the larger $F_{\rm{H\alpha}}/F_{\rm{H\beta}}$. For this region only, we therefore consider the BLR subtraction to provide the best estimate of the SFR, $29\pm1~\rm{M}_\odot$/yr, whereas for all other regions the N\&B subtraction provides the best estimate.

For \VDES, the companion galaxy Region 1 has a SFR of $24\pm2~\rm{M}_\odot$/yr as measured from the N\&B quasar subtracted spectrum, or  51$\substack{ +50\\-22}~\rm{M}_\odot$/yr when corrected for dust attenuation. Using the BLR subtraction, the SFR is similarly $26\pm2~\rm{M}_\odot$/yr. However, this subtraction results in a much lower \ha/\hb\ ratio of $2.91\pm0.3$, due to contamination from the unresolved quasar emission, and so in this case there is minimal change in the SFR when corrected for dust attenuation, 27$\substack{ +15\\-8}~\rm{M}_\odot$/yr .
The companion galaxy Region 3 has a SFR of $18\pm2~ \rm{M}_\odot$/yr, as measured from the N\&B quasar subtracted spectrum, or  70$\substack{ +118\\-40}~\rm{M}_\odot$/yr when corrected for dust attenuation, due to its large \ha/\hb\ ratio.
Region 4 has a SFR of $17\pm1 ~\rm{M}_\odot$/yr, as measured from both the N\&B quasar subtracted and BLR quasar subtracted spectrum.
When corrected for dust-attenuation, this SFR increases to 42$\substack{ +42\\-18 } ~\rm{M}_\odot$/yr.
The subtraction method makes little difference for the companion regions 3 and 4, which are sufficiently spatially separated from the quasar such that the unresolved quasar emission has little contribution (see Table \ref{tab:HostProperties}).

We note that our quasar hosts and companion galaxies have no detected \nii\ emission, implying low metallicities. For our SFR calculation we use the typically used conversion from \ha\ in Equation \ref{eq:Kenn}, which is calculated for solar abundances. 
\citet{Lee2009} recalculated the SFR--$L_{\rm{H}\alpha}$ relation for a range of metallicities using the \citet{bruzual_2003} population synthesis models. Assuming a case B recombination with nebular temperature of $10^4$ K, a density of 100 cm$^{-3}$, and a Salpeter IMF, \citet{Lee2009} found that the ratio $\textrm{SFR}/L_{\rm{H}\alpha}$ in Equation \ref{eq:Kenn} decreases from  $7.9\times 10^{-42}$ at $Z=Z_\odot$, to $6.2\times 10^{-42}$ at $Z=0.2 Z_\odot$ and $5.2\times 10^{-42}$ at $Z=0.02Z_\odot$. This would reduce our calculated SFRs by 22\% and 34\%, respectively.

We do not detect any \nii\ emission from our quasars, their hosts or their companions. To place upper limits on the amount of \niia\ present in the different emission-line regions, we integrate the ERR array at the expected location of the line, assuming a line width equal to that of \oiiia.
%insert artificial lines into the integrated region spectra and measure the mean flux at which the line would have been successfully recovered by our algorithm. We simulate the line to have a Gaussian profile, with line width equal to the \oiiia\ GN line, centred at 50 different wavelength locations within the line-free \ha\ region to account for variations due to the existing noise. We vary the amplitude and thus flux until the line is detected. 
Our \niia\ limits are given in Table \ref{tab:HostFlux}, with the resulting \niia/\ha\ flux ratio limits in Table \ref{tab:HostProperties}. 

\begin{figure}
\begin{center}
\includegraphics[scale=0.8]{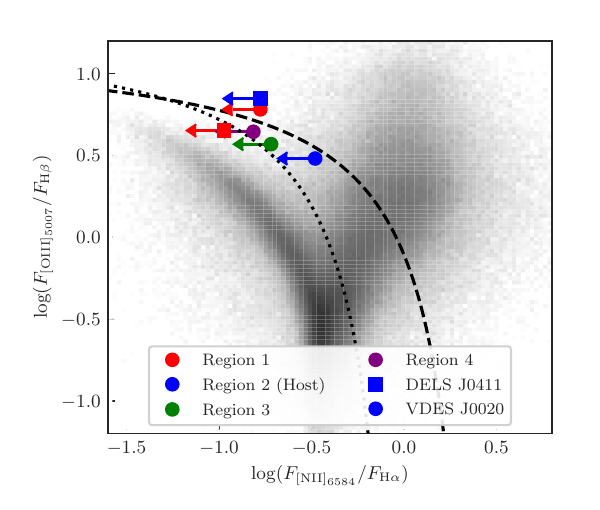}
\caption{The BPT diagnostic diagram showing the flux ratios of the various emission regions for \DELS\ (squares) and \VDES\ (circles); the $F_{\rm{[N\textsc{ii}]{\lambda6584\AA}}}/F_{\rm{H\alpha}}$ ratios are all upper limits due to a non-detection of \niia. The different colours represent the different emission regions,  following the colour scheme in Figures \ref{fig:DELSRegions} and \ref{fig:VDESRegions}. The flux ratios are measured from the BLR-subtracted cube.
For reference, we show the \citet{kauffmann_2003} (black dotted) and \citet{Kewley2001} (black dashed) curves which attempt to differentiate between galaxies primarily ionised by star formation (lower left) and AGN (upper right) at low-$z$.
 We also show local sources from the SDSS \citep{Abazajian2009} as the 2D histogram, for comparison.}
\label{fig:BPT}
\end{center}
\end{figure}

Using our measured flux values and limits, we are able to place our emission line regions on a BPT diagram \citep{Baldwin1981}, which we show in Figure \ref{fig:BPT}.
We only consider regions where both the \ha\ and \hb\ lines are significantly detected (see Table \ref{tab:HostProperties}).
This diagnostic aims to characterise the main photoionization source based on the narrow line flux. For this we use the BLR quasar subtracted cube, as this gives the purest measure of the full narrow line flux, whereas in the N\&B subtraction a large portion of the quasar NLR flux has been removed.

We find that for both quasars, the companion emission regions (Region 1 for \DELS\ and Regions 1, 3 and 4 for \VDES) have $\log(F_{\rm{[O\textsc{iii}]{\lambda5007}}}/F_{\rm{H\beta}})$ between 0.55 and 0.78.
\DELS\ Region 1 lies within the region of the diagram typically populated by galaxies that are primarily ionised by star-formation.
The \VDES\ companion regions could lie in the star-forming regime \emph{or} in the transition region, and Region 1 could also lie within the AGN regime, depending on the ratio $F_{\rm{[N\textsc{ii}]{\lambda6584\AA}}}/F_{\rm{H\alpha}}$, for which we only have upper limits due to a lack of \niia\ detected in our sources.

The quasar host galaxies (Region 2) have $\log(F_{\rm{[O\textsc{iii}]{\lambda5007}}}/F_{\rm{H\beta}})=0.85$ and 0.48 for \DELS\ and \VDES, with $\log(F_{\rm{[N\textsc{ii}]{\lambda6584\AA}}}/F_{\rm{H\alpha}})<-0.77$ and $-0.48$.  
For \DELS\ we find that the host lies in the upper left of the diagram, which is sparsely populated by low-$z$ objects, although lies within the AGN-dominated regime according to the low-$z$ \citet{Kewley2001} demarcation. For \VDES, the quasar host lies within the low-$z$ transition region or star-forming regime. 
Theoretical models predict that high-$z$ AGN will reside in the left side of the BPT diagram, in the AGN regime that is sparsely populated at $z=0$ as well in the star-forming regime, mainly due to their lower metallicities \citep[see e.g.][]{Feltre2016,Strom2017,Nakajima2022,Hirschmann2022}. This is consistent with our measurements for both quasar host galaxies. For further discussion on this topic, see \citet{Uebler2023}.

To accurately determine the primary photoionization mechanism within these various regions, we would require either deeper observations for a \niia\ detection or tighter limit, or alternative line-ratio diagnostic methods. 
We could use a $[\ions{S}{ii}]~ {\lambda\lambda6717,6731}$-based diagnostic \citep[e.g.][]{Veilleux1987}, however $[\ions{S}{ii}]~ {\lambda\lambda6717, 6731}$ emission was not detected for these quasars. 
For the integrated quasar spectrum, by integrating the corresponding ERR array across the expected location of the $[\ions{S}{ii}]~ {\lambda6717}$ line, assuming the same line width as the narrow \ha\ component (Table \ref{tab:QSOLines}), and applying our aperture correction, we find $F_{[S\rm{II}]\lambda6717}<2.8\times10^{-18} ~ \rm{erg/s/cm}^2 (3\sigma)$ for \DELS. 
For \VDES, this line is beyond the wavelength coverage of the detector.
These lines are either very close to or just above the detectable wavelength range for this G395H/F290LP NIRSpec configuration, making accurate measurements difficult if a line was detected.
We have also not detected any $\ions{\rm{He}}{ii}~ {\lambda4686}$ emission, which is particularly useful for distinguishing star formation and AGN dominated regions in the high-$z$ Universe where metallicities are low \citep[see e.g.][]{Nakajima2022,Uebler2023}. 
For the integrated quasar spectrum, following the same process for $[\ions{S}{ii}]~ {\lambda6717}$ but instead assuming the same line width as the narrow \oiiia\ component (Table \ref{tab:QSOLines}), we find $F_{\rm{HeII}\lambda4686}<2.3\times10^{-18} ~ \rm{erg/s/cm}^2 (3\sigma)$ for \DELS\ and $F_{\rm{HeII}\lambda4686}<1.6\times10^{-18}~ \rm{erg/s/cm}^2 (3\sigma)$ for \VDES.

\subsubsection{Dynamical Masses and Black Hole--Dynamical Mass Ratios}

\begin{table*}
\caption{Dynamical masses, and values used in their calculation, for the emission-line regions in \DELS\ (upper rows) and \VDES\ (lower rows), using the N\&B quasar subtraction. \DELS\ Region 3 and \VDES\ Region 1 cannot be well described by a 2D Gaussian, and so we do not measure their dynamical mass.
}
\begin{tabular}{lrrrrrrrrrrrr}
\hline 
\hline 
& $R$
& $\sigma$
& $a/b$
& $i$
 & $M_{\rm{dyn, disp}}$ 
 & $M_{\rm{dyn, rot}}$
 & $M_{\rm{dyn, rot}}$, \tiny{$i=55^\circ$} 
 & $M_{\rm{dyn, vir}}$ \\
 
 & (kpc) & (km/s) & & (deg) & \tiny{($10^{10}\rm{M}_\odot$)}& \tiny{($10^{10}\rm{M}_\odot$)} & \tiny{($10^{10}\rm{M}_\odot$)} & \tiny{($10^{10}\rm{M}_\odot$)}  \\

 \hline \hline
\multicolumn{2}{l}{\DELS}\\
 \hline 
Region 1, GN &0.94$\pm$0.04 &135$\pm{ 51 }$ &0.70 $\pm$ 0.04 &45$\pm3$ &0.6$\substack{ +0.6\\-0.4 }$ & 2.5$\substack{ +3.1\\-1.6 }$ & & 2.0$\substack{ +1.9\\-1.2 }$\\
Region 2, GN &0.75$\pm$0.09 &314$\pm{ 51 }$ &0.81 $\pm$ 0.14 &36$\substack{ +12\\-17 }$ &2.6$\substack{ +1.3\\-1.0 }$ & 16$\substack{ +59\\-9 }$ & $8.0\substack{ +4.1\\-3.0 }$ & 8.6$\substack{ +4.4\\-3.3 }$\\

 \hline \hline
\multicolumn{2}{l}{\VDES}\\
 \hline 
Region 2, GN &0.97$\pm$0.13 &369$\pm{ 51 }$ &0.66 $\pm$ 0.12 &49$\substack{ +8\\-10 }$ &4.6$\substack{ +2.1\\-1.6 }$ & 17$\substack{ +18\\-8 }$ & $14\substack{ +7\\-5 }$ & 15$\substack{ +7\\-5 }$\\
Region 3, GN &2.15$\pm$0.60 &250$\pm{ 51 }$ &0.58 $\pm$ 0.23 &54$\substack{ +15\\-19 }$ &4.7$\substack{ +4.0\\-2.5 }$ & 15$\substack{ +39\\-10 }$ & & 16$\substack{ +13\\-8 }$\\
Region 4, GN &2.46$\pm$0.51 &189$\pm{ 51 }$ &0.46 $\pm$ 0.14 &62$\pm9$ &3.1$\substack{ +2.9\\-1.8 }$ & 8.1$\substack{ +11.2\\-5.1 }$ & & 10$\substack{ +10\\-6 }$\\
\hline

\end{tabular} 
\footnotesize{}
\label{tab:DynamicalMasses}
\end{table*}

To calculate the dynamical masses of the various emission regions, we follow the method as in \citet{Decarli2018} for ALMA quasar host studies. Firstly, we assume each region is dispersion-dominated, with:
\begin{equation}
M_{\rm{dyn, dispersion}}=\frac{3}{2} \frac{R\sigma^2}{G}
\end{equation}
where $R$ is the radius of the region, $\sigma$ the integrated line width, and $G$ the gravitational constant. As an alternative, we assume that the regions are rotation-dominated thin discs, with:
\begin{equation}
M_{\rm{dyn, rotation}}=\frac{R}{G} \left(\frac{0.75~ \rm{FWHM}}{\sin(i)}\right)^2
\end{equation}
where 0.75 FWHM scales the line FWHM to the width of the line at 20\% of the peak as in \citet{Willott2015}, and $i$ is the inclination angle of the thin disc. 
Finally, we also consider the virial approximation 
\begin{equation}
{M}_{\rm{dyn, virial}}\approx \frac{5R{\sigma}^{2}}{G}.
\end{equation}
This commonly-used scaling factor of 5 was found to be the optimal scaling factor when calibrated to a sample of 25 galaxies by \citet{Cappellari2006} \citep[see e.g.][for further discussion]{VanDerWel2022}. This estimate will be a factor of 10/3 larger than $M_{\rm{dyn, dispersion}}$ by construction.
As it is most common for high-$z$ quasar host studies to assume a rotating disc geometry \citep[e.g.][]{Wang2013,Willott2015,Decarli2018}, we choose our best dynamical mass estimate to be $M_{\rm{dyn, rotation}}$.

To measure the radii, we fit a 2D Gaussian to the spatial distribution of the \oiiia\ emission from each region, and take $R$ to be the $\sigma$ of the semi-major axis; these values are described in Section \ref{sec:Structure}. From this Gaussian fit we also measure the axis-ratio, $b/a$, from which we derive the inclination angle, $i$, via $\cos(i) = b/a$.
As above, we note that $R$ and $b/a$ have not been deconvolved by the PSF.
For the emission line widths we take the GN \oiiia\ component, which is the brightest of the lines we have measured. We note that this line width represents the total dynamics of the gas, including rotation as well as non-rotational components such as dispersion and winds \citep[see e.g.][]{Wisnioski2018}.
We consider only the N\&B quasar subtracted cube for this calculation, since otherwise the measurements could be skewed by the unresolved NLR regions around the quasar. The kinematics of this central NLR are not dominated by the larger-scale galaxy potential, thus this emission does not trace the dynamical mass. Considering only the N\&B quasar subtracted cube ensures that we are measuring the spatially extended host regions and not the PSF shape, which is particularly important for the host regions (Region 2). The values used in these calculations and the resulting dynamical mass estimates are listed in Table \ref{tab:DynamicalMasses}. 

For \DELS, for Region 1 we find a dynamical mass between 0.6--2.5$\times10^{10}\rm{M}_\odot$, with a best estimate of $M_{\rm{dyn, rotation}}=(2.5\substack{ +3.1\\-1.6})\times10^{10}\rm{M}_\odot$, from an inclination angle $i=45\pm3$ degrees.
For the host Region 2, we find that the rotational measure for the dynamical mass, $M_{\rm{dyn, rotation}}=(16\substack{ +59\\-9 })\times10^{10}\rm{M}_\odot$, is larger than the two other estimates, $M_{\rm{dyn, dispersion}}=(2.6\substack{ +1.3\\-1.0 })\times10^{10}\rm{M}_\odot$ and  $M_{\rm{dyn, virial}}=(8.6\substack{ +4.4\\-3.3 })\times10^{10}\rm{M}_\odot$, albeit with large uncertainties particularly due to the inclination angle, $i=36\substack{ +12\\-17}$ degrees. This difference is due to our measured inclination angle for this region being relatively low, and so the inclination correction is relatively large. 
For their ALMA quasar host galaxies that are marginally resolved, and thus where their axis ratio cannot be accurately measured, \citet{Willott2015} and \citet{Decarli2018} assume an inclination of 55 degrees, the median inclination of the resolved sources in \citet{Willott2015} and \citet{Wang2013}. As our quasar hosts are difficult to measure, requiring the quasar subtraction, and they are interacting systems which complicates their structure, we are not confident in our inclination measurements. If we instead assume the median $i=55$ degrees, the rotational dynamical mass is $(8.0\substack{ +4.1\\-3.0})\times10^{10}\rm{M}_\odot$, similar to the virial estimate. For Region 2 of \DELS, we therefore assume a best dynamical mass estimate of $M_{\rm{dyn, virial}}=(8.6\substack{ +4.4\\-3.3})\times10^{10}\rm{M}_\odot$.
Region 3 cannot be well described by a 2D Gaussian, and so we cannot perform our fit as for the other regions.

\VDES\ Region 1 also cannot be well described by a 2D Gaussian, due to contamination by the quasar host, and so we do not estimate its dynamical mass.
For the \VDES\ host Region 2, we find a dynamical mass between 4.6--17 $\times10^{10}\rm{M}_\odot$, with a best estimate of $M_{\rm{dyn, rotation}}=(17\substack{ +18\\-8})\times10^{10}\rm{M}_\odot$, from an inclination of $i=49\substack{ +8\\-10}$ degrees. This is similar to the median inclination of the resolved sources in \citet{Willott2015} and \citet{Wang2013}, $i=55$ degrees, which would reduce the value slightly to $M_{\rm{dyn, rotation}}=(14\substack{ +7\\-5})\times10^{10}\rm{M}_\odot$.
Region 3 has dynamical mass estimates of 4.7--15$\times10^{10}\rm{M}_\odot$, with our best estimate $M_{\rm{dyn, rotation}}=(15\substack{ +39\\-10 })\times10^{10}\rm{M}_\odot$ from an inclination of $54\substack{+15\\-19}$ degrees.
For \VDES\ Region 4 we find a dynamical mass between 3.1--10$\times10^{10}\rm{M}_\odot$, with a best estimate from $M_{\rm{dyn, rotation}}$ of $(8.1\substack{ +11.2\\-5.1})\times10^{10}\rm{M}_\odot$ with an inclination of $i=62\pm9$ degrees.

\begin{figure}
\begin{center}
\includegraphics[scale=0.8]{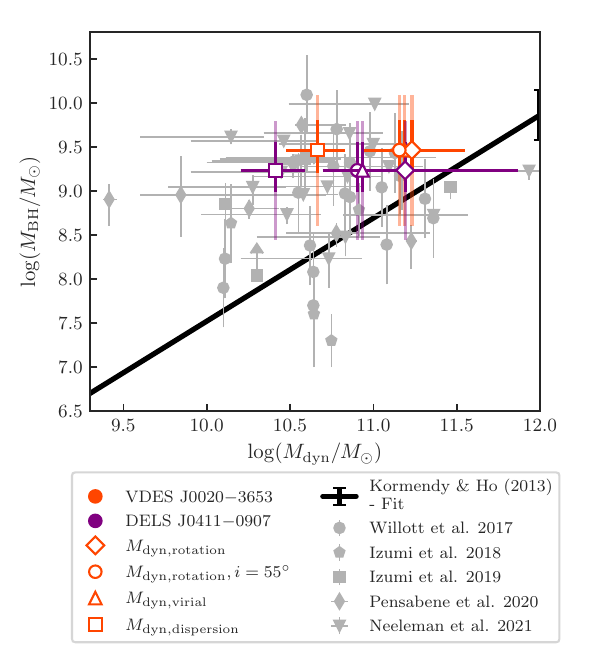}
\caption{The dynamical mass--black hole mass relation. Our two quasars \DELS\ (purple) and \VDES\ (orange) are shown, with the four $M_{\rm{dyn}}$ estimates, $M_{\rm{dyn, rotation}}$, 
$M_{\rm{dyn, rotation}}$ assuming an inclination angle of 55 degrees, $M_{\rm{dyn, dispersion}}$ and $M_{\rm{dyn, virial}}$. 
The solid black hole mass error bars show the 1$\sigma$ uncertainty on our best estimate of the black hole mass, while the transparent error bars very conservatively show the range of the lower limit of the lowest mass estimate to the upper limit of the largest mass estimate.
The black line shows the local \citet{Kormendy2013} relation. The grey points show a range of observations of high-$z$ quasars as measured via their cold gas from ALMA \citep{Willott2017,Izumi2018,Izumi2019,Pensabene2020,Neeleman2021}.}
\label{fig:DynamicalMasses}
\end{center}
\end{figure}

We plot the dynamical mass--black hole mass relation in Figure \ref{fig:DynamicalMasses} for these two quasars. We find that both of our quasars lie above the local \citet{Kormendy2013} black hole--stellar mass relation, although these are consistent within the scatter and uncertainties. 
Previous ALMA observations have found that luminous quasars generally lie above the local black hole--host relation, while lower luminosity quasars lie around or below the relation \citep[see e.g.][]{Willott2017,Izumi2019}. Our luminous quasars lie above the relation, which is consistent with this existing picture, however a larger sample of high-$z$ quasars is required to test this scenario with NIRSpec. For a true comparison to the local black hole--stellar mass relation, we require stellar and not dynamical masses. 

We note that in the current observations the faint host galaxy continuum emission is not detectable, and so we cannot measure their stellar masses. 
Our high spectral resolution ($R\simeq2700$) observations were designed to study the emission lines in detail.
This sacrifices the sensitivity that is required to detect the host continuum, which is expected to be achievable in similar exposure times at lower spectral resolution \citep[either $R\sim100$ IFU spectra, or alternatively images as in e.g.][]{Ding2022,Harikane2023}. 
Future work within the GA-NIFS GTO program will use the $R\sim100$ prism mode to search for the continuum emission of several high-$z$ quasar host galaxies, which would make stellar mass measurements possible.

In Figure \ref{fig:DynamicalMasses} we also find that both of our quasars have dynamical masses and black hole--dynamical mass ratios consistent with ALMA observations of $z\gtrsim6$ quasars. 
However, we note that \DELS\ and \VDES\ do not have published ALMA dynamical mass estimates that could be used for a direct comparison of these methods for our targets. A detailed comparison of ALMA and JWST NIRSpec dynamical mass and JWST stellar mass estimates will be a key aim of future quasar studies, which will allow for a much deeper understanding of the high-$z$ black hole--host mass relations.

\subsection{Outflow properties}
\label{sec:outflows}

\begin{table*}
\caption{The properties of the quasar outflows as measured in Section \ref{sec:outflows}, for \DELS\ and \VDES. The luminosities and velocities are taken from the \hb\ and \oiiia\ GB components from the full integrated quasar spectrum (Figure \ref{fig:QuasarFit} and Table \ref{tab:QSOLines}), while the outflow extent $R_{\rm{out}}$ is measured from the spatial distribution of \oiiia\ where $w_{80}>600$ km/s, in the BLR-subtracted cube.
}
\begin{tabular}{lllllllllll}
\hline 
\hline 
&$L_{\rm{H\beta,GB}}$\hspace{-0.2cm} & $L_{\rm{[OIII],GB}}$ & $v_{\rm{out}}$ & $R_{\rm{out}}$ & $M_{\rm{out,H\beta}}$\hspace{-0.2cm} & $M_{\rm{out,[OIII]}}$ & $\dot{M}_{\rm{out,H\beta}}$\hspace{-0.2cm} & $\dot{M}_{\rm{out,[OIII]}}$& $\dot{P}_{\rm{out,H\beta}}$\hspace{-0.2cm} & $\dot{P}_{\rm{out,[OIII]}}$\\

&\multicolumn{2}{c}{\tiny{($10^{43}$ erg / s)}} &  \tiny{(km / s)}  & \tiny{(kpc)} & \multicolumn{2}{c}{\tiny{($10^{7}M_\odot$)}} & \multicolumn{2}{c}{\tiny{($M_\odot$ / yr)}} &\multicolumn{2}{c}{\tiny{($10^{43}$ erg / s)}}\hspace{-0.2cm}\\
\hline 
\hline

\DELS\ & $1.0 \substack{ +0.8\\-0.7 }$ & $13.5 \pm0.6$ & 
      $1296\pm31$ & $4.0\pm0.4$ & 
      $17\substack{ +13\\-11 }$ & $10.8\substack{ +0.5\\-0.4 }$ 
      & $58\substack{ +44\\-37 }$ & $36.2\substack{ +1.9\\-1.7 }$
      & $3.1\substack{ +2.3\\-2.0 }$ & $1.9\pm0.1$
      \\
      
\vspace{0.2cm}

\VDES\ & $5.7 \substack{ +0.8\\-1.0 }$ & $9.6 \substack{ +0.5\\-0.4 }$ & 
      $1749\pm55$ & $3.3\pm0.4$ & 
      $97\substack{ +14\\-17 }$ & $7.7\substack{ +0.4\\-0.3 }$ 
      & $525\substack{ +75\\-92 }$ & $41.4\substack{ +2.6\\-2.2 }$
      & $51\pm9$ & $4.0\pm0.3$
      \\
\hline
\end{tabular} 

\footnotesize{}
\label{tab:Outflows}
\end{table*}

The clear detection of an outflow component, as traced by the broad kinematic component (GB components in Figure \ref{fig:QuasarFit}) via the analysis of the \oiiia\ velocity and spatial distribution, opens the possibility of investigating the ionized outflow properties in these early quasars. We note that this is the first time that it is possible to properly characterise the ionized outflows in quasars in the Epoch of Reionisation. Previous studies have relied on spatially unresolved information, and in particular on the detection of UV broad and blueshifted absorption features of UV resonant lines \citep[e.g.][]{Wang2018,Ginolfi2018,Onoue2019,Bischetti2022,Bischetti2023}. These features trace primarily the nuclear outflow, which is generally a small fraction of the global, galactic scale outflow, and the lack of spatial information generally prevents a reliable estimate of the outflow parameters such as the outflow rate, kinetic power and momentum rate. We also note that NIRSpec fixed-slit spectroscopic observations of these two quasars covering the rest-frame UV do not reveal any broad absorption features (Willott, priv.comm.), indicating that the presence of ionized outflow is not necessarily traced by broad absorption features in the UV part of the spectrum \citep[see e.g.][]{Rankine2020}.

In this work we have found outflow signatures within \ha, \hb\ and \oiii\ for both quasars (GB components, Figure \ref{fig:QuasarFit}).
In this section we use the flux, velocity and spatial information of the \hb\ and \oiiia\ broad galaxy (GB) components to infer some of the basic properties of the ionized outflow. 
We use the GB fluxes and velocities measured from our full integrated spectral fit for the quasar and host galaxy, prior to quasar subtraction, as described in Section \ref{sec:QuasarLineFitting} and shown in Figure \ref{fig:QuasarFit}. This gives the most reliable measurements of the integrated quantities, as these are not affected by the quasar subtraction technique.
To measure the spatial extent of the outflow, we use the quasar BLR-subtracted cube, from which we can measure the \oiiia\ line width in each spaxel without contamination from the quasar's BLR. For this we consider the full cube, without splitting the emission into regions as in Section \ref{sec:HostProperties} above. The broad flux emanates from the quasar, approximately covering Region 2 for both quasars, as well as extending into Region 3 for \DELS---this can be seen in the $w_{80}$ maps in Figure \ref{fig:MapsBroad}.

To derive the mass of the ionized outflow as inferred from the blueshifted ``outflow'' component of \hb, we use the equation
\begin{equation}
M_{\rm{out}} = 1.7\times10^9 {C}
\left(\frac{L_{\rm{H}\beta}}{10^{44}\rm{erg/s}}\right) 
\left(\frac{\langle n_e\rangle}{500~\rm{cm}^{-3}}\right)^{-1} \rm{M}_{\odot}
\end{equation}
from \citet{Carniani2015}, assuming a density of the gas in the ionized outflow of $\langle n_e\rangle = 500 \rm{cm}^{-3}$, as in \citet{Carniani2015}, and
where $C$ is a factor taking into account the clumpiness of the medium ($C=\langle n_e\rangle ^2/\langle n_e^2\rangle$), which we assume to be unity.
Here we consider the $L_{\rm{H}\beta}$ to be the luminosity of the GB \hb\ component as measured from our full integrated spectral fit for the quasar and host galaxy, as described in Section \ref{sec:QuasarLineFitting} and shown in Figure \ref{fig:QuasarFit}. We note that as in our calculations above, we do not correct for extinction.
We perform the equivalent calculation using the GB component of \oiiia\, using the corresponding equation from \citet{Carniani2015}:
\begin{equation}
M_{\rm{out}} = 0.8\times10^8 \frac{C}{10^{[\rm{O/H}] - [\rm{O/H}]_\odot}}
\left(\frac{L_{[\rm{OIII}]}}{10^{44}\rm{erg/s}}\right) 
\left(\frac{\langle n_e\rangle}{500~\rm{cm}^{-3}}\right)^{-1} \rm{M}_{\odot},
\end{equation}
and assuming a solar oxygen abundance $[\rm{O/H}] = [\rm{O/H}]_\odot$.
These luminosities and calculated outflow masses are given in Table \ref{tab:Outflows}.
By using the luminosities of the ``outflow'' component of H$\beta$ we infer outflow masses of ionized gas of $(17\substack{ +13\\-11 })\times 10^7~\rm{M}_\odot$ and $(97\substack{ +14\\-17 })\times 10^7~\rm{M}_\odot$ for \DELS\ and \VDES, respectively. We note that the masses using the broad, blueshifted component of \oiiia\ are lower than the \hb\ masses, by a factor of approximately 2 for \DELS\ and an order of magnitude for \VDES. This is a well known issue and it is understood to be a consequence of the ionization structure of the clouds in the NLR of AGNs, whereby the \oiiia\ only traces a small part of the mass of each cloud \citep[e.g.][]{Perna2015}.

Once the outflow mass is inferred, the ionized outflow rate can be calculated under the thin shell approximation \citep[which generally provides a better description of galactic outflow rates, relative to the uniformly filled outflows, as discussed in ][]{Lutz2020} by using the equation
\begin{equation}
\dot{M}_{\rm{out}} = \frac{M_{\rm{out}} v_{\rm{out}}}{R_{\rm{out}}}
\end{equation}
where $R_{\rm{out}}$ is size of the outflow and $v_{\rm{out}}$ is its velocity. 
To estimate $R_{\rm{out}}$, we measure the maximum extent from the quasar centre of the \oiiia\ emission with $w_{80}>600$ km/s, in the BLR-subtracted cube. This $w_{80}>600$ km/s is a conservative criterion that has been used in previous studies to classify outflows \citep[e.g.][]{Harrison2015,Kakkad2020}. We note that for \DELS\ this outflow region extends from the host galaxy in Region 2 into Region 3, which is seen to have an outflow GB component in Figure \ref{fig:DELSRegions}. For \VDES, when we measure the maximum extent we exclude the south-southwest direction, avoiding the nearby Region 1 which may be artificially contributing to the large $w_{80}$ values as its velocity is offset from the quasar.
This $R_{\rm{out}}$ is a projected radius and it reproduces the true radius only if the directions of the outflow expansion include the plane of the sky. For the outflow velocity we adopt the same approach as \citet{Fluetsch2018} by defining $v_{\rm{out}}=v_{\rm{peak}}+v_{\rm{FWHM}}/2$, where $v_{\rm{peak}}$ is the peak velocity of the GB component of \oiiia\ (and equivalently \hb\ as they are kinematically tied in our spectral fits) and $v_{\rm{FWHM}}$ its FWHM. These velocities are measured from our full integrated spectral fit for the quasar and host galaxy, which are given in Table \ref{tab:QSOLines}.
Our measured $R_{\rm{out}}$ and $v_{\rm{out}}$ are given in Table \ref{tab:Outflows}.
Based on these measured values we obtain ionized outflow rates from \hb\ of $58\substack{ +44\\-37 }$~$\rm{M}_{\odot}~yr^{-1}$ and $525\substack{ +75\\-92 }$~$\rm{M}_{\odot}~yr^{-1}$ for \DELS\ and \VDES, respectively; the \oiiia\ rates are also quoted in Table \ref{tab:Outflows}.

We can also estimate the kinetic power of the outflows as
\begin{equation}
\dot{P}_{\rm{K,out}} = \frac{1}{2}\dot{M}_{\rm{out}} v^2_{\rm{out}}.
\end{equation}
From the \hb\ outflow, we obtain values of $(3.1\substack{ +2.3\\-2.0 })\times 10^{43}~\rm{erg~s}^{-1}$ and $(51\pm9)\times 10^{43}~\rm{erg~s}^{-1}$ for \DELS\ and \VDES, respectively, as given in Table \ref{tab:Outflows}.

The implications of these findings are discussed in Section \ref{sec:outflowDiscussion}.

\section{Discussion}
\label{sec:discussion}
\subsection{Quasar companions}

Both of our quasars show additional emission-line regions nearby their host galaxies, which are consistent with being companion galaxies.

For \DELS, Region 1 to the southeast of the quasar has a projected offset of $\simeq4.3$ kpc from the quasar, with a widest extent of $\simeq8.8$ kpc. From the integrated spectrum we find that this region has a velocity offset relative to the quasar host of $-98\pm72$ km/s, indicating minimal line-of-sight offset (Figure \ref{fig:DELSRegions} and Table \ref{tab:HostFlux}). From the emission maps in Figures \ref{fig:DELSMapsQSO} and \ref{fig:MapsBroad} we see that the velocity spans about 100 km/s either side of the host velocity. This companion galaxy is likely undergoing a merger with the quasar host. 
Region 3 exhibits some broader \oiii\ emission, indicating an ionized gas outflow in this region, and is likely a tidal tail.

For \VDES, 
Region 1 to the south-southwest of the quasar has a projected offset of $\simeq2.8$ kpc from the quasar, with a widest extent of $\simeq3.6$ kpc. This region has a line-of-sight velocity offset by $-55\pm72$ km/s from the host (Figures  \ref{fig:VDESMapsQSO} and \ref{fig:MapsBroad}).
Region 3 to the northwest of \VDES\ has a projected offset of $\simeq4.2$ kpc from the quasar, with a widest extent of $\simeq4.6$ kpc. This region has a line-of-sight velocity offset by $+83\pm72$ km/s from the host, with the bulk of the gas at a similar offset.
Region 4 to the southwest of the quasar has a projected offset of  $\simeq5.6$ kpc, and a widest extent of  $\simeq4.8$ kpc. This region has a line-of-sight velocty offset by $+218\pm72$ km/s from the host, again with the bulk of this gas at a similar velocity. 

All of these regions satisfy the common merger condition of having a projected distance of $<20/h$ kpc, where $h$ is the dimensionless Hubble constant $h=H_0/(100$ km/s/Mpc), and a velocity difference $\delta V < 500$ km/s \citep[e.g.][]{Patton2000,Conselice2009}.

The presence of quasar companions, or lack thereof, provides important insights into quasar growth mechanisms.
For example, theory predicts that kpc-scale interactions trigger AGN activity \citep[e.g.][]{Sanders1988,Hopkins2006}, however mergers and companion galaxies are not ubiquitously observed around low-$z$ quasars \citep{Cisternas2011, Kocevski2012, Mechtley2016,Marian2019}. However, these early quasars have more intense accretion, and higher merger rates are likely in the early Universe, suggesting that these mergers could be a key driver of this early black hole growth.
ALMA observations have detected companion galaxies around high-$z$ quasars at $\simeq8$--60 kpc separations which are interpreted as major galaxy interactions \citep[e.g.][]{Wagg2012,Decarli2017}, at rates of up to 50\% in some samples \citep{Trakhtenbrot2017,Nguyen2020}.  These companions are less frequently observed in the rest-frame UV \citep{willott_2005}, although some potential companions have been discovered with HST \citep{McGreer2014,Marshall2019c}. The lack of UV-discovered companions may be due to dust obscuration \citep{Trakhtenbrot2017}, however our observations which reveal such close neighbours may indicate that resolution has previously limited our ability to identify merging quasars in the early Universe. It is indeed striking that both of our quasars have these companion galaxies, and it will be insightful to see whether other upcoming JWST observations also reveal similar structures around other high-$z$ quasars, building up a larger statistical sample. Discovering these close companions highlights the ability of JWST to understand these early quasars in significantly greater detail, through detailed measurements of their environments and thus potential quasar triggering mechanisms \citep[see also][]{Perna2023}.

\begin{figure*}
\begin{center}
\includegraphics[scale=0.8]{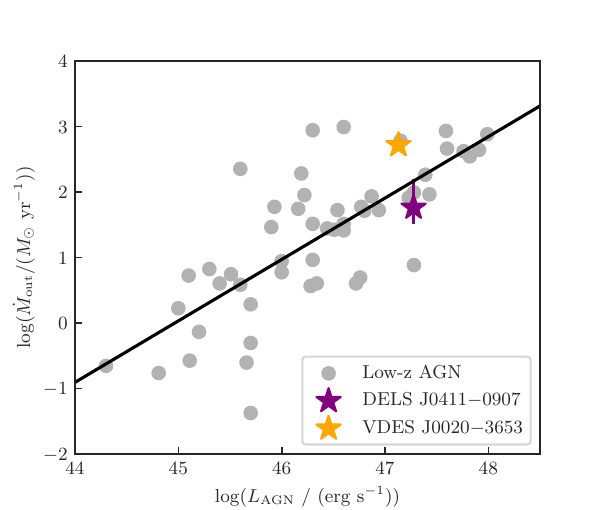}
\includegraphics[scale=0.8]{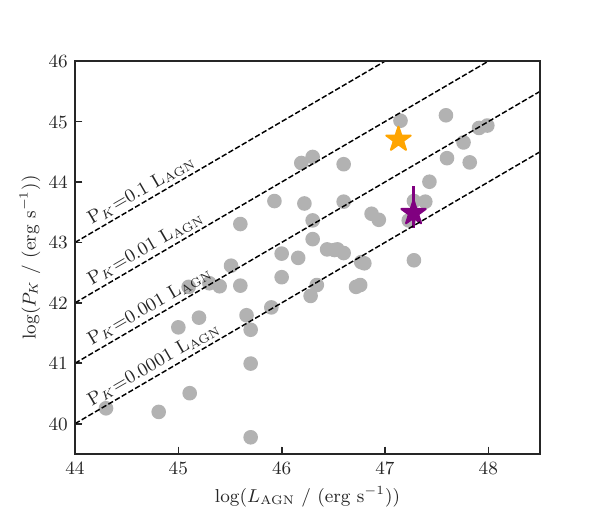}
\caption{Left: Ionized gas outflow mass rate as a function of AGN luminosity, for the two quasars at $z\sim6.8$ presented in this paper, and for various other AGN at lower redshift \citep[$0<z<3$; from][]{Fiore2017}. The solid line is a linear fit (in logarithmic scales) of the latter data. The ionized outflows of the two quasars at $z\sim7$ follow the same relation as quasars at lower redshifts. Right: Kinetic power (P$_K$) of the ionized outflows as a function of AGN luminosity (same symbols as in the left plot). The dashed lines show constant ratios of P$_K$ and L$_{\rm{AGN}}$ as labelled. 
We note that the \citet{Fiore2017} galaxies have been shifted by $\log(5/2)$ for consistency with our density assumption of $500\rm{cm}^{-3}$, as they assume $200\rm{cm}^{-3}$.
}
\label{fig:outflows}
\end{center}
\end{figure*}

\subsection{Quasar driven outflows}
\label{sec:outflowDiscussion}
In Figure \ref{fig:outflows} we show the ionized outflow rate as a function of AGN luminosity for our two quasars at $z\sim6.8$, and for AGN whose ionized outflows had been measured at lower redshift \citep[from the compilation of][]{Fiore2017}. Clearly, our two high-$z$ quasars follow the same relation as lower redshift quasars.

The inferred outflow rates ($\sim58 \rm{M}_{\odot}~\rm{yr}^{-1}$  for \DELS\ and $\sim525 \rm{M}_{\odot}~\rm{yr}^{-1}$ for \VDES) are significantly larger than the upper limits we have measured for the host SFRs ($<33 \rm{M}_{\odot}~\rm{yr}^{-1}$  for \DELS\ and $<54 \rm{M}_{\odot}~\rm{yr}^{-1}$ for \VDES; see Section \ref{sec:SFRs}). These outflow rates are much larger than the SFRs, especially if we consider that the measured outflow rates are only for the ionized component and that the other phases, in particular the molecular phase, may also contribute. This result implies that, if prolonged, the quasar-driven outflow may be the main agent responsible for depleting the galaxy of its gas content. However, one has to take into account that the quasar activity is likely short lived and bursty, hence it may not continue for a time long enough to compete with star formation in depleting the gas mass. 

More quantitative consideration on the potential feedback effect of the quasar driven outflow can be obtained by considering the kinetic power of the outflows. Efficient feedback, with high coupling with the ISM, in many scenarios requires that the kinetic power of the outflow should be about 5--7\% of the AGN bolometric luminosity. The right panel of Figure \ref{fig:outflows} shows the kinetic power of ionized outflows as a function of AGN luminosity, where locii of constant P$_K$/L$_{\rm{AGN}}$ ratio are also indicated with dashed lines. 
Clearly, the ionized outflows of two quasars at z$\sim$6.8 have a kinetic power that is less than 1\% of L$_{\rm{AGN}}$ (in the case of \DELS\ actually <0.1\%), indicating that the outflow may not be an effective feedback mechanism.

Moreover, the momentum rate ($\dot M_{\rm{out}}v_{\rm{out}}$) of the two quasars is 0.07~L$_{\rm{AGN}}/c$ and 1.28~L$_{\rm{AGN}}/c$ for \DELS\ and \VDES, respectively. These values are much lower than expected by the (very feedback-effective) energy-driven outflows, which expect a momentum boost by a factor of $\sim$20 relative to the radiation momentum rate L$_{\rm{AGN}}$/c. Based on these values, these outflows seem to be more consistent with  momentum-driven outflows, possibly resulting from radiation pressure on the dusty clouds \citep{Fabian2012,Costa2014}. However, once again, we caution that with our NIRSpec data we can only probe the ionized phase of these outflows. The other phases of the outflows, and in particular the molecular component, may be much more massive and energetic \citep{Fluetsch2018}, although \citet{Fiore2017} suggest that at such high luminosities the ionized outflows may be comparable or stronger than the other phases. Future deep observations with ALMA may provide constraints on the other phases of these outflows and, therefore, provide additional information on the feedback role of AGN-driven outflows in these two early quasars.

\section{Conclusions}
\label{sec:conclusions}
We have analysed JWST NIRSpec IFU observations of two $z\simeq6.8$ quasars, \DELS\ and \VDES.

By analysing their integrated quasar spectrum, we measured black hole masses from the \hb\ and \ha\ broad lines, which are the most reliable single-epoch mass estimators. These are some of the first observations of Balmer broad emission lines in high-$z$ quasars, making such a measurement possible for the first time.
We derive a best estimate for the black hole mass in \DELS\ of $M_{\rm{BH}}=1.85\substack{ +2\\-0.8}\times10^9\rm{M}_\odot$, implying an Eddington ratio of $\lambda_{\rm{Edd}}=0.8\substack{ +0.7\\-0.4}$. This is more massive than the \ions{Mg}{ii} mass estimates of $M_{\rm{BH, MgII}}=(0.6$-$1)\times10^{9}\rm{M}_\odot$ which implied more extreme Eddington ratios of $\lambda_{\rm{Edd}}=1.3$--2.4. By measuring a more accurate black hole mass, we have revised our understanding of the accretion physics of this early quasar.
For \VDES\ our best estimate of the black hole mass is $M_{\rm{BH}}=2.9\substack{ +3.5\\-1.3}\times10^9\rm{M}_\odot$, corresponding to an Eddington ratio of $\lambda_{\rm{Edd}}=0.4\substack{ +0.3\\-0.2}$. This is more consistent with the \ions{Mg}{ii} measurements of \citet{Reed2019}. Together, these quasars show that the \ions{Mg}{ii} black hole mass can be significantly biased relative to the \hb\ estimate for some high-$z$ quasars, while for others these two techniques may give consistent black hole mass measurements, highlighting the benefit of measuring these more reliable \hb\ masses.

We perform quasar subtraction on our IFU observations, by modelling the PSF using the variation in the flux of the quasar broad \ha\ and \hb\ lines across the spatial domain. From this subtraction, we reveal that the host galaxy of \DELS\ shows emission extending to a radius of 2.5 kpc, described by a 2D Gaussian with semi-major axis $\sigma=0.75$ kpc. This host has narrow \oiii, \hb\ and \ha\ emission, alongside a broader wing emission from an outflow.
We estimate that the host has a SFR of $<33 ~\rm{M}_\odot$/yr, with a dynamical mass in the range of 2.6-16$\times10^{10}\rm{M}_\odot$ with a best estimate of $8.6\substack{+4.4\\-3.3}\times10^{10}~ \rm{M}_\odot$. This host galaxy lies within the AGN-dominated regime of the low-$z$ BPT diagram.

The host galaxy of \VDES\ has emission extending to a radius of 2.1 kpc, described by a 2D Gaussian with semi-major axis $\sigma=1.0$ kpc. Both narrow \oiii, \hb\ and \ha\ and broad-wing emission is detected. We estimate a SFR of $<54~ \rm{M}_\odot$/yr for this host, with a dynamical mass between 4.6--17$\times10^{10}~\rm{M}_\odot$ with a best estimate of $17\substack{+18\\-8}\times10^{10}~\rm{M}_\odot$. The host of \VDES\ lies within the star-forming or transition region of the low-$z$ BPT diagram, however this is likely due to the low metallicity of these quasar hosts, with low-metallicity, high-$z$ AGN predicted to lie in this region.

We find that from our measured black hole and dynamical masses, \DELS\ and \VDES\ both lie slightly above the local black hole--host mass relation \citep{Kormendy2013} and are consistent with the existing observations of $z\gtrsim6$ quasar host galaxies with ALMA. 

Both quasars are found to have additional emission-line regions near their host galaxies, with two regions for \DELS\ and three for \VDES. We study these companion regions, measuring their SFRs, excitation mechanisms, and kinematics. We find that four companion regions are likely to be either galaxies merging with the quasar host galaxy, and one an extended tidal tail. It is striking that both of our high-$z$ quasars are undergoing galaxy mergers, and it will be insightful to see the fraction of high-$z$ quasars that are found to also have companions in future NIRSpec IFU observations.

We detect ionized outflows in \oiii\ and \hb\ from both \VDES\ and \DELS.
We measure the rates of the ionized outflow from \hb\ to be $58\substack{ +44\\-37 }$~$\rm{M}_{\odot}~yr^{-1}$ for \DELS\ and $525\substack{ +75\\-92 }$~$\rm{M}_{\odot}~yr^{-1}$ for \VDES. These outflow rates are much larger than the SFRs of the host galaxies, implying that if this is prolonged, the quasar-driven outflow may be the main driver for gas depletion within the galaxies. We find that these outflows are likely momentum-driven, and are not likely to be effective feedback mechanisms.

This work highlights the exquisite capabilities of the JWST NIRSpec IFU for observing high-$z$ quasars. With the high-resolution spectral coverage in the infrared, we are able to study key rest-frame optical emission lines that have been previously unobservable, such as \hb, \oiii\ and \ha. This allows us to measure the black hole masses and Eddington ratios of these quasars more accurately than has been previously possible. Obtaining spatially resolved IFU spectra also allows us to study the host galaxies, local environments, and outflow properties of these quasars in great detail. Overall, this work shows the possibilities of the NIRSpec IFU, and we look forward to the insights that this instrument will give about the formation and growth of quasars and their host galaxies in the early Universe.

\section*{Acknowledgements}
We thank the anonymous referee for their constructive feedback which greatly improved the manuscript. We thank Bernd Husemann for his help with the QDeblend3D software and sharing his expertise on quasar subtraction.
MM would like to thank Toby Brown for his helpful advice, particularly on creating the kinematic maps.

This work is based on observations made with the NASA/ESA/CSA James Webb Space Telescope. The data were obtained from the Mikulski Archive for Space Telescopes at the Space Telescope Science Institute, which is operated by the Association of Universities for Research in Astronomy, Inc., under NASA contract NAS 5-03127 for JWST. These observations are associated with program \#1222, as part of the Galaxy Assembly with NIRSpec IFS GTO program.

MAM acknowledges the support of a National Research Council of Canada Plaskett Fellowship, and the Australian Research Council Centre of Excellence for All Sky Astrophysics in 3 Dimensions (ASTRO 3D), through project number CE170100013.
SA, BRP and MP acknowledge support from the research project PID2021-127718NB-I00 of the Spanish Ministry of Science and Innovation/State Agency of Research (MICIN/AEI). MP also acknowledges support from the Programa Atracci\'on de Talento de la Comunidad de Madrid via grant 2018-T2/TIC-11715.
RM, JS and FDE acknowledge support by the Science and Technology Facilities Council (STFC), from the ERC Advanced Grant 695671 ``QUENCH''.
RM and JS also acknowledge funding from a research professorship from the Royal Society. 
GC acknowledges the support of the INAF Large Grant 2022 ``The metal circle: a new sharp view of the baryon
cycle up to Cosmic Dawn with the latest generation IFU facilities''.
H{\"U} gratefully acknowledges support by the Isaac Newton Trust and by the Kavli Foundation through a Newton-Kavli Junior Fellowship.
AJB, GCJ and AJC acknowledge funding from the ``FirstGalaxies'' Advanced Grant from the European Research Council (ERC) under the European Union’s Horizon 2020 research and innovation programme (Grant agreement No. 789056).
SC acknowledges support from the European Union (ERC, WINGS,101040227).
PGP-G acknowledges support  from  Spanish  Ministerio  de  Ciencia e Innovaci\'on MCIN/AEI/10.13039/501100011033 through grant PGC2018-093499-B-I00.
The Cosmic Dawn Center (DAWN) is funded by the Danish National Research Foundation under grant no.140

This research has made use of NASA's Astrophysics Data System, QFitsView, and SAOImageDS9, developed by Smithsonian Astrophysical Observatory.

This paper made use of Python packages and software
AstroPy \citep{Astropy2013},
Matplotlib \citep{Matplotlib2007},
NumPy \citep{Numpy2011},
Pandas \citep{reback2020pandas}, 
Photutils \citep{photutils},
Regions \citep{Bradley2022},
SciPy \citep{2020SciPy-NMeth},
Seaborn \citep{Waskom2021},
Spectral Cube \citep{Ginsburg2019}, 
QDeblend3D \citep{Husemann2013,Husemann2014}, and
QubeSpec\footnote{\url{https://github.com/honzascholtz/Qubespec}}.

%%%%%%%%%%%%%%%%%%%%%%%%%%%%%%%%%%%%%%%%%%%%%%%%%%

\begin{appendix}
\section{Measured PSFs}
\begin{figure*}
\begin{center}
\includegraphics[scale=0.8]{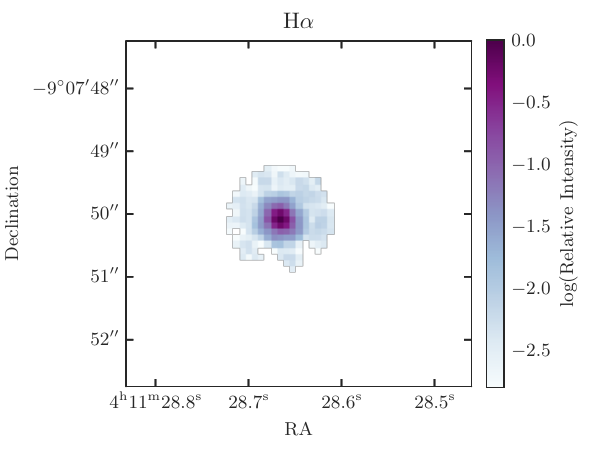}
\includegraphics[scale=0.8]{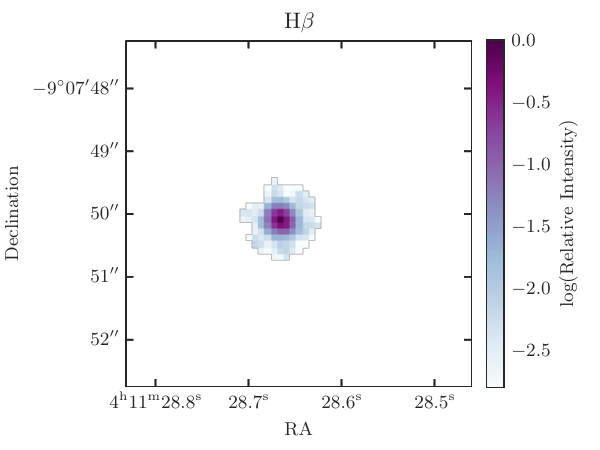}

\includegraphics[scale=0.8]{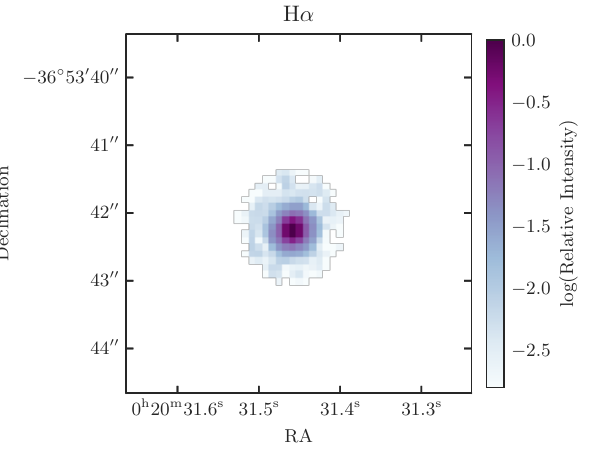}
\includegraphics[scale=0.8]{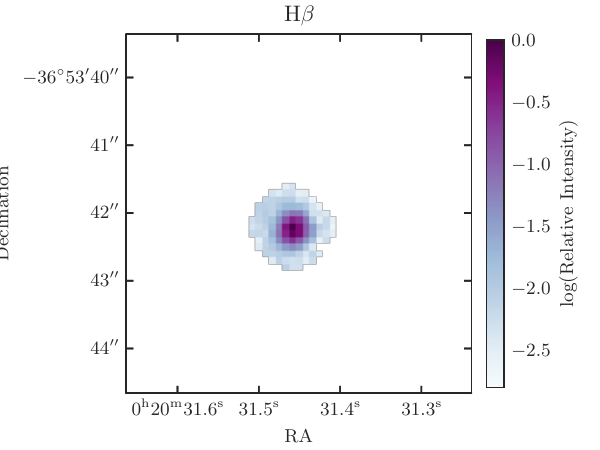}

\caption{The PSFs measured from the broad \ha\ and \hb\ lines from \DELS\ (top) and \VDES\ (bottom), from the N\&B quasar subtraction, as described in Section \ref{sec:QuasarSubtraction}. These are created from the best fit quasar cube, which is the peak quasar spectrum scaled in flux to match the observed broad line flux in each spaxel. The relative intensity here depicts this scaling factor, which is the amount of flux in each spaxel relative to the brightest spaxel. We create two PSFs for each quasar, one from each of \ha\ and \hb, as the PSF varies with wavelength.}
\label{fig:PSFs}
\end{center}
\end{figure*}

The measured PSFs from the N\&B quasar subtraction, for both \ha\ and \hb, are shown for the two quasars in Figure \ref{fig:PSFs}. 
We note that for \VDES\ the PSF shape does not follow the typical expected 6-pointed JWST PSF. This data cube is a combination of two separate observations taken at different position angles, 160.169 and 188.177 degrees, and so the resulting PSF of this combined cube would theoretically be two of the standard 6-point PSFs superimposed, one rotated by 28 degrees.

As the quasar subtraction technique measures the PSF from the data cube itself, and does not rely on PSF models, this does not affect our results. 
We have performed the quasar subtraction separately on both halves of the observation as a test, and verify that the host and companion features appear in both halves and are not distorted by this combination.

\section{Kinematic maps using the BLR subtraction technique}
\begin{figure*}
\begin{center}
\includegraphics[scale=0.7]{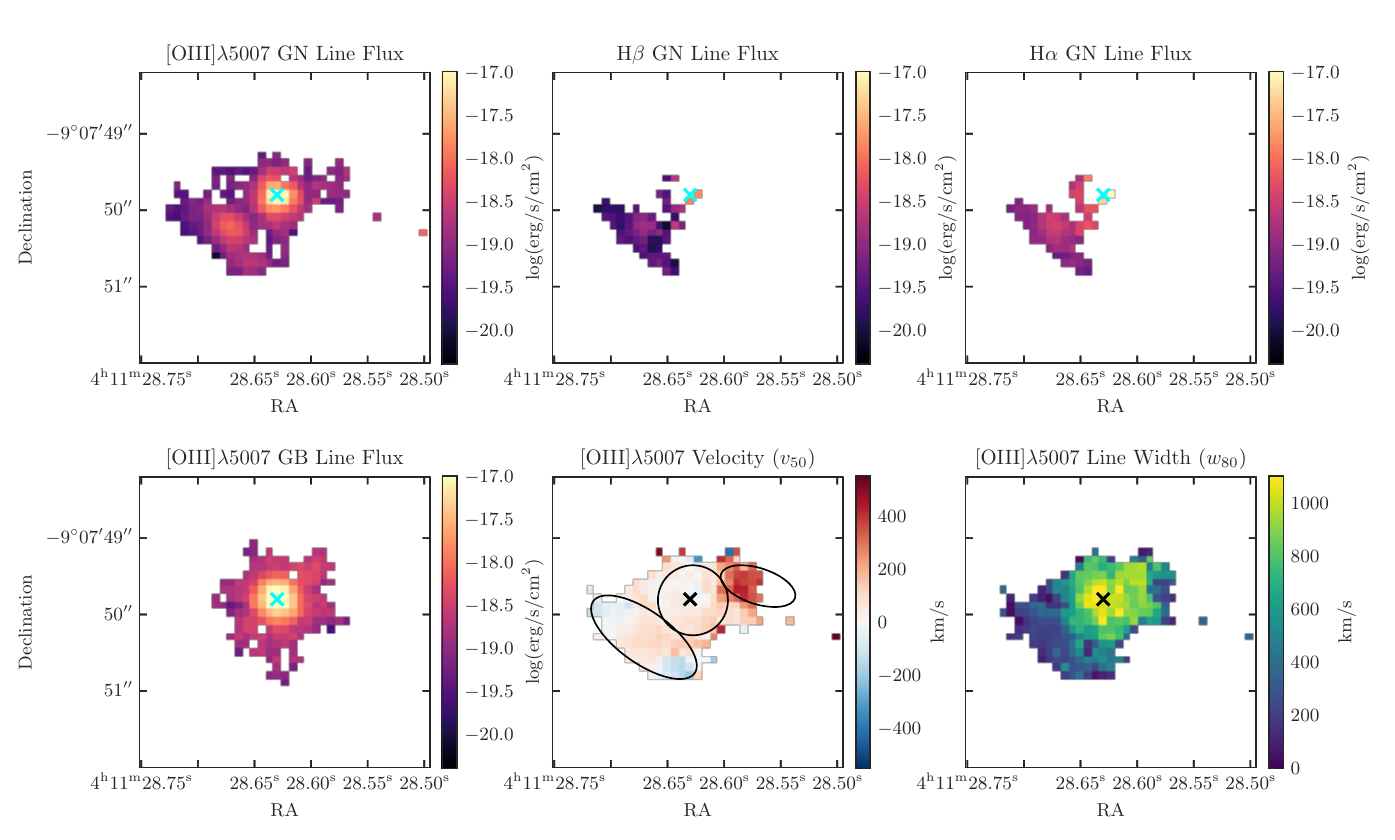}
\vspace*{-10.5cm}  % Tune this to the image height.
\begin{center}
\large{\DELS}
\end{center}
\vspace*{10cm}

\includegraphics[scale=0.7]{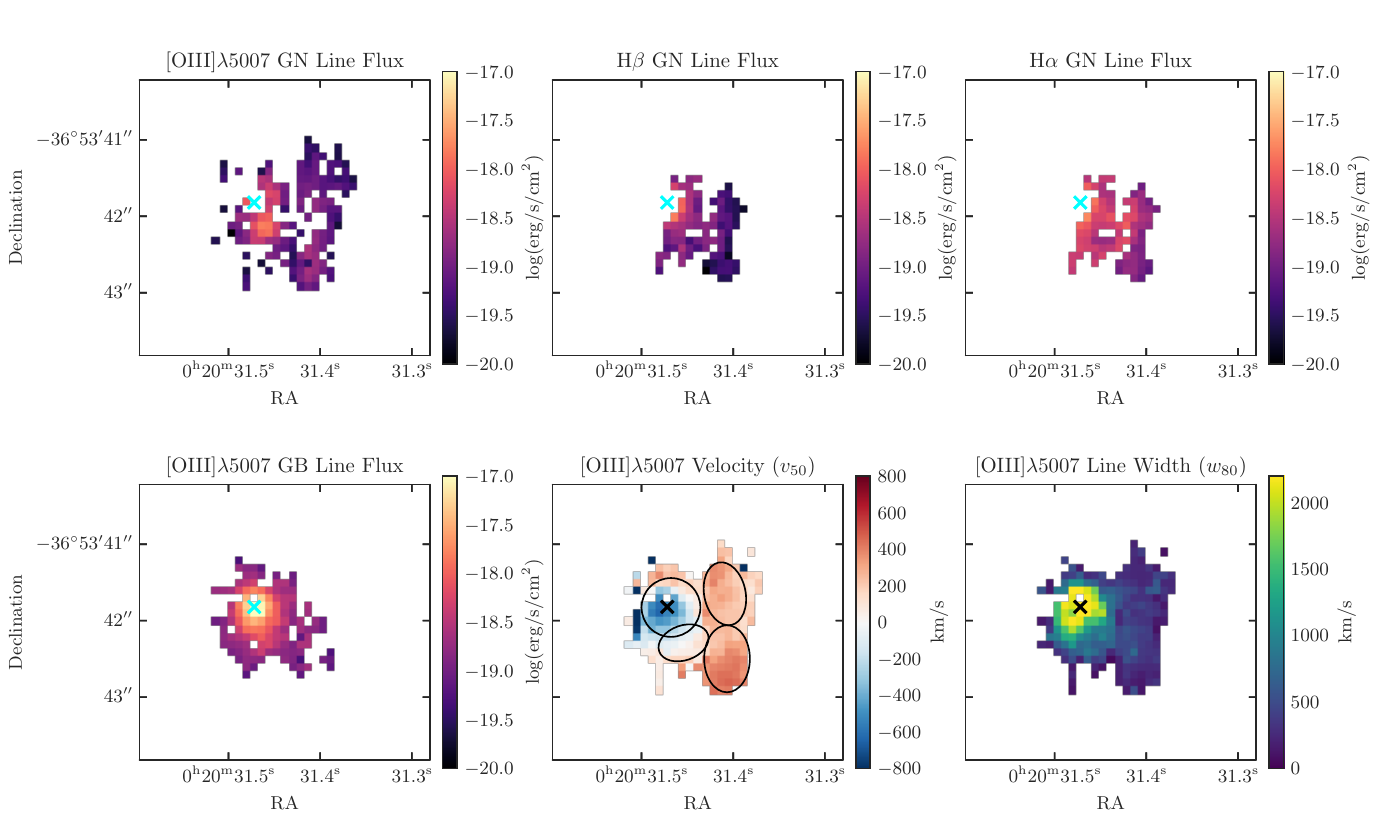}
\vspace*{-10.5cm}  % Tune this to the image height.
\begin{center}
\large{\VDES}
\end{center}
\vspace*{10cm}
\caption{Kinematic maps for \DELS\ (top) and \VDES\ (bottom), after the subtraction of the quasar emission using the BLR subtraction technique. The top panels show the flux of the narrow component of the \oiiia, \hb, and \ha\ lines (GN), as fit by a Gaussian. The \oiiib\ line is not shown as it is constrained to have an amplitude 3 times fainter than that of \oiiia.
The bottom left panel shows the flux of the broader GB component of the \oiiia\ line, a second Gaussian with a larger $\sigma$. 
The lower middle and right panels show our kinematic maps, showing the non-parametric central velocity of the line ($v_{50}$; middle) and the line width ($w_{80}$; right).
As these are non-parametric, this combines both the GN and GB components. 
We describe our method for creating these maps in Section \ref{sec:Maps}.
The black ellipses in the lower middle panels depict the regions as in Figures \ref{fig:DELSRegions} and \ref{fig:VDESRegions}.}
\label{fig:MapsBroad}
\end{center}
\end{figure*}

Our kinematic maps for \DELS\ and \VDES\ as measured from the BLR quasar subtracted cubes are shown in Figure \ref{fig:MapsBroad}. This can be compared to Figures \ref{fig:DELSMapsQSO} and Figures \ref{fig:VDESMapsQSO}, which show the equivalent maps for the N\&B quasar subtracted cubes for \DELS\ and \VDES\ respectively.

\section{Data cubes with pixel sub-sampling}
\label{sec:AppDrizzle}
Throughout this work we use data cubes with the native spaxel spatial size of 0.1''/pix, produced using the EMSM weighting to combine the various dither positions.
With our dithering strategy it is also possible to subsample the data down to a pixel size of 0.05''/pix, to improve the spatial resolution which is limited by the pixel scale. For the best possible spatial resolution, we use cubes with a pixel scale of 0.05'' that have been combined using the \textit{drizzle} weighting. This weighting optimizes the resolution, but is more affected by the PSF effects \citep[see][for details]{Perna2023}.
These 0.05''/pix \textit{drizzle} cubes are highly affected by the PSF oscillations and chromatic effects, which we find result in very poor results for the quasar subtraction. Because the 0.1''/pix pixels optimize the SNR, and, most critically, give significantly improved results for the quasar subtraction that is the focus of our work, we use these 0.1''/pix EMSM cubes throughout this paper.

\begin{figure*}
\begin{center}
\includegraphics[scale=0.8]
{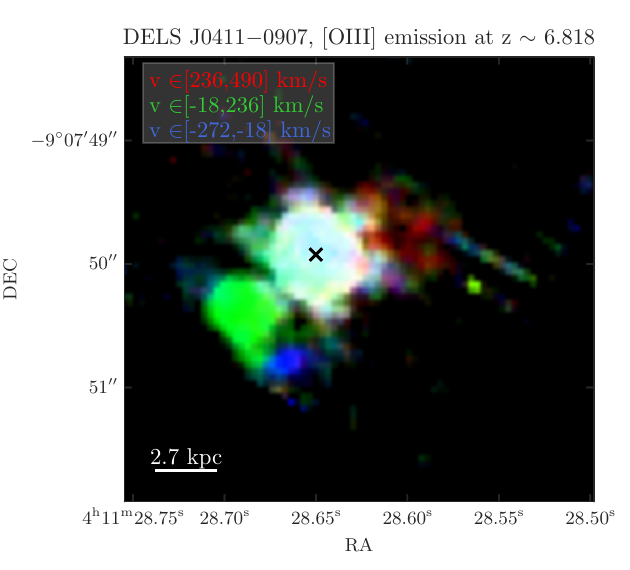}\hspace{0.2cm}
\includegraphics[scale=0.8]{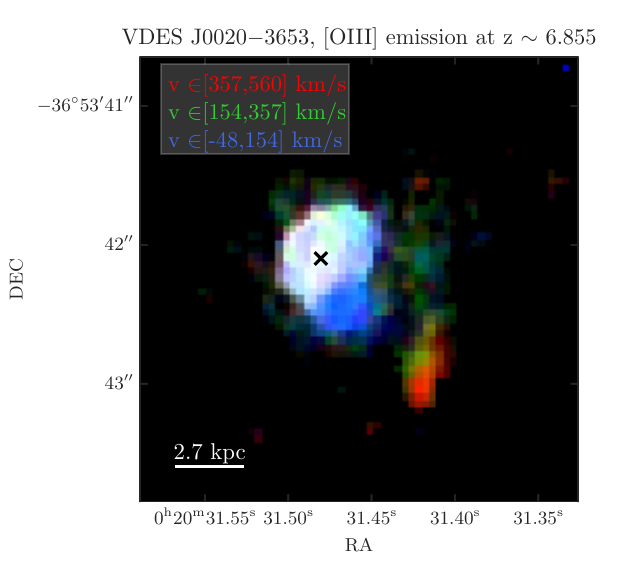}

\caption{\oiiia\ flux maps for \DELS\ (left) and \VDES\ (right) using data cubes that have been produced with 0.05''/pix pixels, combined using the \textit{drizzle} weighting, following the method described in detail in \citet{Perna2023}.
These maps are made using the original, non-continuum subtracted cubes. 
The stacked red, green, and blue images depict the integral of the flux across each of the specified wavelength/velocity windows, relative to the quasar redshift. The quasar peak location is marked as a black cross.
}
\label{fig:OIII0p05Cube}
\end{center}
\end{figure*}

However, the 0.05''/pix drizzle cubes are more optimal for looking at the structural/morphological properties beyond the quasar host galaxy itself.
In Figure \ref{fig:OIII0p05Cube} we show the \oiiia\ flux maps for \DELS\ and \VDES\ using cubes that have been produced with 0.05'' pixels, combined using the \textit{drizzle} weighting, following the method described in detail in \citet{Perna2023}. This shows the companion galaxy morphology in greater detail.  We find that in general, the morphology is consistent with our results from the 0.1''/pix cubes, showing a companion to the southeast of \DELS\ with a potential outflow to the northwest, and the two companion regions to the west of \VDES.

One key difference is that the higher spatial resolution data for \VDES\ reveals a second distinct peak in the \oiiia\ emission, separated by only $\sim0.37''$ or 2 kpc (quasar peak to companion peak) to the south-southwest of the quasar.
This very nearby companion, with a velocity offset by -179 km/s from the host Region 2 (GN), or -30 km/s with respect to the quasar's peak \oiiia\ emission at $z=6.854$, would be significantly interacting with the quasar host galaxy. This is so close to the quasar host that in the 0.1''/pix cubes this appears to be one elongated galaxy. Thus, if this region is not considered separately, and instead one elongated ellipse for this full emission region was chosen, we would overestimate the fluxes of the host galaxy of \VDES.  To account for this, we use the structure seen in Figure \ref{fig:OIII0p05Cube} to define our region ellipses as seen in Figure \ref{fig:VDESRegions}, which we use to extract and thus analyse the individual regions' spectra.

As we cannot perform a reliable quasar subtraction on the higher resolution cubes, we do not use the cubes throughout this work. Our results focus on the host detection and thus require an accurate quasar subtraction.

\end{appendix}

\bibliographystyle{aa}
\bibliography{manuscript.bib} 

\begin{thebibliography}{143}
\expandafter\ifx\csname natexlab\endcsname\relax\def\natexlab#1{#1}\fi

\bibitem[{Abazajian {et~al.}(2009)Abazajian, Adelman-McCarthy, Agüeros, Allam,
  Prieto, An, Anderson, Anderson, Annis, Bahcall, Bailer-Jones, Barentine,
  Bassett, Becker, Beers, Bell, Belokurov, Berlind, Berman, Bernardi,
  Bickerton, Bizyaev, Blakeslee, Blanton, Bochanski, Boroski, Brewington,
  Brinchmann, Brinkmann, Brunner, Budav{\'{a}}ri, Carey, Carliles, Carr,
  Castander, Cinabro, Connolly, Csabai, Cunha, Czarapata, Davenport, de~Haas,
  Dilday, Doi, Eisenstein, Evans, Evans, Fan, Friedman, Frieman, Fukugita,
  Gänsicke, Gates, Gillespie, Gilmore, Gonzalez, Gonzalez, Grebel, Gunn,
  Györy, Hall, Harding, Harris, Harvanek, Hawley, Hayes, Heckman, Hendry,
  Hennessy, Hindsley, Hoblitt, Hogan, Hogg, Holtzman, Hyde, ichi Ichikawa,
  Ichikawa, Im, Ivezi{\'{c}}, Jester, Jiang, Johnson, Jorgensen, Juri{\'{c}},
  Kent, Kessler, Kleinman, Knapp, Konishi, Kron, Krzesinski, Kuropatkin,
  Lampeitl, Lebedeva, Lee, Lee, Leger, L{\'{e}}pine, Li, Lima, Lin, Long,
  Loomis, Loveday, Lupton, Magnier, Malanushenko, Malanushenko, Mandelbaum,
  Margon, Marriner, Mart{\'{\i}}nez-Delgado, Matsubara, McGehee, McKay,
  Meiksin, Morrison, Mullally, Munn, Murphy, Nash, Nebot, Neilsen, Newberg,
  Newman, Nichol, Nicinski, Nieto-Santisteban, Nitta, Okamura, Oravetz,
  Ostriker, Owen, Padmanabhan, Pan, Park, Pauls, Peoples, Percival, Pier, Pope,
  Pourbaix, Price, Purger, Quinn, Raddick, Fiorentin, Richards, Richmond,
  Riess, Rix, Rockosi, Sako, Schlegel, Schneider, Scholz, Schreiber, Schwope,
  Seljak, Sesar, Sheldon, Shimasaku, Sibley, Simmons, Sivarani, Smith, Smith,
  Smol{\v{c}}i{\'{c}}, Snedden, Stebbins, Steinmetz, Stoughton, Strauss,
  SubbaRao, Suto, Szalay, Szapudi, Szkody, Tanaka, Tegmark, Teodoro, Thakar,
  Tremonti, Tucker, Uomoto, Berk, Vandenberg, Vidrih, Vogeley, Voges, Vogt,
  Wadadekar, Watters, Weinberg, West, White, Wilhite, Wonders, Yanny, Yocum,
  York, Zehavi, Zibetti, \& Zucker}]{Abazajian2009}
Abazajian, K.~N., Adelman-McCarthy, J.~K., Agüeros, M.~A., {et~al.} 2009,
  ApJS, 182, 543

\bibitem[{{Astropy Collaboration} {et~al.}(2013){Astropy Collaboration},
  Robitaille, Tollerud, Greenfield, Droettboom, Bray, Aldcroft, Davis,
  Ginsburg, Price-Whelan, Kerzendorf, Conley, Crighton, Barbary, Muna,
  Ferguson, Grollier, Parikh, Nair, Günther, Deil, Woillez, Conseil, Kramer,
  Turner, Singer, Fox, Weaver, Zabalza, Edwards, Bostroem, Burke, Casey,
  Crawford, Dencheva, Ely, Jenness, Labrie, Lim, Pierfederici, Pontzen, Ptak,
  Refsdal, Servillat, \& Streicher}]{Astropy2013}
{Astropy Collaboration}, Robitaille, T.~P., Tollerud, E.~J., {et~al.} 2013,
  A{\&}A, 558, A33

\bibitem[{Baldwin {et~al.}(1981)Baldwin, Phillips, \& Terlevich}]{Baldwin1981}
Baldwin, J.~A., Phillips, M.~M., \& Terlevich, R. 1981, PASP, 93, 5

\bibitem[{Banados {et~al.}(2022)Banados, Schindler, Venemans, Connor, Decarli,
  Farina, Mazzucchelli, Meyer, Stern, Walter, Fan, Hennawi, Khusanova, Morrell,
  Nanni, Noirot, Pensabene, Rix, Simon, Kleijn, Xie, Yang, \&
  Connor}]{Banados2022}
Banados, E., Schindler, J.-T., Venemans, B.~P., {et~al.} 2022, ApJS, 265, 29

\bibitem[{Ba{\~{n}}ados {et~al.}(2016)Ba{\~{n}}ados, Venemans, Decarli, Farina,
  Mazzucchelli, Walter, Fan, Stern, Schlafly, Chambers, Rix, Jiang, McGreer,
  Simcoe, Wang, Yang, Morganson, Rosa, Greiner, Balokovi{\'{c}}, Burgett,
  Cooper, Draper, Flewelling, Hodapp, Jun, Kaiser, Kudritzki, Magnier,
  Metcalfe, Miller, Schindler, Tonry, Wainscoat, Waters, \& Yang}]{Banados2016}
Ba{\~{n}}ados, E., Venemans, B.~P., Decarli, R., {et~al.} 2016, ApJS, 227, 11

\bibitem[{Ba{\~{n}}ados {et~al.}(2018)Ba{\~{n}}ados, Venemans, Mazzucchelli,
  Farina, Walter, Wang, Decarli, Stern, Fan, Davies, Hennawi, Simcoe, Turner,
  Rix, Yang, Kelson, Rudie, \& Winters}]{Banados2017}
Ba{\~{n}}ados, E., Venemans, B.~P., Mazzucchelli, C., {et~al.} 2018, Nat, 553,
  473

\bibitem[{Bertoldi {et~al.}(2003)Bertoldi, Cox, Neri, Carilli, Walter, Omont,
  Beelen, Henkel, Fan, Strauss, \& Menten}]{Bertoldi2003}
Bertoldi, F., Cox, P., Neri, R., {et~al.} 2003, A\&A, 409, L47

\bibitem[{Bischetti {et~al.}(2022)Bischetti, Feruglio, D'Odorico, Arav,
  Ba{\~{n}}ados, Becker, Bosman, Carniani, Cristiani, Cupani, Davies, Eilers,
  Farina, Ferrara, Maiolino, Mazzucchelli, Mesinger, Meyer, Onoue, Piconcelli,
  Ryan-Weber, Schindler, Wang, Yang, Zhu, \& Fiore}]{Bischetti2022}
Bischetti, M., Feruglio, C., D'Odorico, V., {et~al.} 2022, Nature, 605, 244

\bibitem[{Bischetti {et~al.}(2023)Bischetti, Fiore, Feruglio, D'Odorico, Arav,
  Zubovas, Becker, Cupani, Davies, Eilers, Farina, Ferrara, Gaspari,
  Mazzucchelli, Onoue, Piconcelli, Zanchettin, \& Zhu}]{Bischetti2023}
Bischetti, M., Fiore, F., Feruglio, C., {et~al.} 2023, ApJ, 952, 44

\bibitem[{Bradley {et~al.}(2022)Bradley, Deil, Ginsburg, {Sushobhana Patra},
  Robitaille, Sipőcz, King, {P. L. Lim}, Homeier, Singer, De~Val-Borro,
  Jenness, Baumann, {Yash Gondhalekar}, Donath, Tollerud, {Jae-Joon Lee},
  Leinweber, \& {Zé Vinícius}}]{Bradley2022}
Bradley, L., Deil, C., Ginsburg, A., {et~al.} 2022, astropy/regions: v0.7

\bibitem[{Bradley {et~al.}(2018)Bradley, Sipocz, Robitaille, Vinícius,
  Tollerud, Deil, Barbary, Günther, Cara, Busko, Conseil, Droettboom,
  Bostroem, Bray, Bratholm, Craig, Barentsen, Pascual, Donath, Greco, Perren,
  Kerzendorf, de~Val-Borro, StuartLittlefair, Ogaz, Lim, Ferreira, D'Eugenio,
  \& Weaver}]{photutils}
Bradley, L., Sipocz, B., Robitaille, T., {et~al.} 2018, astropy/photutils: v0.5

\bibitem[{Bruzual \& Charlot(2003)}]{bruzual_2003}
Bruzual, G. \& Charlot, S. 2003, MNRAS, 344, 1000

\bibitem[{Bushouse {et~al.}(2022)Bushouse, Eisenhamer, Dencheva, Davies,
  Greenfield, Morrison, Hodge, Simon, Grumm, Droettboom, Slavich, Sosey, Pauly,
  Miller, Jedrzejewski, Hack, Davis, Crawford, Law, Gordon, Regan, Cara,
  MacDonald, Bradley, Shanahan, Jamieson, Teodoro, \& Williams}]{Bushouse2022}
Bushouse, H., Eisenhamer, J., Dencheva, N., {et~al.} 2022, JWST Calibration
  Pipeline

\bibitem[{Böker {et~al.}(2022)Böker, Arribas, Lützgendorf, de~Oliveira,
  Beck, Birkmann, Bunker, Charlot, de~Marchi, Ferruit, Giardino, Jakobsen,
  Kumari, L{\'{o}}pez-Caniego, Maiolino, Manjavacas, Marston, Moseley,
  Muzerolle, Ogle, Pirzkal, Rauscher, Rawle, Rix, Sabbi, Sargent, Sirianni,
  te~Plate, Valenti, Willott, \& Zeidler}]{Boeker2022}
Böker, T., Arribas, S., Lützgendorf, N., {et~al.} 2022, A\&A, 661, A82

\bibitem[{Böker {et~al.}(2023)Böker, Beck, Birkmann, Giardino, Keyes, Kumari,
  Muzerolle, Rawle, Zeidler, Abul-Huda, de~Oliveira, Arribas, Bechtold,
  Bhatawdekar, Bonaventura, Bunker, Cameron, Carniani, Charlot, Curti,
  Espinoza, Ferruit, Franx, Jakobsen, Karakla, L{\'{o}}pez-Caniego,
  Lützgendorf, Maiolino, Manjavacas, Marston, Moseley, Ogle, Perna,
  Pe{\~{n}}a-Guerrero, Pirzkal, Plesha, Proffitt, Rauscher, Rix, del Pino,
  Rustamkulov, Sabbi, Sing, Sirianni, te~Plate, {\'{U}}beda, Wahlgren,
  Wislowski, Wu, \& Willott}]{Boeker2023}
Böker, T., Beck, T.~L., Birkmann, S.~M., {et~al.} 2023, Publications of the
  Astronomical Society of the Pacific, 135, 038001

\bibitem[{{Calzetti} {et~al.}(2000){Calzetti}, {Armus}, {Bohlin}, {Kinney},
  {Koornneef}, \& {Storchi-Bergmann}}]{Calzetti2000}
{Calzetti}, D., {Armus}, L., {Bohlin}, R.~C., {et~al.} 2000, ApJ, 533, 682

\bibitem[{Cappellari {et~al.}(2006)Cappellari, Bacon, Bureau, Damen, Davies,
  Zeeuw, Emsellem, Falcon-Barroso, Krajnovic, Kuntschner, McDermid, Peletier,
  Sarzi, Bosch, \& Ven}]{Cappellari2006}
Cappellari, M., Bacon, R., Bureau, M., {et~al.} 2006, MNRAS, 366, 1126

\bibitem[{Carniani {et~al.}(2015)Carniani, Marconi, Maiolino, Balmaverde,
  Brusa, Cano-D{\'{\i}}az, Cicone, Comastri, Cresci, Fiore, Feruglio, Franca,
  Mainieri, Mannucci, Nagao, Netzer, Piconcelli, Risaliti, Schneider, \&
  Shemmer}]{Carniani2015}
Carniani, S., Marconi, A., Maiolino, R., {et~al.} 2015, A\&A, 580, A102

\bibitem[{Cisternas {et~al.}(2011)Cisternas, Jahnke, Inskip, Kartaltepe,
  Koekemoer, Lisker, Robaina, Scodeggio, Sheth, Trump, Andrae, Miyaji, Lusso,
  Brusa, Capak, Cappelluti, Civano, Ilbert, Impey, Leauthaud, Lilly, Salvato,
  Scoville, \& Taniguchi}]{Cisternas2011}
Cisternas, M., Jahnke, K., Inskip, K.~J., {et~al.} 2011, ApJ, 726, 57

\bibitem[{Coatman {et~al.}(2016)Coatman, Hewett, Banerji, \&
  Richards}]{Coatman2016}
Coatman, L., Hewett, P.~C., Banerji, M., \& Richards, G.~T. 2016, MNRAS, 461,
  647

\bibitem[{Conselice {et~al.}(2009)Conselice, Yang, \& Bluck}]{Conselice2009}
Conselice, C.~J., Yang, C., \& Bluck, A. F.~L. 2009, MNRAS, 394, 1956

\bibitem[{Costa {et~al.}(2014)Costa, Sijacki, \& Haehnelt}]{Costa2014}
Costa, T., Sijacki, D., \& Haehnelt, M.~G. 2014, MNRAS, 444, 2355

\bibitem[{Cresci {et~al.}(2015)Cresci, Mainieri, Brusa, Marconi, Perna,
  Mannucci, Piconcelli, Maiolino, Feruglio, Fiore, Bongiorno, Lanzuisi,
  Merloni, Schramm, Silverman, \& Civano}]{Cresci2015}
Cresci, G., Mainieri, V., Brusa, M., {et~al.} 2015, ApJ, 799, 82

\bibitem[{{Decarli} {et~al.}(2017){Decarli}, {Walter}, {Venemans},
  {Ba{\~n}ados}, {Bertoldi}, {Carilli}, {Fan}, {Farina}, {Mazzucchelli},
  {Riechers}, {Rix}, {Strauss}, {Wang}, \& {Yang}}]{Decarli2017}
{Decarli}, R., {Walter}, F., {Venemans}, B.~P., {et~al.} 2017, Nature, 545, 457

\bibitem[{Decarli {et~al.}(2018)Decarli, Walter, Venemans, Ba{\~{n}}ados,
  Bertoldi, Carilli, Fan, Farina, Mazzucchelli, Riechers, Rix, Strauss, Wang,
  \& Yang}]{Decarli2018}
Decarli, R., Walter, F., Venemans, B.~P., {et~al.} 2018, ApJ, 854, 97

\bibitem[{Decarli {et~al.}(2012)Decarli, Walter, Yang, Carilli, Fan, Hennawi,
  Kurk, Riechers, Rix, Strauss, \& Venemans}]{Decarli2012}
Decarli, R., Walter, F., Yang, Y., {et~al.} 2012, ApJ, 756, 150

\bibitem[{Ding {et~al.}(2022)Ding, Onoue, Silverman, Matsuoka, Izumi, Strauss,
  Jahnke, Andika, Aoki, Baba, Bieri, Bosman, Eilers, Fujimoto, Habouzit,
  Haiman, Imanishi, Inayoshi, Iwasawa, Kashikawa, Kawaguchi, Kohno, Lee, Li,
  Lupi, Lyu, Nagao, Overzier, Phillips, Schindler, Schramm, Shimasaku, Toba,
  Trakhtenbrot, Trebitsch, Treu, Umehata, Venemans, Vestergaard, Volonteri,
  Walter, Wang, \& Yang}]{Ding2022}
Ding, X., Onoue, M., Silverman, J.~D., {et~al.} 2022, Nature
  [\eprint[arXiv]{2211.14329}]

\bibitem[{Dom{\'{\i}}nguez {et~al.}(2013)Dom{\'{\i}}nguez, Siana, Henry,
  Scarlata, Bedregal, Malkan, Atek, Ross, Colbert, Teplitz, Rafelski, McCarthy,
  Bunker, Hathi, Dressler, Martin, \& Masters}]{Dominguez2013}
Dom{\'{\i}}nguez, A., Siana, B., Henry, A.~L., {et~al.} 2013, ApJ, 763, 145

\bibitem[{Dunlop {et~al.}(2003)Dunlop, McLure, Kukula, Baum, O'Dea, \&
  Hughes}]{dunlop_2003}
Dunlop, J.~S., McLure, R.~J., Kukula, M.~J., {et~al.} 2003, MNRAS, 340, 1095

\bibitem[{Eilers {et~al.}(2022)Eilers, Simcoe, Yue, Mackenzie, Matthee,
  Durovcikova, Kashino, Bordoloi, \& Lilly}]{Eilers2022}
Eilers, A.-C., Simcoe, R.~A., Yue, M., {et~al.} 2022, ApJ, 950, 68

\bibitem[{Fabian(2012)}]{Fabian2012}
Fabian, A. 2012, ARA\&A, 50, 455

\bibitem[{Fan {et~al.}(2001)Fan, Narayanan, Lupton, Strauss, Knapp, Becker,
  White, Pentericci, Leggett, Haiman, Gunn, Ivezi{\'{c}}, Schneider, Anderson,
  Brinkmann, Bahcall, Connolly, Csabai, Doi, Fukugita, Geballe, Grebel,
  Harbeck, Hennessy, Lamb, Miknaitis, Munn, Nichol, Okamura, Pier, Prada,
  Richards, Szalay, \& York}]{fan_2001}
Fan, X., Narayanan, V.~K., Lupton, R.~H., {et~al.} 2001, AJ, 122, 2833

\bibitem[{Fan {et~al.}(2003)Fan, Strauss, Schneider, Becker, White, Haiman,
  Gregg, Pentericci, Grebel, Narayanan, Loh, Richards, Gunn, Lupton, Knapp,
  Ivezi{\'{c}}, Brandt, Collinge, Hao, Harbeck, Prada, Schaye, Strateva,
  Zakamska, Anderson, Brinkmann, Bahcall, Lamb, Okamura, Szalay, \&
  York}]{Fan2003}
Fan, X., Strauss, M.~A., Schneider, D.~P., {et~al.} 2003, AJ, 125, 1649

\bibitem[{Fan {et~al.}(2000)Fan, White, Davis, Becker, Strauss, Haiman,
  Schneider, Gregg, Gunn, Knapp, Lupton, Anderson, Anderson, Annis, Bahcall,
  Boroski, Brunner, Chen, Connolly, Csabai, Doi, Fukugita, Hennessy, Hindsley,
  Ichikawa, Ivezić, Loveday, Meiksin, McKay, Munn, Newberg, Nichol, Okamura,
  Pier, Sekiguchi, Shimasaku, Stoughton, Szalay, Szokoly, Thakar, Vogeley, \&
  York}]{fan_2000}
Fan, X., White, R.~L., Davis, M., {et~al.} 2000, AJ, 120, 1167

\bibitem[{Farina {et~al.}(2022)Farina, Schindler, Walter, Ba{\~{n}}ados,
  Davies, Decarli, Eilers, Fan, Hennawi, Mazzucchelli, Meyer, Trakhtenbrot,
  Volonteri, Wang, Worseck, Yang, Gutcke, Venemans, Bosman, Costa, Rosa, Drake,
  \& Onoue}]{Farina2022}
Farina, E.~P., Schindler, J.-T., Walter, F., {et~al.} 2022, ApJ, 941, 106

\bibitem[{Feltre {et~al.}(2016)Feltre, Charlot, \& Gutkin}]{Feltre2016}
Feltre, A., Charlot, S., \& Gutkin, J. 2016, MNRAS, 456, 3354

\bibitem[{Fiore {et~al.}(2017)Fiore, Feruglio, Shankar, Bischetti, Bongiorno,
  Brusa, Carniani, Cicone, Duras, Lamastra, Mainieri, Marconi, Menci, Maiolino,
  Piconcelli, Vietri, \& Zappacosta}]{Fiore2017}
Fiore, F., Feruglio, C., Shankar, F., {et~al.} 2017, A\&A, 601, A143

\bibitem[{Fluetsch {et~al.}(2019)Fluetsch, Maiolino, Carniani, Marconi, Cicone,
  Bourne, Costa, Fabian, Ishibashi, \& Venturi}]{Fluetsch2018}
Fluetsch, A., Maiolino, R., Carniani, S., {et~al.} 2019, MNRAS

\bibitem[{Gardner {et~al.}(2023)Gardner, Mather, Abbott, Abell, Abernathy,
  Abney, Abraham, Abraham, Abul-Huda, Acton, Adams, Adams, Adler, Adriaensen,
  Aguilar, Ahmed, Ahmed, Ahmed, Albat, Albert, Alberts, Aldridge, Allen, Allen,
  Altenburg, Altunc, Alvarez, Álvarez Márquez, de~Oliveira, Ambrose,
  Anandakrishnan, Andersen, Anderson, Anderson, Anderson, Anderson, Aprea,
  Archer, Arenberg, Argyriou, Arribas, Artigau, Arvai, Atcheson, Atkinson,
  Averbukh, Aymergen, Bacinski, Baggett, Bagnasco, Baker, Balzano, Banks,
  Baran, Barker, Barrett, Barringer, Barto, Bast, Baudoz, Baum, Beatty,
  Beaulieu, Bechtold, Beck, Beddard, Beichman, Bellagama, Bely, Berger,
  Bergeron, Darveau-Bernier, Bertch, Beskow, Betz, Biagetti, Birkmann,
  Bjorklund, Blackwood, Blazek, Blossfeld, Bluth, Boccaletti, Boegner, Bohlin,
  Boia, Böker, Bonaventura, Bond, Bosley, Boucarut, Bouchet, Bouwman, Bower,
  Bowers, Bowers, Boyce, Boyer, Boyer, Boyer, Boyer, Bradley, Brady, Brandl,
  Brannen, Breda, Bremmer, Brennan, Bresnahan, Bright, Broiles, Bromenschenkel,
  Brooks, Brooks, Brown, Brown, Brown, Bruce, Bryson, Bujanda, Bullock, Bunker,
  Bureo, Burt, Bush, Bushouse, Bussman, Cabaud, Cale, Calhoon, Calvani, Canipe,
  Caputo, Cara, Carey, Case, Cesari, Cetorelli, Chance, Chandler, Chaney,
  Chapman, Charlot, Chayer, Cheezum, Chen, Chen, Cherinka, Chichester, Chilton,
  Chittiraibalan, Clampin, Clark, Clark, Clark, Claybrooks, Cleveland, Cohen,
  Cohen, Colón, Coleman, Colina, Comber, Comeau, Comer, Reis, Connolly,
  Conroy, Contos, Contreras, Cook, Cooper, Cooper, Correia, Correnti, Cossou,
  Costanza, Coulais, Cox, Coyle, Cracraft, Noriega-Crespo, Crew, Curtis,
  Cusveller, Maciel, Dailey, Daugeron, Davidson, Davies, Davis, Davis, Day,
  de~Chambure, de~Jong, De~Marchi, Dean, Decker, Delisa, Dell, Dellagatta,
  Dembinska, Demosthenes, Dencheva, Deneu, DePriest, Deschenes, Dethienne,
  Detre, Diaz, Dicken, DiFelice, Dillman, Disharoon, van Dishoeck, Dixon,
  Doggett, Dominguez, Donaldson, Doria-Warner, Santos, Doty, Douglas, Doyon,
  Dressler, Driggers, Driggers, Dunn, DuPrie, Dupuis, Durning, Dutta, Earl,
  Eccleston, Ecobichon, Egami, Ehrenwinkler, Eisenhamer, Eisenhower,
  Eisenstein, Hamel, Elie, Elliott, Elliott, Engesser, Espinoza, Etienne,
  Etxaluze, Evans, Fabreguettes, Falcolini, Falini, Fatig, Feeney, Feinberg,
  Fels, Ferdous, Ferguson, Ferrarese, Ferreira, Ferruit, Ferry, Filippazzo,
  Firre, Fix, Flagey, Flanagan, Fleming, Florian, Flynn, Foiadelli, Fontaine,
  Fontanella, Forshay, Fortner, Fox, Framarini, Francisco, Franck, Franx,
  Franz, Friedman, Friend, Frost, Fu, Fullerton, Gaillard, Galkin, Gallagher,
  Galyer, Marín, Gardner, Garland, Garrett, Gasman, Gáspár, Gastaud,
  Gaudreau, Gauthier, Geers, Geithner, Gennaro, Gerber, Gereau, Giampaoli,
  Giardino, Gibbons, Gilbert, Gilman, Girard, Giuliano, Gkountis, Glasse,
  Glassmire, Glauser, Glazer, Goldberg, Golimowski, Gonzaga, Gordon, Gordon,
  Goudfrooij, Gough, Graham, Grau, Green, Greene, Greene, Greenfield,
  Greenhouse, Greve, Greville, Grimaldi, Groe, Groebner, Grumm, Grundy, Güdel,
  Guillard, Guldalian, Gunn, Gurule, Gutman, Guy, Guyot, Hack, Haderlein,
  Hagan, Hagedorn, Hainline, Haley, Hami, Hamilton, Hammann, Hammel, Hanley,
  Hansen, Hardy, Harnisch, Harr, Harris, Hart, Hartig, Hasan, Hashim,
  Hashimoto, Haskins, Hawkins, Hayden, Hayden, Healy, Hecht, Heeg, Hejal, Helm,
  Hengemihle, Henning, Henry, Henry, Henshaw, Hernandez, Herrington, Heske,
  Hesman, Hickey, Hilbert, Hines, Hinz, Hirsch, Hitcho, Hodapp, Hodge, Hoffman,
  Holfeltz, Holler, Hoppa, Horner, Howard, Howard, Huber, Hunkeler, Hunter,
  Hunter, Hurd, Hurst, Hutchings, Hylan, Ignat, Illingworth, Irish, Isaacs,
  Jackson, Jaffe, Jahic, Jahromi, Jakobsen, James, James, James, Jamieson,
  Jandra, Jayawardhana, Jedrzejewski, Jeffers, Jensen, Joanne, Johns, Johnson,
  Johnson, Johnson, Johnson, Johnson, Johnson, Johnstone, Jollet, Jones, Jones,
  Jones, Jones, Jones, Jordan, Jordan, Jue, Jurkowski, Justis, Justtanont,
  Kaleida, Kalirai, Kalmanson, Kaltenegger, Kammerer, Kan, Kanarek, Kao,
  Karakla, Karl, Kassin, Kauffman, Kavanagh, Kelley, Kelly, Kendrew, Kennedy,
  Kenny, Keski-Kuha, Keyes, Khan, Kidwell, Kimble, King, King, Kinzel, Kirk,
  Kirkpatrick, Klaassen, Klingemann, Klintworth, Knapp, Knight, Knollenberg,
  Knutsen, Koehler, Koekemoer, Kofler, Kontson, Kovacs, Kozhurina-Platais,
  Krause, Kriss, Krist, Kristoffersen, Krogel, Krueger, Kulp, Kumari, Kwan,
  Kyprianou, Labador, Labiano, Lafrenière, Lagage, Laidler, Laine, Laird,
  Lajoie, Lallo, Lam, LaMassa, Lambros, Lampenfield, Lander, Langston, Larson,
  Larson, LaVerghetta, Law, Lawrence, Lee, Lee, Lee, Leisenring, Leveille,
  Levenson, Levi, Levine, Lewis, Lewis, Lewis, Libralato, Lidon, Liebrecht,
  Lightsey, Lilly, Lim, Lim, Ling, Link, Link, Lipinski, Liu, Lo, Lobmeyer,
  Logue, Long, Long, Long, Long, López-Caniego, Lotz, Love-Pruitt, Lubskiy,
  Luers, Luetgens, Luevano, Lui, Lund, Lundquist, Lunine, Lützgendorf, Lynch,
  MacDonald, MacDonald, Macias, Macklis, Maghami, Maharaja, Maiolino,
  Makrygiannis, Malla, Malumuth, Manjavacas, Marini, Marrione, Marston, Martel,
  Martin, Martin, Martinez, Maschmann, Masci, Masetti, Maszkiewicz, Matthews,
  Matuskey, McBrayer, McCarthy, McCaughrean, McClare, McClare, McCloskey,
  McClurg, McCoy, McElwain, McGregor, McGuffey, McKay, McKenzie, McLean,
  McMaster, McNeil, De~Meester, Mehalick, Meixner, Meléndez, Menzel, Menzel,
  Merz, Mesterharm, Meyer, Meyett, Meza, Midwinter, Milam, Miller, Miller,
  Miskey, Misselt, Mitchell, Mohan, Montoya, Moran, Morishita, Moro-Martín,
  Morrison, Morrison, Morse, Moschos, Moseley, Mosier, Mosner, Mountain,
  Muckenthaler, Mueller, Mueller, Muhiem, Mühlmann, Mullally, Mullen, Munger,
  Murphy, Murray, Muzerolle, Mycroft, Myers, Myers, Myers, Myers, Myrick,
  Nagle, Nayak, Naylor, Neff, Nelan, Nella, Nguyen, Nguyen, Nickson, Nidhiry,
  Niedner, Nieto-Santisteban, Nikolov, Nishisaka, Noriega-Crespo, Nota, O'Mara,
  Oboryshko, O'Brien, Ochs, Offenberg, Ogle, Ohl, Olmsted, Osborne,
  O'Shaughnessy, Östlin, O'Sullivan, Otor, Ottens, Ouellette, Outlaw, Owens,
  Pacifici, Page, Paranilam, Park, Parrish, Paschal, Patapis, Patel, Patrick,
  Pattishall, Paul, Paul, Pauly, Pavlovsky, Peña-Guerrero, Pedder, Peek,
  Pelham, Penanen, Perriello, Perrin, Perrine, Perrygo, Peslier, Petach,
  Peterson, Pfarr, Pierson, Pietraszkiewicz, Pilchen, Pipher, Pirzkal, Pitman,
  Player, Plesha, Plitzke, Pohner, Poletis, Pollizzi, Polster, Pontius,
  Pontoppidan, Porges, Potter, Prescott, Proffitt, Pueyo, Neira, Radich, Rager,
  Rameau, Ramey, Alarcon, Rampini, Rapp, Rashford, Rauscher, Ravindranath,
  Rawle, Rawlings, Ray, Regan, Rehm, Rehm, Reid, Reis, Renk, Reoch, Ressler,
  Rest, Reynolds, Richon, Richon, Ridgaway, Riedel, Rieke, Rieke, Rifelli,
  Rigby, Riggs, Ringel, Ritchie, Rix, Robberto, Robinson, Robinson, Rock,
  Rodriguez, del Pino, Roellig, Rohrbach, Roman, Romelfanger, Romo, Rosales,
  Rose, Roteliuk, Roth, Rothwell, Rouzaud, Rowe, Rowlands, Roy, Royer, Rui,
  Rumler, Rumpl, Russ, Ryan, Ryan, Saad, Sabata, Sabatino, Sabbi, Sabelhaus,
  Sabia, Sahu, Saif, Salvignol, Samara-Ratna, Samuelson, Sanders, Sappington,
  Sargent, Sauer, Savadkin, Sawicki, Schappell, Scheffer, Scheithauer, Scherer,
  Schiff, Schlawin, Schmeitzky, Schmitz, Schmude, Schneider, Schreiber,
  Schroeven-Deceuninck, Schultz, Schwab, Schwartz, Scoccimarro, Scott, Scott,
  Seaton, Seely, Seery, Seidleck, Sembach, Shanahan, Shaughnessy, Shaw, Shay,
  Sheehan, Sheth, Shih, Shivaei, Siegel, Sienkiewicz, Simmons, Simon, Sirianni,
  Sivaramakrishnan, Slade, Sloan, Slocum, Slowinski, Smith, Smith, Smith,
  Smith, Smith, Smith, Smolik, Soderblom, Sohn, Sokol, Sonneborn, Sontag, Sooy,
  Soummer, Southwood, Spain, Sparmo, Speer, Spencer, Sprofera, Stallcup,
  Stanley, Stansberry, Stark, Starr, Stassi, Steck, Steeley, Stephens,
  Stephenson, Stewart, Stiavelli, Stockman, Strada, Straughn, Streetman,
  Strickland, Strobele, Stuhlinger, Stys, Such, Sukhatme, Sullivan, Sullivan,
  Sumner, Sun, Sunnquist, Swade, Swam, Swenton, Swoish, Litten, Tamas, Tao,
  Taylor, Taylor, Plate, Van~Tea, Teague, Telfer, Temim, Texter, Thatte,
  Thompson, Thompson, Thomson, Thronson, Tierney, Tikkanen, Tinnin, Tippet,
  Todd, Tran, Trauger, Trejo, Truong, Tsukamoto, Tufail, Tumlinson, Tustain,
  Tyra, Ubeda, Underwood, Uzzo, Vaclavik, Valenduc, Valenti, Van~Campen, van~de
  Wetering, Van Der~Marel, van Haarlem, Vandenbussche, Vanterpool, Vernoy,
  Costas, Volk, Voorzaat, Voyton, Vydra, Waddy, Waelkens, Wahlgren, Walker,
  Wander, Warfield, Warner, Wasiak, Wasiak, Wehner, Weiler, Weilert, Weiss,
  Wells, Welty, Wheate, Wheeler, White, Whitehouse, Whiteleather, Whitman,
  Williams, Willmer, Willott, Willoughby, Wilson, Wilson, Wilson, Windhorst,
  Wislowski, Wolfe, Wolfe, Wolff, Wondel, Woo, Woods, Worden, Workman, Wright,
  Wu, Wu, Wun, Wymer, Yadetie, Yan, Yang, Yates, Yeager, Yerger, Young, Young,
  Yu, Yu, Zak, Zeidler, Zepp, Zhou, Zincke, Zonak, \& Zondag}]{Gardner2023}
Gardner, J.~P., Mather, J.~C., Abbott, R., {et~al.} 2023, PASP, 135, 068001

\bibitem[{Gardner {et~al.}(2006)Gardner, Mather, Clampin, Doyon, Greenhouse,
  Hammel, Hutchings, Jakobsen, Lilly, Long, Lunine, Mccaughrean, Mountain,
  Nella, Rieke, Rieke, Rix, Smith, Sonneborn, Stiavelli, Stockman, Windhorst,
  \& Wright}]{Gardner2006}
Gardner, J.~P., Mather, J.~C., Clampin, M., {et~al.} 2006, SSRv, 123, 485

\bibitem[{Ginolfi {et~al.}(2018)Ginolfi, Maiolino, Carniani, Battaia,
  Cantalupo, \& Schneider}]{Ginolfi2018}
Ginolfi, M., Maiolino, R., Carniani, S., {et~al.} 2018, MNRAS, 476, 2421

\bibitem[{Ginsburg {et~al.}(2019)Ginsburg, Koch, Robitaille, Beaumont,
  {Adamginsburg}, Sipőcz, ZuHone, {Sushobhana Patra}, Jones, {P. L. Lim},
  Stern, Rosolowsky, Earl, Val-Borro, {Jrobbfed}, {Shuokong}, Kepley, {Vlas
  Sokolov}, Badger, Maret, Garrido, Booker, \& Tollerud}]{Ginsburg2019}
Ginsburg, A., Koch, E., Robitaille, T., {et~al.} 2019,
  radio-astro-tools/spectral-cube: Release v0.4.5

\bibitem[{Greene \& Ho(2005)}]{Greene2005}
Greene, J.~E. \& Ho, L.~C. 2005, ApJ, 630, 122

\bibitem[{Harikane {et~al.}(2023)Harikane, Zhang, Nakajima, Ouchi, Isobe, Ono,
  Hatano, Xu, \& Umeda}]{Harikane2023}
Harikane, Y., Zhang, Y., Nakajima, K., {et~al.} 2023
  [\eprint[arXiv]{2303.11946}]

\bibitem[{Harrison {et~al.}(2015)Harrison, Alexander, Mullaney, Stott,
  Swinbank, Arumugam, Bauer, Bower, Bunker, \& Sharples}]{Harrison2015}
Harrison, C.~M., Alexander, D.~M., Mullaney, J.~R., {et~al.} 2015, MNRAS, 456,
  1195

\bibitem[{{Hinshaw} {et~al.}(2013){Hinshaw}, {Larson}, {Komatsu}, {Spergel},
  {Bennett}, {Dunkley}, {Nolta}, {Halpern}, {Hill}, {Odegard}, {Page}, {Smith},
  {Weiland}, {Gold}, {Jarosik}, {Kogut}, {Limon}, {Meyer}, {Tucker}, {Wollack},
  \& {Wright}}]{Hinshaw2013}
{Hinshaw}, G., {Larson}, D., {Komatsu}, E., {et~al.} 2013, ApJS, 208, 19

\bibitem[{Hirschmann {et~al.}(2022)Hirschmann, Charlot, Feltre, Curtis-Lake,
  Somerville, Chevallard, Choi, Nelson, Morisset, Plat, \&
  Vidal-Garcia}]{Hirschmann2022}
Hirschmann, M., Charlot, S., Feltre, A., {et~al.} 2022
  [\eprint[arXiv]{2212.02522}]

\bibitem[{Hopkins {et~al.}(2006)Hopkins, Hernquist, Cox, Di~Matteo, Robertson,
  \& Springel}]{Hopkins2006}
Hopkins, P.~F., Hernquist, L., Cox, T.~J., {et~al.} 2006, ApJS, 163, 1

\bibitem[{Hunter(2007)}]{Matplotlib2007}
Hunter, J.~D. 2007, CiSE, 9, 90

\bibitem[{Husemann {et~al.}(2014)Husemann, Jahnke, S{\'{a}}nchez, Wisotzki,
  Nugroho, Kupko, \& Schramm}]{Husemann2014}
Husemann, B., Jahnke, K., S{\'{a}}nchez, S.~F., {et~al.} 2014, MNRAS, 443, 755

\bibitem[{Husemann {et~al.}(2013)Husemann, Wisotzki, S{\'{a}}nchez, \&
  Jahnke}]{Husemann2013}
Husemann, B., Wisotzki, L., S{\'{a}}nchez, S.~F., \& Jahnke, K. 2013, A\&A,
  549, A43

\bibitem[{Hutchings(2003)}]{hutchings_2003}
Hutchings, J.~B. 2003, AJ, 125, 1053

\bibitem[{Izumi {et~al.}(2019)Izumi, Onoue, Matsuoka, Nagao, Strauss, Imanishi,
  Kashikawa, Fujimoto, Kohno, Toba, Umehata, Goto, Ueda, Shirakata, Silverman,
  Greene, Harikane, Hashimoto, Ikarashi, Iono, Iwasawa, Lee, Minezaki,
  Nakanishi, Tamura, Tang, \& Taniguchi}]{Izumi2019}
Izumi, T., Onoue, M., Matsuoka, Y., {et~al.} 2019, PASJ, 71, 6, 111

\bibitem[{Izumi {et~al.}(2018)Izumi, Onoue, Shirakata, Nagao, Kohno, Matsuoka,
  Imanishi, Strauss, Kashikawa, Schulze, Silverman, Fujimoto, Harikane, Toba,
  Umehata, Nakanishi, Greene, Tamura, Taniguchi, Yamaguchi, Goto, Hashimoto,
  Ikarashi, Iono, Iwasawa, Lee, Makiya, Minezaki, \& Tang}]{Izumi2018}
Izumi, T., Onoue, M., Shirakata, H., {et~al.} 2018, PASJ, 70, 3, 36

\bibitem[{Jakobsen {et~al.}(2022)Jakobsen, Ferruit, de~Oliveira, Arribas,
  Bagnasco, Barho, Beck, Birkmann, Böker, Bunker, Charlot, de~Jong, de~Marchi,
  Ehrenwinkler, Falcolini, Fels, Franx, Franz, Funke, Giardino, Gnata, Holota,
  Honnen, Jensen, Jentsch, Johnson, Jollet, Karl, Kling, Köhler, Kolm, Kumari,
  Lander, Lemke, L{\'{o}}pez-Caniego, Lützgendorf, Maiolino, Manjavacas,
  Marston, Maschmann, Maurer, Messerschmidt, Moseley, Mosner, Mott, Muzerolle,
  Pirzkal, Pittet, Plitzke, Posselt, Rapp, Rauscher, Rawle, Rix, Rödel,
  Rumler, Sabbi, Salvignol, Schmid, Sirianni, Smith, Strada, te~Plate, Valenti,
  Wettemann, Wiehe, Wiesmayer, Willott, Wright, Zeidler, \&
  Zincke}]{Jakobsen2022}
Jakobsen, P., Ferruit, P., de~Oliveira, C.~A., {et~al.} 2022, A\&A, 661, A80

\bibitem[{Kakkad {et~al.}(2020)Kakkad, Mainieri, Vietri, Carniani, Harrison,
  Perna, Scholtz, Circosta, Cresci, Husemann, Bischetti, Feruglio, Fiore,
  Marconi, Padovani, Brusa, Cicone, Comastri, Lanzuisi, Mannucci, Menci,
  Netzer, Piconcelli, Puglisi, Salvato, Schramm, Silverman, Vignali, Zamorani,
  \& Zappacosta}]{Kakkad2020}
Kakkad, D., Mainieri, V., Vietri, G., {et~al.} 2020, A\&A, 642, A147

\bibitem[{Kashikawa {et~al.}(2015)Kashikawa, Ishizaki, Willott, Onoue, Im,
  Furusawa, Toshikawa, Ishikawa, Niino, Shimasaku, Ouchi, \&
  Hibon}]{Kashikawa2015}
Kashikawa, N., Ishizaki, Y., Willott, C.~J., {et~al.} 2015, ApJ, 798, 28

\bibitem[{Kashino {et~al.}(2022)Kashino, Lilly, Matthee, Eilers, Mackenzie,
  Bordoloi, \& Simcoe}]{Kashino2022}
Kashino, D., Lilly, S.~J., Matthee, J., {et~al.} 2022, ApJ, 950, 66

\bibitem[{Kaspi {et~al.}(2021)Kaspi, Brandt, Maoz, Netzer, Schneider, Shemmer,
  \& Grier}]{Kaspi2021}
Kaspi, S., Brandt, W.~N., Maoz, D., {et~al.} 2021, ApJ, 915, 129

\bibitem[{Kaspi {et~al.}(2000)Kaspi, Smith, Netzer, Maoz, Jannuzi, \&
  Giveon}]{Kaspi2000}
Kaspi, S., Smith, P.~S., Netzer, H., {et~al.} 2000, ApJ, 533, 631

\bibitem[{Kauffmann {et~al.}(2003)Kauffmann, Heckman, Tremonti, Brinchmann,
  Charlot, White, Ridgway, Brinkmann, Fukugita, Hall, IveziÄ, Richards, \&
  Schneider}]{kauffmann_2003}
Kauffmann, G., Heckman, T.~M., Tremonti, C., {et~al.} 2003, MNRAS, 346, 1055

\bibitem[{Kennicutt(1998)}]{kennicutt_1998}
Kennicutt, R.~C. 1998, ARA\&A, 36, 189

\bibitem[{Kennicutt \& Evans(2012)}]{Kennicutt2012}
Kennicutt, R.~C. \& Evans, N.~J. 2012, ARA\&A, 50, 531

\bibitem[{Kewley {et~al.}(2001)Kewley, Dopita, Sutherland, Heisler, \&
  Trevena}]{Kewley2001}
Kewley, L.~J., Dopita, M.~A., Sutherland, R.~S., Heisler, C.~A., \& Trevena, J.
  2001, ApJ, 556, 121

\bibitem[{Kocevski {et~al.}(2012)Kocevski, Faber, Mozena, Koekemoer, Nandra,
  Rangel, Laird, Brusa, Wuyts, Trump, Koo, Somerville, Bell, Lotz, Alexander,
  Bournaud, Conselice, Dahlen, Dekel, Donley, Dunlop, Finoguenov, Georgakakis,
  Giavalisco, Guo, Grogin, Hathi, Juneau, Kartaltepe, Lucas, McGrath, McIntosh,
  Mobasher, Robaina, Rosario, Straughn, van~der Wel, \&
  Villforth}]{Kocevski2012}
Kocevski, D.~D., Faber, S.~M., Mozena, M., {et~al.} 2012, ApJ, 744, 148

\bibitem[{Kormendy \& Ho(2013)}]{Kormendy2013}
Kormendy, J. \& Ho, L.~C. 2013, ARA\&A, 51, 511

\bibitem[{Kovacevic {et~al.}(2010)Kovacevic, Popovic, \&
  Dimitrijevic}]{Kovacevic2010}
Kovacevic, J., Popovic, L.~C., \& Dimitrijevic, M.~S. 2010, ApJS, 189, 15

\bibitem[{Kuraszkiewicz {et~al.}(2002)Kuraszkiewicz, Green, Forster, Aldcroft,
  Evans, \& Koratkar}]{Kuraszkiewicz2002}
Kuraszkiewicz, J.~K., Green, P.~J., Forster, K., {et~al.} 2002, ApJS, 143, 257

\bibitem[{Law {et~al.}(2023)Law, Morrison, Argyriou, Patapis, Alvarez-Marquez,
  Labiano, \& Vandenbussche}]{Law2023}
Law, D.~R., Morrison, J.~E., Argyriou, I., {et~al.} 2023, A 3D Drizzle
  Algorithm for JWST and Practical Application to the MIRI Medium Resolution
  Spectrometer

\bibitem[{Lee {et~al.}(2009)Lee, de~Paz, Tremonti, Kennicutt, Salim, Bothwell,
  Calzetti, Dalcanton, Dale, Engelbracht, Funes, Johnson, Sakai, Skillman, van
  Zee, Walter, \& Weisz}]{Lee2009}
Lee, J.~C., de~Paz, A.~G., Tremonti, C., {et~al.} 2009, ApJ, 706, 599

\bibitem[{Lutz {et~al.}(2020)Lutz, Sturm, Janssen, Veilleux, Aalto, Cicone,
  Contursi, Davies, Feruglio, Fischer, Fluetsch, Garcia-Burillo, Genzel,
  Gonz{\'{a}}lez-Alfonso, Graci{\'{a}}-Carpio, Herrera-Camus, Maiolino,
  Schruba, Shimizu, Sternberg, Tacconi, \& Wei{\ss}}]{Lutz2020}
Lutz, D., Sturm, E., Janssen, A., {et~al.} 2020, A\&A, 633, A134

\bibitem[{Marian {et~al.}(2019)Marian, Jahnke, Mechtley, Cohen, Husemann,
  Jones, Koekemoer, Schulze, van~der Wel, Villforth, \& Windhorst}]{Marian2019}
Marian, V., Jahnke, K., Mechtley, M., {et~al.} 2019, ApJ, 882, 141

\bibitem[{Marshall {et~al.}(2020)Marshall, Mechtley, Windhorst, Cohen, Jansen,
  Jiang, Jones, Wyithe, Fan, Hathi, Jahnke, Keel, Koekemoer, Marian, Ren,
  Robinson, Röttgering, Ryan, Scannapieco, Schneider, Schneider, Smith, \&
  Yan}]{Marshall2019c}
Marshall, M.~A., Mechtley, M., Windhorst, R.~A., {et~al.} 2020, ApJ, 900, 21

\bibitem[{Marshall {et~al.}(2021)Marshall, Wyithe, Windhorst, Matteo, Ni,
  Wilkins, Croft, \& Mechtley}]{MarshallBTpsfMC}
Marshall, M.~A., Wyithe, J. S.~B., Windhorst, R.~A., {et~al.} 2021, MNRAS, 506,
  1209

\bibitem[{Matsuoka {et~al.}(2018)Matsuoka, Onoue, Kashikawa, Iwasawa, Strauss,
  Nagao, Imanishi, Lee, Akiyama, Asami, Bosch, Foucaud, Furusawa, Goto, Gunn,
  Harikane, Ikeda, Izumi, Kawaguchi, Kikuta, Kohno, Komiyama, Lupton, Minezaki,
  Miyazaki, Morokuma, Murayama, Niida, Nishizawa, Oguri, Ono, Ouchi, Price,
  Sameshima, Schulze, Shirakata, Silverman, Sugiyama, Tait, Takada, Takata,
  Tanaka, Tang, Toba, Utsumi, \& Wang}]{Matsuoka2018}
Matsuoka, Y., Onoue, M., Kashikawa, N., {et~al.} 2018, PASJ, 70

\bibitem[{McElwain {et~al.}(2023)McElwain, Feinberg, Perrin, Clampin, Mountain,
  Lallo, Lajoie, Kimble, Bowers, Stark, Acton, Atkinson, Barinek, Barto,
  Basinger, Beck, Bergkoetter, Bluth, Boucarut, Brady, Brooks, Brown, Byard,
  Carey, Carrasquilla, Chae, Chaney, Chayer, Chonis, Cohen, Cole, Comeau, Coon,
  Coppock, Coyle, Dean, Dziak, Eisenhower, Flagey, Franck, Gallagher, Gilman,
  Glassman, Green, Grieco, Haase, Hadjimichael, Hagopian, Hahn, Hartig, Havey,
  Hayden, Hellekson, Hicks, Holfeltz, Howard, Huguet, Jahne, Johnson, Johnston,
  Jurling, Kegley, Kennard, Keski-Kuha, Knight, Kulp, Levi, Levine, Lightsey,
  Luetgens, Mather, Matthews, McKay, Mehalick, Mel{\'{e}}ndez, Mosier, Murphy,
  Nelan, Niedner, Nol, Ohara, Ohl, Olczak, Osborne, Park, Perrygo, Pueyo,
  Redding, Regan, Reynolds, Rifelli, Rigby, Sabatke, Saif, Scorse, Seo, Shi,
  Sigrist, Smith, Smith, Smith, Sohn, Stahl, Telfer, Terlecki, Texter, Buren,
  Campen, Vila, Voyton, Waldman, Walker, Weiser, Wells, West, Whitman, Wolf, \&
  Zielinski}]{McElwain2023}
McElwain, M.~W., Feinberg, L.~D., Perrin, M.~D., {et~al.} 2023, PASP, 135,
  058001

\bibitem[{McGreer {et~al.}(2014)McGreer, Fan, Strauss, Haiman, Richards, Jiang,
  Bian, \& Schneider}]{McGreer2014}
McGreer, I.~D., Fan, X., Strauss, M.~A., {et~al.} 2014, AJ, 148, 73

\bibitem[{McLeod \& Rieke(1994)}]{mcleod_1994}
McLeod, K.~K. \& Rieke, G.~H. 1994, ApJ, 420, 58

\bibitem[{Mechtley {et~al.}(2016)Mechtley, Jahnke, Windhorst, Andrae,
  Cisternas, Cohen, Hewlett, Koekemoer, Schramm, Schulze, Silverman, Villforth,
  van~der Wel, \& Wisotzki}]{Mechtley2016}
Mechtley, M., Jahnke, K., Windhorst, R.~A., {et~al.} 2016, ApJ, 830, 156

\bibitem[{Mechtley {et~al.}(2012)Mechtley, Windhorst, Ryan, Schneider, Cohen,
  Jansen, Fan, Hathi, Keel, Koekemoer, Röttgering, Scannapieco, Schneider,
  Strauss, \& Yan}]{Mechtley2012}
Mechtley, M., Windhorst, R.~A., Ryan, R.~E., {et~al.} 2012, ApJ, 756, L38

\bibitem[{Nagao {et~al.}(2006)Nagao, Marconi, \& Maiolino}]{Nagao2006}
Nagao, T., Marconi, A., \& Maiolino, R. 2006, A\&A, 447, 157

\bibitem[{Nakajima \& Maiolino(2022)}]{Nakajima2022}
Nakajima, K. \& Maiolino, R. 2022, MNRAS, 513, 5134

\bibitem[{Neeleman {et~al.}(2021)Neeleman, Novak, Venemans, Walter, Decarli,
  Kaasinen, Schindler, Ba{\~{n}}ados, Carilli, Drake, Fan, \&
  Rix}]{Neeleman2021}
Neeleman, M., Novak, M., Venemans, B.~P., {et~al.} 2021, ApJ, 911, 141

\bibitem[{Nguyen {et~al.}(2020)Nguyen, Lira, Trakhtenbrot, Netzer, Cicone,
  Maiolino, \& Shemmer}]{Nguyen2020}
Nguyen, N.~H., Lira, P., Trakhtenbrot, B., {et~al.} 2020, ApJ, 895, 74

\bibitem[{Onoue {et~al.}(2019)Onoue, Kashikawa, Matsuoka, Kato, Izumi, Nagao,
  Strauss, Harikane, Imanishi, Ito, Iwasawa, Kawaguchi, Lee, Noboriguchi, Suh,
  Tanaka, \& Toba}]{Onoue2019}
Onoue, M., Kashikawa, N., Matsuoka, Y., {et~al.} 2019, ApJ, 880, 77

\bibitem[{Osterbrock(1989)}]{Osterbrock1989}
Osterbrock, D.~E. 1989, Astrophysics of gaseous nebulae and active galactic
  nuclei (University Science Books)

\bibitem[{{Pandas Development Team}(2020)}]{reback2020pandas}
{Pandas Development Team}. 2020, pandas-dev/pandas: Pandas

\bibitem[{Patton {et~al.}(2000)Patton, Carlberg, Marzke, Pritchet, da~Costa, \&
  Pellegrini}]{Patton2000}
Patton, D.~R., Carlberg, R.~G., Marzke, R.~O., {et~al.} 2000, ApJ, 536, 153

\bibitem[{Pensabene {et~al.}(2020)Pensabene, Carniani, Perna, Cresci, Decarli,
  Maiolino, \& Marconi}]{Pensabene2020}
Pensabene, A., Carniani, S., Perna, M., {et~al.} 2020, A\&A, 637, A84

\bibitem[{Perna {et~al.}(2023)Perna, Arribas, Marshall, D'Eugenio, Übler,
  Bunker, Charlot, Carniani, Jakobsen, Maiolino, Del~Pino, Willott, Böker,
  Circosta, Cresci, Curti, Husemann, Kumari, Lamperti, Pérez-González, \&
  Scholtz}]{Perna2023}
Perna, M., Arribas, S., Marshall, M., {et~al.} 2023, The ultradense,
  interacting environment of a dual AGN at z $\sim$ 3.3 revealed by
  JWST/NIRSpec IFS

\bibitem[{Perna {et~al.}(2015)Perna, Brusa, Cresci, Comastri, Lanzuisi, Lusso,
  Marconi, Salvato, Zamorani, Bongiorno, Mainieri, Maiolino, \&
  Mignoli}]{Perna2015}
Perna, M., Brusa, M., Cresci, G., {et~al.} 2015, A\&A, 574, A82

\bibitem[{Peterson(2009)}]{Peterson2009}
Peterson, B.~M. 2009, Proc. Int. Astron. Union, 5, 151

\bibitem[{Pons {et~al.}(2019)Pons, McMahon, Simcoe, Banerji, Hewett, \&
  Reed}]{Pons2019a}
Pons, E., McMahon, R.~G., Simcoe, R.~A., {et~al.} 2019, MNRAS, 484, 5142

\bibitem[{Rankine {et~al.}(2020)Rankine, Hewett, Banerji, \&
  Richards}]{Rankine2020}
Rankine, A.~L., Hewett, P.~C., Banerji, M., \& Richards, G.~T. 2020, MNRAS,
  492, 4553

\bibitem[{Reed {et~al.}(2019)Reed, Banerji, Becker, Hewett, Martini, McMahon,
  Pons, Rauch, Abbott, Allam, Annis, Avila, Bertin, Brooks, Buckley-Geer,
  Rosell, Kind, Carretero, Castander, Cunha, D'Andrea, da~Costa, Vicente,
  Desai, Diehl, Doel, Evrard, Flaugher, Frieman, Garc{\'{\i}}a-Bellido,
  Gaztanaga, Gruen, Gschwend, Gutierrez, Hollowood, Honscheid, Hoyle, James,
  Kuehn, Lahav, Lima, Maia, Marshall, Miquel, Ogando, Plazas, Roodman, Sanchez,
  Scarpine, Schubnell, Serrano, Sevilla-Noarbe, Smith, Smith, Sobreira,
  Suchyta, Swanson, Tarle, Thomas, Tucker, \& Vikram}]{Reed2019}
Reed, S.~L., Banerji, M., Becker, G.~D., {et~al.} 2019, MNRAS, 487, 1874

\bibitem[{Riechers {et~al.}(2007)Riechers, Walter, Carilli, \&
  Bertoldi}]{Riechers2007}
Riechers, D.~A., Walter, F., Carilli, C.~L., \& Bertoldi, F. 2007, ApJL, 671,
  L13

\bibitem[{Rigby {et~al.}(2022)Rigby, Perrin, McElwain, Kimble, Friedman, Lallo,
  Doyon, Feinberg, Ferruit, Glasse, Rieke, Rieke, Wright, Willott, Colon,
  Milam, Neff, Stark, Valenti, Abell, Abney, Abul-Huda, Acton, Adams, Adler,
  Aguilar, Ahmed, Albert, Alberts, Aldridge, Allen, Altenburg, de~Oliveira,
  Anderson, Anderson, Anderson, Argyriou, Armstrong, Arribas, Artigau, Arvai,
  Atkinson, Bacon, Bair, Banks, Barrientes, Barringer, Bartosik, Bast, Baudoz,
  Beatty, Bechtold, Beck, Bergeron, Bergkoetter, Bhatawdekar, Birkmann, Blazek,
  Blome, Boccaletti, Boeker, Boia, Bonaventura, Bond, Bosley, Boucarut,
  Bourque, Bouwman, Bower, Bowers, Boyer, Brady, Braun, Breda, Bresnahan,
  Bright, Britt, Bromenschenkel, Brooks, Brooks, Brown, Brown, Brown, Bunker,
  Burger, Bushouse, Cale, Cameron, Cameron, Canipe, Caplinger, Caputo, Carey,
  Carniani, Carrasquilla, Carruthers, Case, Chance, Chapman, Charlot, Charlow,
  Chayer, Chen, Cherinka, Chichester, Chilton, Chonis, Clark, Clark, Coe,
  Coleman, Comber, Comeau, Connolly, Cooper, Cooper, Coppock, Correnti, Cossou,
  Coulais, Coyle, Cracraft, Curti, Cuturic, Davis, Davis, Dean, DeLisa,
  deMeester, Dencheva, Dencheva, DePasquale, Deschenes, Örs Hunor~Detre, Diaz,
  Dicken, DiFelice, Dillman, Dixon, Doggett, Donaldson, Douglas, DuPrie,
  Dupuis, Durning, Easmin, Eck, Edeani, Egami, Ehrenwinkler, Eisenhamer,
  Eisenhower, Elie, Elliott, Elliott, Ellis, Engesser, Espinoza, Etienne,
  Etxaluze, Falini, Feeney, Ferry, Filippazzo, Fincham, Fix, Flagey, Florian,
  Flynn, Fontanella, Ford, Forshay, Fox, Franz, Fu, Fullerton, Galkin, Galyer,
  Marin, Gardner, Gardner, Garland, Gasman, Gaspar, Gaudreau, Gauthier, Geers,
  Geithner, Gennaro, Giardino, Girard, Giuliano, Glassmire, Glauser, Glazer,
  Godfrey, Golimowski, Gollnitz, Gong, Gonzaga, Gordon, Gordon, Goudfrooij,
  Greene, Greenhouse, Grimaldi, Groebner, Grundy, Guillard, Gutman, Ha,
  Haderlein, Hagedorn, Hainline, Haley, Hami, Hamilton, Hammel, Hansen,
  Harkins, Harr, Hart, Hart, Hartig, Hashimoto, Haskins, Hathaway, Havey,
  Hayden, Hecht, Heller-Boyer, Henry, Hermann, Hernandez, Hesman, Hicks,
  Hilbert, Hines, Hoffman, Holfeltz, Holler, Hoppa, Hott, Howard, Hunter,
  Hunter, Hurst, Husemann, Hustak, Ignat, Irish, Jackson, Jahromi, Jakobsen,
  James, James, Januszewski, Jenkins, Jirdeh, Johnson, Johnson, Jones, Jones,
  Jones, Jones, Jordan, Jordan, Jurczyk, Jurling, Kaleida, Kalmanson, Kammerer,
  Kang, Kao, Karakla, Kavanagh, Kelly, Kendrew, Kennedy, Kenny, Keski-kuha,
  Keyes, Kidwell, Kinzel, Kirk, Kirkpatrick, Kirshenblat, Klaassen, Knapp,
  Knight, Knollenberg, Koehler, Koekemoer, Kovacs, Kulp, Kumari, Kyprianou,
  Massa, Labador, Ortega, Lagage, Lajoie, Lallo, Lam, Lamb, Lambros,
  Lampenfield, Langston, Larson, Law, Lawrence, Lee, Leisenring, Lepo,
  Leveille, Levenson, Levine, Levy, Lewis, Lewis, Libralato, Lightsey, Link,
  Liu, Lo, Lockwood, Logue, Long, Long, Loomis, Lopez-Caniego, Alvarez,
  Love-Pruitt, Lucy, Luetzgendorf, Maghami, Maiolino, Major, Malla, Malumuth,
  Manjavacas, Mannfolk, Marrione, Marston, Martel, Maschmann, Masci,
  Masciarelli, Maszkiewicz, Mather, McKenzie, McLean, McMaster, Melbourne,
  Meléndez, Menzel, Merz, Meyett, Meza, Miskey, Misselt, Moller, Morrison,
  Morse, Moseley, Mosier, Mountain, Mueckay, Mueller, Mullally, Murphy, Murray,
  Murray, Muzerolle, Mycroft, Myers, Myrick, Nanavati, Nance, Nayak, Naylor,
  Nelan, Nickson, Nielson, Nieto-Santisteban, Nikolov, Noriega-Crespo,
  O'Shaughnessy, O'Sullivan, Ochs, Ogle, Oleszczuk, Olmsted, Osborne, Ottens,
  Owens, Pacifici, Pagan, Page, Parrish, Patapis, Pauly, Pavlovsky, Pedder,
  Peek, Pena-Guerrero, Pennanen, Perez, Perna, Perriello, Phillips,
  Pietraszkiewicz, Pinaud, Pirzkal, Pitman, Piwowar, Platais, Player, Plesha,
  Pollizi, Polster, Pontoppidan, Porterfield, Proffitt, Pueyo, Pulliam, Quirt,
  Neira, Alarcon, Ramsay, Rapp, Rapp, Rauscher, Ravindranath, Rawle, Regan,
  Reichard, Reis, Ressler, Rest, Reynolds, Rhue, Richon, Rickman, Ridgaway,
  Ritchie, Rix, Robberto, Robinson, Robinson, Robinson, Rock, Rodriguez, Pino,
  Roellig, Rohrbach, Roman, Romelfanger, Rose, Roteliuk, Roth, Rothwell,
  Rowlands, Roy, Royer, Royle, Rui, Rumler, Runnels, Russ, Rustamkulov, Ryden,
  Ryer, Sabata, Sabatke, Sabbi, Samuelson, Sappington, Sargent, Sauer,
  Scheithauer, Schlawin, Schlitz, Schmitz, Schneider, Schreiber, Schulze,
  Schwab, Scott, Sembach, Shaughnessy, Shaw, Shawger, Shay, Sheehan, Shen,
  Sherman, Shiao, Shih, Shivaei, Sienkiewicz, Sing, Sirianni, Sivaramakrishnan,
  Skipper, Sloan, Slocum, Slowinski, Smith, Smith, Smith, Smith, Snyder, Soh,
  Sohn, Soto, Spencer, Stallcup, Stansberry, Starr, Starr, Stewart, Stiavelli,
  Straughn, Strickland, Stys, Summers, Sun, Sunnquist, Swade, Swam, Swaters,
  Swoish, Taylor, Taylor, Plate, Tea, Teague, Telfer, Temim, Thatte, Thompson,
  Thompson, Thomson, Tikkanen, Tippet, Todd, Toolan, Tran, Trejo, Truong,
  Tsukamoto, Tustain, Tyra, Ubeda, Underwood, Uzzo, Campen, Vandal,
  Vandenbussche, Vila, Volk, Wahlgren, Waldman, Walker, Wander, Warfield,
  Warner, Wasiak, Watkins, Weilert, Weiser, Weiss, Weissman, Welty, West,
  Wheate, Wheatley, Wheeler, White, Whiteaker, Whitehouse, Whiteleather,
  Whitman, Williams, Willmer, Willoughby, Wilson, Wirth, Wislowski, Wolf,
  Wolfe, Wolff, Workman, Wright, Wu, Wu, Wymer, Yates, Yates, Yeager, Yerger,
  Yoon, Young, Yu, Zak, Zeidler, Zhou, Zielinski, Zincke, \& Zonak}]{Rigby2022}
Rigby, J., Perrin, M., McElwain, M., {et~al.} 2022, PASP, 135, 048001

\bibitem[{Sanders {et~al.}(1988)Sanders, Soifer, Elias, Madore, Matthews,
  Neugebauer, \& Scoville}]{Sanders1988}
Sanders, D.~B., Soifer, B.~T., Elias, J.~H., {et~al.} 1988, ApJ, 325, 74

\bibitem[{Schmidt(1963)}]{schmidt_1963}
Schmidt, M. 1963, Nature, 197, 1040

\bibitem[{Shen {et~al.}(2008)Shen, Greene, Strauss, Richards, \&
  Schneider}]{shen2008}
Shen, Y., Greene, J.~E., Strauss, M.~A., Richards, G.~T., \& Schneider, D.~P.
  2008, ApJ, 680, 169

\bibitem[{Shen \& Kelly(2012)}]{Shen2012}
Shen, Y. \& Kelly, B.~C. 2012, ApJ, 746, 169

\bibitem[{Shen {et~al.}(2011)Shen, Richards, Strauss, Hall, Schneider, Snedden,
  Bizyaev, Brewington, Malanushenko, Malanushenko, Oravetz, Pan, \&
  Simmons}]{Shen2011}
Shen, Y., Richards, G.~T., Strauss, M.~A., {et~al.} 2011, ApJS, 194, 45

\bibitem[{Sijacki {et~al.}(2009)Sijacki, Springel, \& Haehnelt}]{Sijacki2009}
Sijacki, D., Springel, V., \& Haehnelt, M.~G. 2009, MNRAS, 400, 100

\bibitem[{Strom {et~al.}(2017)Strom, Steidel, Rudie, Trainor, Pettini, \&
  Reddy}]{Strom2017}
Strom, A.~L., Steidel, C.~C., Rudie, G.~C., {et~al.} 2017, ApJ, 836, 164

\bibitem[{Sun(2020)}]{SunGithub}
Sun, J. 2020, Sun\_Astro\_Tools, GitHub

\bibitem[{Sun {et~al.}(2018)Sun, Leroy, Schruba, Rosolowsky, Hughes, Kruijssen,
  Meidt, Schinnerer, Blanc, Bigiel, Bolatto, Chevance, Groves, Herrera, Hygate,
  Pety, Querejeta, Usero, \& Utomo}]{Sun2018}
Sun, J., Leroy, A.~K., Schruba, A., {et~al.} 2018, ApJ, 860, 172

\bibitem[{{Trakhtenbrot} {et~al.}(2017){Trakhtenbrot}, {Lira}, {Netzer},
  {Cicone}, {Maiolino}, \& {Shemmer}}]{Trakhtenbrot2017}
{Trakhtenbrot}, B., {Lira}, P., {Netzer}, H., {et~al.} 2017, ApJ, 836, 8

\bibitem[{Trevese {et~al.}(2014)Trevese, Perna, Vagnetti, Saturni, \&
  Dadina}]{Trevese2014}
Trevese, D., Perna, M., Vagnetti, F., Saturni, F.~G., \& Dadina, M. 2014, ApJ,
  795, 164

\bibitem[{van~der Walt {et~al.}(2011)van~der Walt, Colbert, \&
  Varoquaux}]{Numpy2011}
van~der Walt, S., Colbert, S.~C., \& Varoquaux, G. 2011, CiSE, 13, 22

\bibitem[{van~der Wel {et~al.}(2022)van~der Wel, van Houdt, Bezanson, Franx,
  D'Eugenio, Straatman, Bell, Muzzin, Sobral, Maseda, de~Graaff, \&
  Holden}]{VanDerWel2022}
van~der Wel, A., van Houdt, J., Bezanson, R., {et~al.} 2022, ApJ, 936, 9

\bibitem[{van Dokkum(2001)}]{Dokkum2001}
van Dokkum, P.~G. 2001, PASP, 113, 1420

\bibitem[{Vayner {et~al.}(2016)Vayner, Wright, Do, Larkin, Armus, \&
  Gallagher}]{Vayner2014}
Vayner, A., Wright, S.~A., Do, T., {et~al.} 2016, ApJ, 821:64, 2016
  [\eprint{1410.4229}]

\bibitem[{Veilleux \& Osterbrock(1987)}]{Veilleux1987}
Veilleux, S. \& Osterbrock, D.~E. 1987, ApJS, 63, 295

\bibitem[{Venemans {et~al.}(2017)Venemans, Walter, Decarli, Ba{\~{n}}ados,
  Hodge, Hewett, McMahon, Mortlock, \& Simpson}]{Venemans2017a}
Venemans, B.~P., Walter, F., Decarli, R., {et~al.} 2017, ApJ, 837, 146

\bibitem[{Venemans {et~al.}(2015)Venemans, Walter, Zschaechner, Decarli, Rosa,
  Findlay, McMahon, \& Sutherland}]{Venemans2015}
Venemans, B.~P., Walter, F., Zschaechner, L., {et~al.} 2015, ApJ, 816, 37

\bibitem[{Vestergaard \& Osmer(2009)}]{Vestergaard2009}
Vestergaard, M. \& Osmer, P.~S. 2009, ApJ, 699, 800

\bibitem[{Vestergaard \& Peterson(2006)}]{Vestergaard2006}
Vestergaard, M. \& Peterson, B.~M. 2006, ApJ, 641, 689

\bibitem[{{Virtanen} {et~al.}(2020){Virtanen}, {Gommers}, {Oliphant},
  {Haberland}, {Reddy}, {Cournapeau}, {Burovski}, {Peterson}, {Weckesser},
  {Bright}, {van der Walt}, {Brett}, {Wilson}, {Jarrod Millman}, {Mayorov},
  {Nelson}, {Jones}, {Kern}, {Larson}, {Carey}, {Polat}, {Feng}, {Moore}, {Vand
  erPlas}, {Laxalde}, {Perktold}, {Cimrman}, {Henriksen}, {Quintero}, {Harris},
  {Archibald}, {Ribeiro}, {Pedregosa}, {van Mulbregt}, \&
  {Contributors}}]{2020SciPy-NMeth}
{Virtanen}, P., {Gommers}, R., {Oliphant}, T.~E., {et~al.} 2020, Nat. Methods,
  17, 261

\bibitem[{Volonteri \& Rees(2005)}]{Volonteri2005}
Volonteri, M. \& Rees, M.~J. 2005, ApJ, 633, 624

\bibitem[{Volonteri {et~al.}(2015)Volonteri, Silk, \& Dubus}]{Volonteri2015}
Volonteri, M., Silk, J., \& Dubus, G. 2015, ApJ, 804, 148

\bibitem[{Wagg {et~al.}(2012)Wagg, Wiklind, Carilli, Espada, Peck, Riechers,
  Walter, Wootten, Aravena, Barkats, Cortes, Hills, Hodge, Impellizzeri, Iono,
  Leroy, Mart{\'{\i}}n, Rawlings, Maiolino, McMahon, Scott, Villard, \&
  Vlahakis}]{Wagg2012}
Wagg, J., Wiklind, T., Carilli, C.~L., {et~al.} 2012, ApJ, 752, L30

\bibitem[{Walter {et~al.}(2003)Walter, Bertoldi, Carilli, Cox, Lo, Neri, Fan,
  Omont, Strauss, \& Menten}]{Walter2003}
Walter, F., Bertoldi, F., Carilli, C., {et~al.} 2003, Nature, 424, 406

\bibitem[{Walter {et~al.}(2004)Walter, Carilli, Bertoldi, Menten, Cox, Lo, Fan,
  \& Strauss}]{Walter2004}
Walter, F., Carilli, C., Bertoldi, F., {et~al.} 2004, ApJ, 615, L17

\bibitem[{Walter {et~al.}(2022)Walter, Neeleman, Decarli, Venemans, Meyer,
  Weiss, Ba{\~{n}}ados, Bosman, Carilli, Fan, Riechers, Rix, \&
  Thompson}]{Walter2022}
Walter, F., Neeleman, M., Decarli, R., {et~al.} 2022, ApJ, 927, 21

\bibitem[{Wandel {et~al.}(1999)Wandel, Peterson, \& Malkan}]{Wandel1999}
Wandel, A., Peterson, B.~M., \& Malkan, M.~A. 1999, ApJ, 526, 579

\bibitem[{Wang {et~al.}(2019)Wang, Yang, Fan, Wu, Yue, Li, Bian, Jiang,
  Ba{\~{n}}ados, Schindler, Findlay, Davies, Decarli, Farina, Green, Hennawi,
  Huang, Mazzuccheli, McGreer, Venemans, Walter, Dye, Lyke, Myers, \&
  Nunez}]{Wang2019}
Wang, F., Yang, J., Fan, X., {et~al.} 2019, ApJ, 884, 30

\bibitem[{Wang {et~al.}(2018)Wang, Yang, Fan, Yue, Wu, Schindler, Bian, Li,
  Farina, Ba{\~{n}}ados, Davies, Decarli, Green, Jiang, Hennawi, Huang,
  Mazzucchelli, McGreer, Venemans, Walter, \& Beletsky}]{Wang2018}
Wang, F., Yang, J., Fan, X., {et~al.} 2018, ApJ, 869, L9

\bibitem[{Wang {et~al.}(2010)Wang, Carilli, Neri, Riechers, Wagg, Walter,
  Bertoldi, Menten, Omont, Cox, \& Fan}]{Wang2010}
Wang, R., Carilli, C.~L., Neri, R., {et~al.} 2010, ApJ, 714, 699

\bibitem[{Wang {et~al.}(2011)Wang, Wagg, Carilli, Neri, Walter, Omont,
  Riechers, Bertoldi, Menten, Cox, Strauss, Fan, \& Jiang}]{Wang2011}
Wang, R., Wagg, J., Carilli, C.~L., {et~al.} 2011, AJ, 142, 101

\bibitem[{Wang {et~al.}(2013)Wang, Wagg, Carilli, Walter, Lentati, Fan,
  Riechers, Bertoldi, Narayanan, Strauss, Cox, Omont, Menten, Knudsen, Neri, \&
  Jiang}]{Wang2013}
Wang, R., Wagg, J., Carilli, C.~L., {et~al.} 2013, ApJ, 773, 44

\bibitem[{Waskom(2021)}]{Waskom2021}
Waskom, M. 2021, JOSS, 6, 3021

\bibitem[{Willott {et~al.}(2010)Willott, Albert, Arzoumanian, Bergeron,
  Crampton, Delorme, Hutchings, Omont, Reyl{\'{e}}, \& Schade}]{Willott2010}
Willott, C.~J., Albert, L., Arzoumanian, D., {et~al.} 2010, AJ, 140, 546

\bibitem[{Willott {et~al.}(2015)Willott, Bergeron, \& Omont}]{Willott2015}
Willott, C.~J., Bergeron, J., \& Omont, A. 2015, ApJ, 801, 123

\bibitem[{Willott {et~al.}(2017)Willott, Bergeron, \& Omont}]{Willott2017}
Willott, C.~J., Bergeron, J., \& Omont, A. 2017, ApJ, 850, 108

\bibitem[{Willott {et~al.}(2009)Willott, Delorme, ReylÃ©, Albert, Bergeron,
  Crampton, Delfosse, Forveille, Hutchings, McLure, Omont, \&
  Schade}]{willott_2009}
Willott, C.~J., Delorme, P., ReylÃ©, C., {et~al.} 2009, AJ, 137, 3541

\bibitem[{Willott {et~al.}(2005)Willott, Percival, McLure, Crampton, Hutchings,
  Jarvis, Sawicki, \& Simard}]{willott_2005}
Willott, C.~J., Percival, W.~J., McLure, R.~J., {et~al.} 2005, ApJ, 626, 657

\bibitem[{Wisnioski {et~al.}(2018)Wisnioski, Mendel, Schreiber, Genzel, Wilman,
  Wuyts, Belli, Beifiori, Bender, Brammer, Chan, Davies, Davies, Fabricius,
  Fossati, Galametz, Lang, Lutz, Nelson, Momcheva, Rosario, Saglia, Tacconi,
  Tadaki, Übler, \& van Dokkum}]{Wisnioski2018}
Wisnioski, E., Mendel, J.~T., Schreiber, N. M.~F., {et~al.} 2018, ApJ, 855, 97

\bibitem[{Wu {et~al.}(2015)Wu, Wang, Fan, Yi, Zuo, Bian, Jiang, McGreer, Wang,
  Yang, Yang, Thompson, \& Beletsky}]{wu_2015}
Wu, X.-B., Wang, F., Fan, X., {et~al.} 2015, Nature, 518, 512

\bibitem[{Yang {et~al.}(2023{\natexlab{a}})Yang, Fan, Gupta, Myers,
  Palanque-Delabrouille, Wang, Yèche, Aguilar, Ahlen, Alexander, Brooks,
  Dawson, de~la Macorra, Dey, Dhungana, Fanning, Font-Ribera, Gontcho, Guy,
  Honscheid, Juneau, Kisner, Kremin, Guillou, Levi, Magneville, Martini,
  Meisner, Miquel, Moustakas, Nie, Percival, Poppett, Prada, Schlafly, Tarlé,
  Magana, Weaver, Wechsler, Zhou, Zhou, \& Zou}]{Yang2023}
Yang, J., Fan, X., Gupta, A., {et~al.} 2023{\natexlab{a}}
  [\eprint[arXiv]{2302.01777}]

\bibitem[{Yang {et~al.}(2021)Yang, Wang, Fan, Barth, Hennawi, Nanni, Bian,
  Davies, Farina, Schindler, Ba{\~{n}}ados, Decarli, Eilers, Green, Guo, Jiang,
  Li, Venemans, Walter, Wu, \& Yue}]{Yang2021a}
Yang, J., Wang, F., Fan, X., {et~al.} 2021, ApJ, 923, 262

\bibitem[{Yang {et~al.}(2023{\natexlab{b}})Yang, Wang, Fan, Hennawi, Barth,
  Bañados, Sun, Liu, Cai, Jiang, Li, Onoue, Schindler, Shen, Wu, Bhowmick,
  Bieri, Blecha, Bosman, Champagne, Colina, Connor, Costa, Davies, Decarli,
  De~Rosa, Drake, Egami, Eilers, Evans, Farina, Habouzit, Haiman, Jin, Jun,
  Kakiichi, Khusanova, Kulkarni, Loiacono, Lupi, Mazzucchelli, Pan, Rojas-Ruiz,
  Strauss, Tee, Trakhtenbrot, Trebitsch, Venemans, Vestergaard, Volonteri,
  Walter, Xie, Yue, Zhang, Zhang, \& Zou}]{Yang2023a}
Yang, J., Wang, F., Fan, X., {et~al.} 2023{\natexlab{b}}, A SPectroscopic
  survey of biased halos In the Reionization Era (ASPIRE): A First Look at the
  Rest-frame Optical Spectra of $z > 6.5$ Quasars Using JWST

\bibitem[{Zakamska \& Greene(2014)}]{Zakamska2014}
Zakamska, N.~L. \& Greene, J.~E. 2014, MNRAS, 442, 784

\bibitem[{Übler {et~al.}(2023)Übler, Maiolino, Curtis-Lake, Pérez-González,
  Curti, Arribas, Charlot, Perna, Marshall, D'Eugenio, Scholtz, Bunker,
  Carniani, Ferruit, Jakobsen, Rix, Pino, Willott, Böker, Cresci, Jones,
  Kumari, \& Rawle}]{Uebler2023}
Übler, H., Maiolino, R., Curtis-Lake, E., {et~al.} 2023, A\&A
  [\eprint[arXiv]{2302.06647}]

\end{thebibliography}

% Don't change these lines
%\bsp	% typesetting comment
%\label{lastpage}
\end{document}